\documentclass[11pt,cite,epsfig,psfrag]{article}
\usepackage[utf8]{inputenc}
\usepackage{placeins}
\usepackage{geometry}
\usepackage{graphicx}
\usepackage{subfig}
\usepackage{amsmath} 
\graphicspath{{Images_wgc/}}
\usepackage[usenames, dvipsnames]{color}
\usepackage{wrapfig}
\usepackage{boxedminipage}
\usepackage{multirow}

\usepackage{fullpage}
\usepackage{amsmath}
\usepackage{amssymb}
\usepackage{graphicx}
\usepackage{mathrsfs}
\usepackage{wrapfig}
\usepackage{boxedminipage}
\usepackage{epsfig}
\usepackage{setspace}
\usepackage{float}
\usepackage{hyperref}
\usepackage{enumerate}
\usepackage{cite}
\usepackage[font=footnotesize,labelfont=rm]{caption}

\usepackage[numbers,sort&compress]{natbib}

\def\be{\begin{equation}}
	\def\ee{\end{equation}}
\def\bea{\begin{eqnarray}}
	\def\eea{\end{eqnarray}}

\numberwithin{equation}{section}

\newcommand{\RN}[1]{%
	\textup{\uppercase\expandafter{\romannumeral#1}}%
}
\begin{document}
	
	\thispagestyle{empty}

	\vskip 2cm

	\begin{center}
		{\Large \bf Off-shell phase diagram of BPS black holes in AdS\textsubscript{5} }
	\end{center}

	\vskip .2cm
	
	\vskip 1.2cm


\centerline{ \bf   Debabrata Sahu \footnote{a22ph09006@iitbbs.ac.in} and  Chandrasekhar Bhamidipati\footnote{chandrasekhar@iitbbs.ac.in}}

\vskip 7mm 

	\begin{center}{ Department of Physics, School of Basic Sciences\\ 
			Indian Institute of Technology Bhubaneswar \\ Bhubaneswar, Odisha, 752050, India}
	\end{center}

	\vskip 1.2cm
	\vskip 1.2cm
	\centerline{\bf Abstract}
	\vskip 0.5cm

The off-shell free energy of supersymmetric black holes in AdS$_5$ is constructed and the phase diagram in various limiting cases is investigated, including in the presence of higher derivative corrections, with particular emphasis on BPS thermodynamics.  Starting from Landau's theory, an exact method is systematically developed to construct the off shell BPS free energy, which in certain limiting cases, can be rearranged in terms of an effective energy and entropy of the system, with the latter being conjugate to an effective BPS temperature. The off-shell BPS phase diagram shows features which resemble the phases of general AdS Schwarzschild black holes, with some nuances in the asymptotic structure, modified by four-derivative corrections. Using AdS/CFT, phenomenological effective potentials in the boundary gauge theory are proposed, dual to both general black holes and their BPS counterparts. The saddle points of the effective potential capture the various locally stable and unstable phases of the gauge theory at finite temperature and chemical potential.

	\newpage
	\setcounter{footnote}{0}
	\noindent
	
	\baselineskip 15pt
	\section{Introduction}

Investigations of black hole thermodynamics and the microscopic interpretation of entropy of BPS black holes has been a challenging arena of research with impressive advances being made from time to time, and receiving continuous attention~\cite{Bekenstein:1973ur,Hawking:1975vcx,Hawking:1982dh,Strominger:1996sh,Chamblin:1999tk,Caldarelli:1999xj,Silva:2006xv,Sen:2007qy,Benini:2015noa,Almheiri:2016fws,Hosseini:2016tor,Benini:2015eyy,Hosseini:2017mds,Choi:2018vbz,Choi:2019miv,Larsen:2019oll,Copetti:2020dil,  Hosseini:2018dob,Hosseini:2018usu,Crichigno:2020ouj,Zaffaroni:2019dhb,Fu:2016vas,Iliesiu:2020qvm,Heydeman:2020hhw,Choi:2021nnq,Gutowski:2004ez,Gutowski:2004yv,Sen:2008yk,Sen:2008vm,Dabholkar:2010rm,Chowdhury:2024ngg}. Driven by AdS/CFT correspondence, significant work is going on in asymptotically anti-de Sitter black holes, which has also fueled the study of superconformal field theories with their holographic duals, which might be the key to a deeper understanding of microscopics of black holes in AdS~\cite{Benini:2016hjo,Benini:2016rke,Choi:2018hmj,Cabo-Bizet:2018ehj,Benini:2018ywd,ArabiArdehali:2019tdm,Honda:2019cio,Cabo-Bizet:2019osg,Kim:2019yrz,Cabo-Bizet:2019eaf,Amariti:2019mgp,GonzalezLezcano:2019nca,Lanir:2019abx,Goldstein:2019gpz,ArabiArdehali:2019orz,Murthy:2020scj,Agarwal:2020zwm,Larsen:2020lhg,Benini:2020gjh,Cabo-Bizet:2020nkr,Cabo-Bizet:2021plf,Cassani:2021fyv,Jejjala:2021hlt,Jejjala:2022lrm,Aharony:2021zkr,Cabo-Bizet:2021jar,Goldstein:2020yvj, Choi:2021rxi,Choi:2023tiq,Cassani:2024tvk}. Some recent insights from the gravity side of the development correspond to realizing the importance of a non-linear relation on the BPS charges and a complex constraint on the dual potentials, which give novel formulations of the BPS free energy \cite{Hosseini:2017mds,Cabo-Bizet:2018ehj}. The phase diagram emerging from the free energy is quite intriguing, as the phases of the system closely mimic the standard structure known from black holes in AdS, but on a closer look reveal considerable differences as well, especially due to the special role played by the rotation parameter~\cite{Ezroura:2021vrt}. Supersymmetry plays a crucial role in getting closed expressions for thermodynamic quantities in a number of cases, particularly because studying the system in the presence of both charge and rotation parameters at generic point in the parameter space is quite non-trivial. There are also new limiting cases which emerge at special points in the parameter space where the potentials dual to charges take certain values leading to new phases~\cite{Larsen:2019oll,Larsen:2020lhg,Choi:2018vbz,Choi:2018hmj,Ezroura:2021vrt}. \\

\noindent
An interesting situation arises for supersymmetric black holes, when one tries to study the phase diagram in the BPS limit. Naively, one expects the free energy to vanish identically as the variables on which it depends, such as, the temperature and potentials dual to the charges, are all constrained and fixed to certain critical numerical values. However, there is a possibility to study the system in the near-BPS limit~\cite{Almheiri:2014cka,Kitaev,Maldacena:2016upp, Jensen:2016pah,Engelsoy:2016xyb,Larsen:2018iou,Nayak:2018qej,Kolekar:2018sba,Moitra:2018jqs,Iliesiu:2020qvm,Heydeman:2020hhw,Castro:2021fhc,Castro:2021wzn,Larsen:2021wnu}, where either the temperature or the potentials are close (but not exactly equal) to their critical values, leading to the construction of a novel free energy, which is intrinsic to the BPS surface. 
The thermodynamics ensuing from this BPS free energy can be studied by varying a certain ``BPS temperature", assembled out of the BPS potentials and can be treated as an analogue of the physical temperature. The phase structure from the BPS free energy follows the familiar pattern of AdS Schwarzschild black holes, with a small and large black hole branch, separated by Hawking-Page like transition point~\cite{Hawking:1982dh,Kubiznak:2016qmn}. However,  there are some differences, e.g., as the BPS temperature is varied, the asymptotic behavior of the free energy for the large black hole branch is different when compared to AdS Schwarzschild black holes. Thus, the connection between the BPS temperature and the physical temperature is not precise, though the analogy is quite intriguing and useful\cite{Choi:2018vbz,Choi:2018hmj,Ezroura:2021vrt}. It is natural to ponder whether the thermodynamic conjugate to the ``BPS temperature", namely, an effective ``entropy'' exists, together with an effective energy, and their possible connection with the actual BPS entropy and mass of the black hole. This might be helpful in our efforts in seeking a precise connection of thermodynamics in the BPS limit to the traditional black hole thermodynamics. \\

\noindent
On the dual gauge theory side, the weak coupling limit of SU(N) Yang-Mills theory on S$^3 \times R$ and AdS black holes have thermodynamical features common to each other, with some qualitative and quantitative differences~\cite{Hawking:1982dh, Witten:1998zw, Sundborg:1999ue,  Aharony:2003sx}. A natural question which has been pursued is, whether the physics of BPS black holes in AdS can be understood from the dual gauge theory side, in particular in a weakly coupled CFT, e.g. with the help of supersymmetric indices. There are however some unresolved puzzles which have received continuous attention\cite{Kinney:2005ej,Romelsberger:2005eg,Closset:2013vra,Assel:2014paa,Cabo-Bizet:2018ehj, Berkooz:2006wc,Grant:2008sk, Chang:2013fba}. In this context, the HHZ free energy~\cite{Hosseini:2017mds} has led to considerable progress in the microscopic understanding of BPS black hole physics in AdS~\cite{Cabo-Bizet:2018ehj,Choi:2018hmj,Benini:2018ywd,Choi:2018vbz,Choi:2019miv}. Although, the BPS free energy of AdS black holes and the HHZ free energy of the gauge theory have very different origins, it is intriguing to seek the connection between the two. Some challenges in this case arise because the latter is a genuine function of complex potentials (which satisfy a constraint) of the boundary CFT, and the former is real quantity with a definite spacetime interpretation. It is thus not straightforward to relate the two quantities, but they do seem to match when the constraint satisfied by the complex potentials takes a special value, which is also when the BPS partition function matches the superconformal index\cite{Choi:2018vbz,Choi:2018hmj,Ezroura:2021vrt}.  Thus,  the extrapolation of results at weak coupling to strong coupling CFT can be trusted, with the latter being related by duality to the semiclassical black hole thermodynamics. There are other proposals on matching the bulk and boundary free energies~\cite{Hosseini:2016tor,Liu:2017vbl,Choi:2018vbz,Choi:2018hmj,Larsen:2019oll,Copetti:2020dil}(see also e.g., the review in~\cite{Zaffaroni:2019dhb}). \\

\noindent
When the gravitational physics is more accessible, the AdS/CFT correspondence has been exploited several times in the past to compute the partition function of the dual theory in the strongly coupled regimes, which are otherwise not accessible by other methods\cite{Larsen:2019oll, Larsen:2020lhg, Drukker:2010nc, Aharony:2003sx}.
In this work, our intent is to follow this route and strengthen the connection between the gravitational thermodynamics and phases of boundary CFT at strong coupling, by using certain off-shell methods. In particular, the plan is to obtain an effective potential for the boundary CFT at finite temperature and chemical potential, which on-shell yields the various phases one expects from the AdS/CFT correspondence. In the current set up, the gauge theories are dual to supersymmetric charged and rotating black holes in AdS$_5$. Considering the non-availability of a direct method of working with strongly coupled gauge theories, the plan is to first construct an off-shell free energy from the bulk gravitational thermodynamics, and then use the AdS/CFT rules to arrive at an effective potential in the boundary gauge theories. Off-shell methods have been known to give the connections between thermodynamical entropy of black holes defined by the first law, and the statistical mechanical entropy. At a general finite temperature, which differs from the Hawking temperature, it is known that a conical singularity appears in the classical solutions of gravitational field equations. The Euclidean gravitational path integral can be evaluated in the saddle point approximation taking the euclidean black hole instanton (with a conical singularity) as a the classical solution, leading to consistent approach to the computation of entropy\cite{Susskind:1993ws,Banados:1993qp,Carlip:1993sa,Teitelboim:1994is,Fursaev:1995ef,Mann:1996bi,Solodukhin:1994yz,Fursaev:1994te,Solodukhin:1995ak,Frolov:1995xe,Frolov:1996hd,Bytsenko:1997ru,Solodukhin:2011gn}. The same method also gives rise to an off-shell free energy for the black hole, but the temperature is now a free arbitrary parameter. There is a special value of the temperature, namely the Hawking temperature, where the deficit angle and conical singularity disappear, connecting the off-shell and the on-shell formalisms for free energies. \\

\noindent
Instead of the gravitational analysis broached above, a convenient route which expeditiously leads to an off-shell formalism is a mean field approach motivated from the study of phase transitions in general thermodynamical systems. The idea is as follows. Within the Grand canonical ensemble, one can construct an off-shell free energy, by taking temperature and the other chemical potentials as free variables, with one of the parameters taken as the order parameter. This is the Bragg-Williams formalism, and works for any thermodynamic system with phase transitions, particularly, when the order parameter is large and jumps discontinuously~\cite{cha95,kubo,bw1,bw2}. Though, the free energy obtained this way is not unique, it quite effectively captures all the equilibrium  phases of the system. In the case of black holes, the relevant order parameter is the horizon radius.  The equilibrium value of the order parameter minimizes the free energy and gives rise to the required phase structure. 
While the Bragg-Williams formalism has been applied earlier to study instabilities of a variety of non-supersymmetric black holes in AdS~\cite{Dey:2007vt,Dey:2006ds,nayak2008bragg,Banerjee:2010ve,Banerjee:2010ng}, with interesting recent applications~\cite{Yerra:2022coh,Wei:2022dzw}, analogous construction for supersymmetric black holes in AdS$_5$ with charge and rotation parameters has not been investigated thus far. This is important due to possibility of investigating the BPS phase diagram in more detail. In this work, we plan to fill this gap by proposing off-shell free energy for such systems. We also put forward proposals for an effective potential in the gauge theory dual to the supersymmetric black hole in bulk, with particular emphasis on the BPS black hole thermodynamics~\cite{Silva:2006xv,Ezroura:2021vrt}. \\

\noindent
Following the above investigations, in this paper we study the phase diagram of general black holes in AdS$_5$, as well as, cases resulting from taking the BPS limit, in the presence of four-derivative corrections to entropy and other charges coming from a modified action~\cite{Cassani_2022,Bobev_2022}. Theories of quantum gravity at low energies are expected to be represented by an action with a universal two-derivative term, corrected by series of higher derivative terms. Investigating the effect of such corrections to black hole thermodynamics is important  and may assist us in getting insights into the full UV complete theory\cite{Adams:2006sv,Arkani-Hamed:2006emk}. 
Several steps have been taken in this direction, with some recent ones focusing on higher derivative corrections to the thermodynamics of AdS black holes in minimal five-dimensional gauged supergravity~\cite{Baggio:2014hua,Bobev:2021qxx,Liu:2022sew,Melo:2020amq,Bobev:2020egg,Bobev:2021oku,Genolini:2021urf,Cassani_2022,Bobev_2022}. As mentioned above, much progress has been achieved by analysing the field theory partition function, which counts the microstates of the AdS black holes, in several different methods~\cite{Cabo-Bizet:2018ehj,Choi:2018hmj,Benini:2018ywd,Zaffaroni:2019dhb}, via superconformal index. The large $N$ expansion of the index should give us information about the higher derivative corrections to the two-derivative action, at least, order by order, though this is not easy\cite{Benini:2018ywd,Cabo-Bizet:2019eaf,ArabiArdehali:2019orz,Cabo-Bizet:2020nkr,Aharony:2021zkr,Cabo-Bizet:2018ehj,Choi:2018hmj,Kim:2019yrz,Cabo-Bizet:2019osg,Cassani:2021fyv,ArabiArdehali:2021nsx,GonzalezLezcano:2020yeb,Amariti:2021ubd,Ohmori_2022,Hosseini:2017mds}. We follow the recent approach of~\cite{Cassani_2022,Bobev_2022}, and consider their higher derivative corrected on-shell action to write an off-shell free energy of general non-supersymmetric black holes in AdS$_5$\cite{Chong:2005hr}. It was argued in~\cite{Cassani_2022,Bobev_2022}, along lines of ~\cite{Reall:2019sah} that, the full corrected black holes solution of four-derivative theory is not necessary and it is sufficient to obtain the corrections to on-shell action by evaluating on the uncorrected black hole solution following from the two-derivative theory. This has the effect of correcting the charges to linear order in the higher derivative coupling $\alpha$, while leaving the chemical potentials, as well as the temperature uncorrected. Though this is convenient, due to its inherent limitations in the way higher derivative terms are treated, our analysis, following the set up in~\cite{Cassani_2022,Bobev_2022} is not rigorous, and will only qualitatively capture  the corrections to the phase diagrams, coming from small variations in the higher derivative couplings. The general expressions for the charges were computed in ~\cite{Cassani_2022,Bobev_2022} and for the case of equal angular momenta are summarized in appendix-(\ref{A}).   \\

\noindent
Rest of the paper is organized as follows. In section-(\ref{two}), we present our construction of off-shell free energy for general black holes in AdS$_5$, which follow from the analysis of two derivative theory. Subsection-(\ref{2.1}) is introductory and contains key formulas of black hole thermodynamics in AdS$_5$~\cite{Chong:2005hr}, necessary for our purposes.  The main result of off-shell free energy for the two derivative case is given in subsection-(\ref{2.2}). Subsequent subsections contain results for the phase diagram in various limiting cases, corresponding to: non-rotating black holes  (subsection-\ref{2.2.1}), subcritical electric potential (subsection-\ref{2.2.2}), vanishing electric potential  (subsection-\ref{2.2.3}),  critical electric potential (subsection-\ref{2.2.4}) and maximal rotational velocity (subsection-\ref{2.2.5}).  In section-(\ref{three}), we study the corrections to the off-shell free energy of general black holes in AdS$_5$ coming from four derivative corrected minimal five-dimensional gauged supergravity action studied in~\cite{Chong:2005hr,Cassani_2022,Bobev_2022}. We also qualitatively study the changes to phase diagram. The following section-(\ref{offshell}) is devoted to the study of phase diagram of BPS black holes in AdS$_5$. After a brief introduction of relevant expressions of on-shell free energy in subsection-(\ref{onshell}), we attempt the construction of an off-shell BPS free energy starting from Landau's phenomenological approach to phase transitions in subsection-(\ref{alternative1}). In the following
subsection-(\ref{alternative}) we present an exact method to obtain the off-shell BPS free energy, which is then used in the construction of off-shell BPS phase diagram for both two-derivative and four derivative cases, in subsections-(\ref{2d}) and (\ref{4d}), respectively.  Section-(\ref{2.4}) contains our proposal for the boundary effective potential followed by a discussion of its equilibrium phases, which are constructed from the bulk free energy of general black holes in AdS$_5$, as well as, in their BPS limits, including in the presence of four-derivative corrections. Appendices-(\ref{A}) and (\ref{B}) contain the four-derivative corrected expressions for charges, the on-shell phase diagram for general black holes, and their BPS counterparts. The results in the appendices serve as a consistency check of the phase diagrams obtained through our off-shell methods in the main text. We end with a discussion of results and conclusions in section-(\ref{untilnexttime}).

	\section{ Off-shell Free energy of AdS$_5$ black holes } \label{two}
	
\noindent 
In subsection-(\ref{2.1}), we start with a brief review of AdS$_5$ black hole solution to minimal five-dimensional gauged supergravity given in~\cite{Chong:2005hr}. We collect the features  important for our discussion, referring to the original works for further details~\cite{Cabo-Bizet:2018ehj,Ezroura:2021vrt}. The main construction of off-shell free energy is given in subsection-(\ref{2.2}) and phase diagram in various limiting cases, corresponding to various values taken by the electric potential and angular velocity is presented in the sequel. 
	
	\subsection{ General AdS\textsubscript{5} black hole thermodynamics } \label{2.1}

The two-derivative action of minimal five-dimensional gauged supergravity is taken to be~\cite{Chong:2005hr}
	\begin{equation}\label{eq:2daction}
		S \,=\, \frac{1}{16\pi G}\int d^5x \,e \left[R+12g^2-\frac{1}{4}F^2-\frac{1}{12\sqrt{3}}\epsilon^{\mu\nu\rho\sigma\lambda}F_{\mu\nu}F_{\rho\sigma} A_{\lambda}\right]\, ,
	\end{equation}
where $A_\mu$ stands for the gauge field, with $F_{\mu\nu}=2\partial_{[\mu}A_{\nu]}$, and $F^2=F_{\mu\nu}F^{\mu\nu}$. The parameter $g$ controls the cosmological constant and is normalized in such way that the AdS solution is defined with radius $1/g$.
The details of the above theory can be found in\cite{Gauntlett:2007ma,Cassani:2019vcl} and the most general known asymptotically AdS$_5$ black hole solution was obtained in\cite{Chong:2005hr}. The solution is known to depend on four parameters, given as $m, a, b, q$ (where $a^2g^2<1, b^2g^2<1$). These respectively control  the four independent conserved charges, i.e., the energy $E$, the angular momenta $J_1, J_2\,$ which correspond to rotations, and the electric charge denoted as $Q$. The expressions of these charges are:
 \begin{equation}
    \begin{aligned}\label{CCLPcharges}
		 E &= \frac{m\pi (2\Xi_a +2\Xi_b - \Xi_a\,\Xi_b) +2\pi qabg^2(\Xi_a+\Xi_b)}{4G\Xi_a^2\,\Xi_b^2}\ ,\qquad Q = \frac{\sqrt 3\pi q}{4G \Xi_a\, \Xi_b}\ ,\\[2mm]
		&\ \  J_1 = \frac{\pi[2am + qb(1+a^2 g^2) ]}{4G \Xi_a^2\, \Xi_b}\ ,\qquad
		J_2 = \frac{\pi[2bm + qa(1+b^2 g^2) ]}{4G \Xi_b^2\, \Xi_a}\ .   
		\end{aligned} 
 \end{equation}
 where $\Xi_a = 1 - a^2 g^2$, $\Xi_b = 1 - b^2 g^2$.
 The dimensionless parameter that sets their scale is
 \begin{equation*}
     \frac{\pi \ell_5^3}{4 G} = \frac{1}{2}N^2,
      \end{equation*}
with $N$ being the gauge group of dual $SU(N)$ $\mathcal{N} = 4$ SYM. The only other dimensionful parameter that enters the formulas is the length scale of AdS\textsubscript{5}, which is connected to the coupling of gauged supergravity,  $g = \ell_5^{-1}$. In the following sections, we shall set $\ell_5=g=1$, while remembering that, upon restoring dimensions, $a,b$ go as lengths, with $E,Q$ going as inverse lengths, and $m,q$ being length squared. \\

\noindent
The position of the outer event horizon, $r=r_+$ can be seen to be the largest positive root of
	\begin{equation}\label{r_m}
		\Delta_r(r)= \frac{(r^2 + a^2)(r^2 + b^2)(1+r^2)+q^2+2abq}{r^2} - 2m = 0 \, ,
	\end{equation}
	which leads us to the expressions of the thermodynamic potentials, such as, the inverse temperature $T$, the angular velocities $\Omega_1,\Omega_2$, and the electrostatic potential $\Phi$, which are given respectively as
	\be\label{temperature_CCLP}
	T \equiv\beta^{-1} = \frac{r_+^4[(1+ g^2(2r_+^2 + a^2+b^2)] -(ab + q)^2}{2\pi\,
		r_+\, [(r_+^2+a^2)(r_+^2+b^2) + abq]}\ ,
	\ee
	\be\label{angular_velocities_CCLP}
	\Omega_1 = \frac{a(r_+^2+ b^2)(1+g^2 r_+^2) + b q}{
		(r_+^2+a^2)(r_+^2+b^2)  + ab q}\ ,\qquad
	\Omega_2 = \frac{b(r_+^2+ a^2)(1+g^2 r_+^2) + a q}{
		(r_+^2+a^2)(r_+^2+b^2)  + ab q}\ ,
	\ee
	\be\label{electrostatic_pot_CCLP}
	\Phi = \frac{\sqrt{3}\, q \,r_+^2}{(r_+^2 + a^2)(r_+^2 + b^2)+abq}\, .
	\ee
The Bekenstein-Hawking entropy is
	\be\label{entropyCCLP}
	{\cal S}= \frac{\rm Area}{4G} =\frac{\pi^2 [(r_+^2 +a^2)(r_+^2 + b^2) +a b q]}{2G\Xi_a \Xi_b r_+}
	\ .
	\ee
	The thermodynamic quantities given above obey the first law,
	\begin{equation}\label{2.9}
	   dE = T d{\cal S} + \Omega_1\, dJ_1+ \Omega_2 \,dJ_2 + \Phi\, dQ\ ,  
	\end{equation}
	and also the quantum statistical relation
	\be\label{2.10}
	I = \beta E - {\cal S} - \beta \Omega_1J_1 - \beta\Omega_2J_2 - \beta\Phi Q\ ,
	\ee
	with $I$ standing for  the Euclidean on-shell action~ \cite{PhysRevD.15.2752}, which when appropriately renormalised~\cite{chen2006mass,de_Haro_2001,Bianchi_2002}, takes the form~\cite{Cassani_2019}:
		\be\label{2.11}
	I = \frac{\pi\beta}{4G\Xi_a\Xi_b}
	\Big[m - g^2 (r_+^2 + a^2)(r_+^2 + b^2) -
	\frac{q^2 r_+^2}{(r_+^2 + a^2)(r_+^2 + b^2)+abq}\Big]\ .
	\ee	
	From the two equations in (\ref{2.9}) and (\ref{2.10}), one gets the relations,
	\be\label{2.12}
	E = \frac{\partial I}{\partial\beta}\ ,\qquad J_1 = -\frac{1}{\beta}\frac{\partial I}{\partial\Omega_1}\ ,\qquad J_2 = -\frac{1}{\beta}\frac{\partial I}{\partial \Omega_2} \ ,\qquad Q = -\frac{1}{\beta}\frac{\partial I}{\partial\Phi}\ .
	\ee
Thus, the saddle of the grand-canonical partition function is related to the on-shell action as, $I = -\log Z_{\rm grand}$ (which is a function of the thermodynamic potentials, i.e.,  $\beta,\Omega_1, \Omega_2,\Phi$).
  The solution turns out to be supersymmetric when the parameters are related as:\cite{Chong:2005hr}
 \be\label{qm}
 q = \frac{m}{1+ag+bg}
\ee	
For supersymmetric black holes, there is an additional condition which needs to be satisfied, given as
\be\label{qa}
q = \frac{1}{g}(a+b)(1+ag)(1+bg)
\ee
Only after one imposes both the conditions given in eqns. (\ref{qm}) and (\ref{qa}) the black holes become both supersymmetric as well as extremal, called as ``BPS" black holes, and there are various ways to achieve extremality~\cite{Cabo-Bizet:2018ehj}. In the BPS limit, for convenience, the quantities are denoted by a $*$ symbol. For instance, the horizon radius in the BPS limit is 
\be
r_* = \sqrt{\frac{1}{g}(a+b+abg)} \, ,
\ee
and the corresponding chemical potentials are:
\be\label{leadingterms_chempot}
\beta \to \infty\ ,\qquad \Omega_1\to \Omega^*_1 = g\ ,\qquad \Omega_2\to \Omega^*_2 = g\ ,\qquad \Phi \to \Phi^* = \sqrt{3}\ .
\ee
One notes that the special values taken by potentials, namely, $\Omega_1^*,\Omega_2^*$ and $\Phi^*$ appear precisely in the superalgebra,
\be\label{superalgebra}
\{\mathcal{Q},\overline{\mathcal{Q}}\} \,\propto\, E - g J_1 - g J_2 - \sqrt{3}\, Q\,. 
\ee
One further notes from eqn. \eqref{superalgebra}, that in supersymmetric solutions a linear relation is satisfied by the conserved charges as
\be\label{susyrel_charges}
E^* - g J_1^* - g J_2^* - \sqrt{3}\, Q^* = 0\,.
\ee
One can now study the phase diagram in the grand canonical ensemble, by writing the Gibb's free energy as a function of temperature ($T$), electric potential ($\Phi$), and angular velocities ($\Omega_{a,b}$) as $ G (T,\Omega_a,\Omega_b,\Phi) = M - TS - \Phi Q - \Omega_a J_a - \Omega_b J_b$. 
This is the on-shell formalism and the phase structure of general black holes in AdS, as well as BPS thermodynamics has been investigated in detail for various values of the potentials~\cite{Ezroura:2021vrt}. We now go on to present the off-shell Bragg-Williams (BW) formalism for black holes in AdS$_5$.

\subsection{Off-shell free energy in the grand canonical ensemble} \label{2.2}

In this subsection we set up the Bragg-William (BW) formalism in the grand canonical ensemble, i.e. as a
function of thermodynamic potentials. For clarity, we study the simplified AdS\textsubscript{5} black hole solutions with only one independent angular momenta by equating rotational parameters $a = b$,  and hence $J_1 = J_2 = J$, and $\Omega_1 = \Omega_2 = \Omega$. The said off-shell free energy can be constructed using the thermodynamic quantities in the previous subsection as \footnote{Throughout the paper we present free energy $\frac{F}{N^2}$ with $N =1$ for simplification.} 
\begin{equation}\label{2.19}
    F = E - {T} {\cal S} - {\Phi} Q - 2{\Omega} J
\end{equation}
Compared to the Gibb's free energy, now the thermodynamic functions in the grand canonical ensemble, such as temperature $T$, electric potential $\Phi$, and angular velocity $\Omega$ are treated as free parameters, whose values are not yet fixed to the ones given eqn. (\ref{temperature_CCLP} -  \ref{electrostatic_pot_CCLP}). In this formalism, the on-shell values of the potentials given in 
eqn. (\ref{temperature_CCLP} - \ref{electrostatic_pot_CCLP}) are obtained from the free energy in eqn. (\ref{2.19}) through a minimisation procedure, which we discuss shortly. The free energy in eqn. (\ref{2.19}) is off-shell in this sense, and its on-shell value is expected to agree with the Gibb's free energy upon using the on-shell values of potentials. In the present case, we consider the horizon radius $r_+$ as the order parameter and we seek a function of the form $F(r_+,{T}, {\Phi}, {\Omega})$~\cite{Banerjee:2010ve,Yerra:2022coh,Yerra_2024}, the saddle points of which are expected to show the equilibrium phases of the system\cite{Banerjee:2010ve,Banerjee:2010ng}. 
To do so let us first trade $m$ for $r_+$ using (\ref{r_m}) as  
\begin{equation}\label{2.20}
		m = \frac{(r_+^2 + a^2)^2(1+r_+^2)+q^2+2a^2 q}{2 r_+^2}  \, .
	\end{equation}
The extensive variables $E,Q,J,{\cal S}$ presented in eqns. (\ref{CCLPcharges}) and (\ref{entropyCCLP}) now become functions of ($r_+, q, a )$.  After using the expressions of these extensive variables, the free energy in eqn. (\ref{2.19}) becomes
\begin{eqnarray}\label{2.21}
  &&      F(r_+, T, \Phi, \Omega) = \left(\frac{\pi}{8 \left(a^2-1\right)^3 G r_+^2}\right)  \Biggl(-\left(a^6 \left(r_+^2+4 \pi  r_+ T+1\right)\right)-3 q^2+r_+^4 \left(-3 r_+^2+4 \pi  r_+ T-3\right)  \nonumber \\
  &&  \hskip 2.5cm \left. + 4 a^5 \left(r_+^2+1\right) \text{$\Omega $}+4 a^3 \text{$\Omega $} \left((q+2) r_+^2+2 q+2 r_+^4\right)+4 a \text{$\Omega $} \left(q^2+q r_+^2+r_+^6+r_+^4\right)  \right. \nonumber \\
   && \hskip 2.5cm \left. + 2 \sqrt{3} q r_+^2 \text{$\Phi $}  -a^2 \left(q^2+2 q \left(r_+ \left(\sqrt{3} r_+ \text{$\Phi $}+4 r_+-2 \pi  T\right)+3\right) \right. \right. \nonumber \\
    &&  \hskip 2.5cm  \left. \left.  + r_+^2 \left(r_+^4 +4 \pi  \left(r_+^2-2\right) r_+ T+7 r_+^2+6\right)\right)  \right. \nonumber \\
    &&  \hskip 2.5cm  -  a^4 \left(4 \pi  r_+ T \left(q+2 r_+^2-1\right)+2 q+2 r_+^4+5 r_+^2+3\right)\Biggr) \, ,
\end{eqnarray}
 where we assumed that $q$ and $a$ can be eliminated in favour of $r_+$ and the potentials $\Phi$ and $\Omega$, using:
 \begin{equation}\label{2.23}
     q = \frac{\Phi(r_+^2 +a^2)^2}{\sqrt{3}r_+^2- a^2 \Phi}\, , \qquad 
    1- \Omega = \frac{1-a}{r_+^2+a^2}\left[r_+^2-r_*^2+a(1+a)(1-\frac{\Phi}{\sqrt{3}})\right] \, .
\end{equation}
The function $F(r_+,{T}, {\Phi}, {\Omega}) $ along with $r_+$ and the free parameters, take in general non-equilibrium values, and its saddle points represent  the equilibrium phases.  The second of the equations in (\ref{2.23}) is not analytically solvable for $a$ in most of the situations. We will thus illustrate the equations leading to the phase diagram  explicitly in a few cases, where the expressions are analytically tractable. In other cases, we use numerics and directly give the final results in plots. Following~\cite{Ezroura:2021vrt}, we now consider various cases, depending on the values of the potentials, such as,  non-rotating black holes ($\Omega=0$) with subcritical electric potential $ 0 < {\Phi} < \sqrt{3}$ (subsection-\ref{2.2.1}), non-zero $\Omega$ with subcritical electric potential $ 0 < {\Phi} < \sqrt{3}$, (subsection-\ref{2.2.2}), vanishing electric potential $\Phi=0$  (subsection-\ref{2.2.3}),  critical electric potential $\Phi = \sqrt{3}~ $(subsection-\ref{2.2.4}) and maximal rotational velocity ($\Omega=1$ )(subsection-\ref{2.2.5}). 



\subsubsection{Nonrotating ($\Omega = 0$) black holes, subcritical electric potential : $ 0 < {\Phi} < \sqrt{3}$} \label{2.2.1}

\begin{figure}[!htbp]
    \centering
   \subfloat[]{\includegraphics[scale = 0.62]{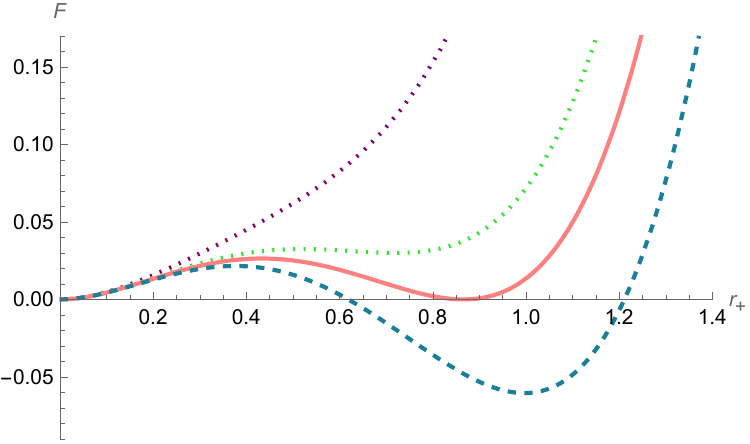}}~~~~~~~
    \subfloat[]{\includegraphics[scale = 0.38]{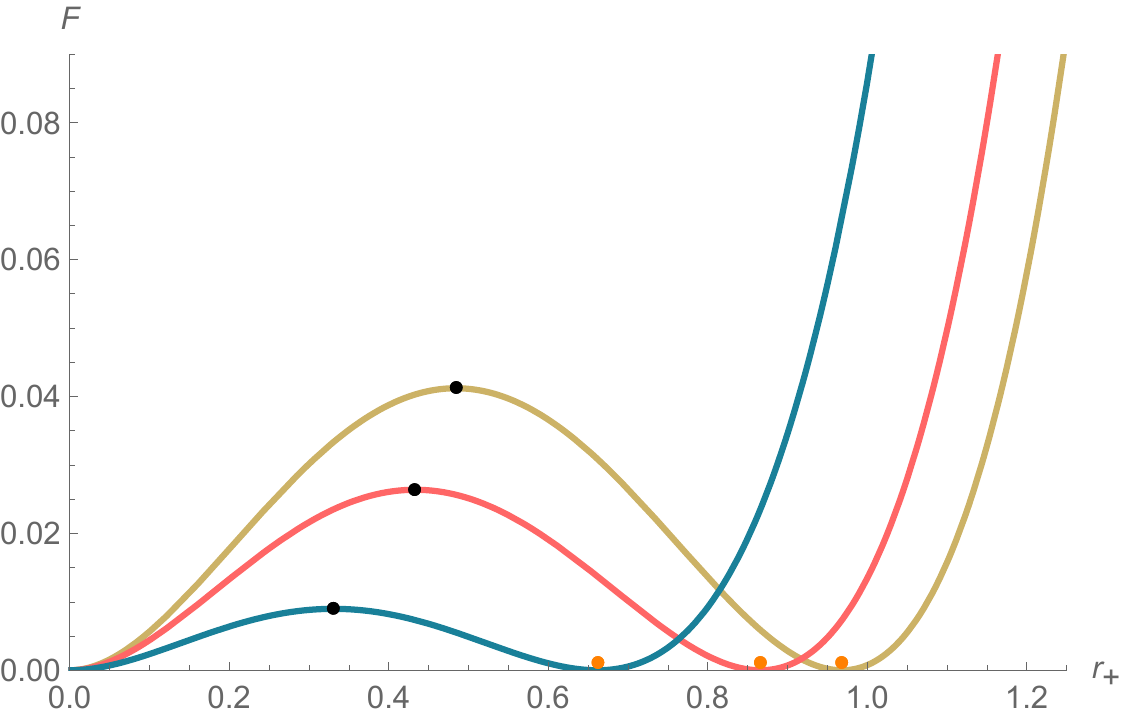}}
   \caption{ (a) For subcritical electric potential ${\Phi} = \frac{\sqrt{3}}{2}, {\Omega} = 0$ : Behaviour of BW free energy as a function of the order parameter $r_+$ at different temperatures ${T}$. Temperature of the curves increases from top to bottom. Hawking-Page transition happens at the temperature ${T}_{HP} = 0.413$ (solid red curve). (b) Behaviour of BW free energy as
a function of the order parameter $r_+$ for $\Omega = 0 $: The yellow, red, blue curves represent ${\Phi } = \frac{\sqrt{3}}{4}, \frac{\sqrt{3}}{2}, \frac{3\sqrt{3}}{4}$ with respective Hawking-Page temperatures $ {T}_{HP} = 0.462, 0.413, 0.315  $. }
   \label{fig1}
\end{figure}

For non rotating black holes with $\Omega = 0$, the Free energy in eqn. (\ref{2.21}) simplifies to 
\begin{equation} \label{offphi}
 F(r_+,{T}, {\Phi})= \frac{1}{4}r_+^2(3 +3 r_+^2-4 \pi r_+ T - \Phi^2)  \, ,
\end{equation}
where we used $a = 0$ and $q = \frac{\Phi r_+^2}{\sqrt{3}}$. Solving the minimisation condition $\partial_r\,F(r_+,T,\Phi)=0$ gives the on-shell value of temperature as
\begin{equation} \label{Tonphi}
    T = \frac{6 r_+^2-\Phi ^2+3}{6 \pi  r_+}\, ,
\end{equation}
which matches the expression in eqn. (\ref{temperature_CCLP}) in the appropriate limit. As a consistency check,  using this on-shell value of temperature in eqn. (\ref{offphi}), we get the  on-shell free energy of the black hole as
\be \label{fonphi}
F(r_+, {\Phi}) = -\frac{1}{12} r_+^2 \left(3 r_+^2+\Phi ^2-3\right) \, ,
\ee
The on-shell expressions for temperature and free energy in eqns. (\ref{Tonphi})  and (\ref{fonphi}) obtained from our off-shell formalism, match with the ones given in~\cite{Ezroura:2021vrt} (see their eqns. (2.39) and (2.40)), after appropriate scaling of $\Phi$ by a factor of $\sqrt{3}$. The first order Hawking-Page transition appears when the following two conditions
\bea
&& F =0 \, ,  \label{2.24} \\
&& \frac{\partial F}{\partial r_+} = 0\, ,  \label{dFzero}
\eea
are satisfied simultaneously. This happens at
\be \label{min1}
r_+= \frac{\sqrt{3-\Phi ^2}}{\sqrt{3}}\, , \qquad T_{\rm HP} = \frac{\sqrt{3} \sqrt{3-\Phi ^2}}{2 \pi }\, .
\ee
The phase diagram following from the off-shell free energy in eqn. (\ref{offphi}) is shown in figure-(\ref{fig1}).  
From figure-(\ref{fig1}(a)), the solid red line is at a critical temperature and has two degenerate minima representing the coexistence of AdS phase (with $r_+ = 0$) and  the black hole phase with minimum of $r_+$ given in eqn. (\ref{min1}) at $T=T_{\rm HP}$. While for $T>T_{\rm HP}$ the black hole phase is preferred (dashed blue curve), AdS is the preferred phase for $T<T_{\rm HP}$ (dotted green curve). From eq (\ref{Tonphi}) it is easy to show that the on shell temperature has a minima ($T_{min}$= 0.39 for $\Phi = \frac{\sqrt{3}}{2}$). Hence, for $T < T_{min}$ there is no black hole solution, which corresponds to the dotted violet curve of figure (\ref{fig1}(a)) as there are no local maxima or minima. 
\FloatBarrier
\noindent
From figure-(\ref{fig1}(b)), we note that the free energy plotted for increasing $\Phi$ (at their respective $T_{HP}$), the size of both small and large black holes are decreasing. 






\subsubsection{Subcritical electric potential :  $ 0 < {\Phi} < \sqrt{3}$ with $\Omega \neq 0$} \label{2.2.2}


\noindent
We now study the most general off-shell phase diagram of black holes with non-zero electric potential and angular velocity in various limits, following from the free energy in eqn. (\ref{2.21}). As the expressions are not analytically tractable, we present the plots of phase diagram directly. For the case 
of the sub critical limit $0 < {\Phi} < \sqrt{3}$ with a fixed value of $\Omega$ in the range $0 < {\Omega} < 1$, the phase diagram and discussion are qualitatively similar to the one given above from figure-(\ref{fig1}a) and hence, is not discussed separately. However, there are new features once the variation of $\Omega$ is considered. \\

\begin{figure}[!htbp]
    \centering
  {\includegraphics[width=0.5\linewidth]{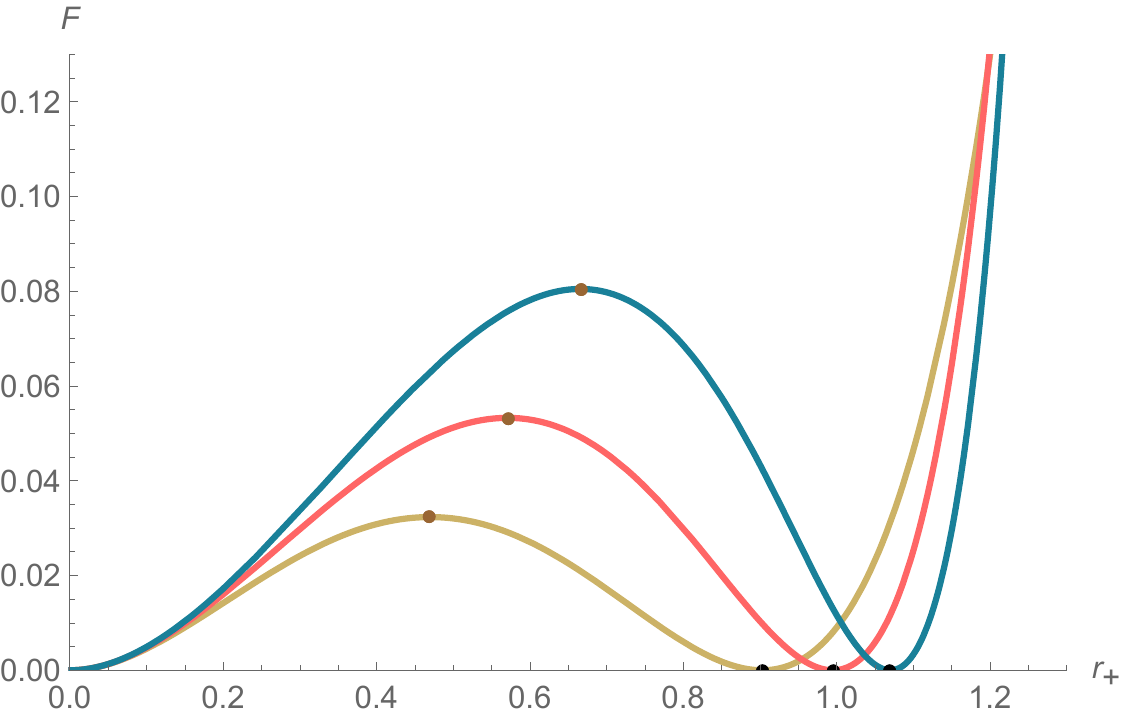}}
   \caption{Free energy as
a function of the order parameter $r_+$ for ${\Phi} = \frac{\sqrt{3}}{2} $: The yellow, red, blue curves represent ${\Omega} = 0.5, 0.8, 0,9$ with respective Hawking-Page temperatures $ {T}_{HP} = 0.372, 0.291, 0.242 $ .}
   \label{fig4}
\end{figure}

\begin{figure}[!htbp]
    \centering
  \subfloat[]{\includegraphics[width=0.5\linewidth]{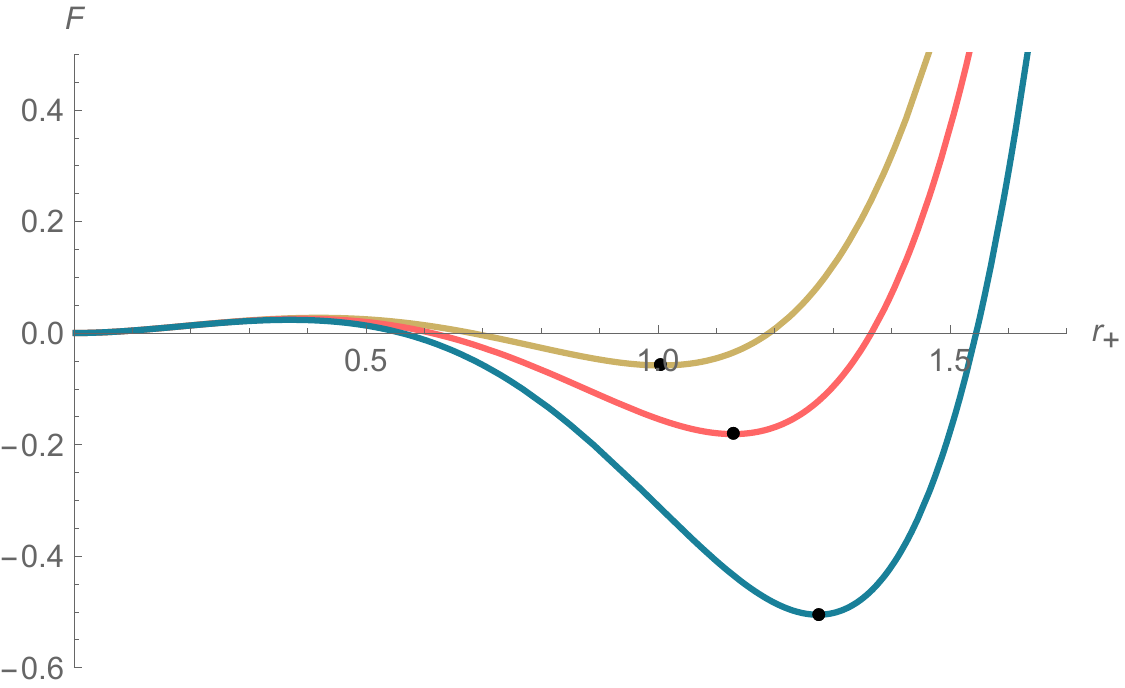}}~~~~
    \subfloat[]{\includegraphics[scale = .4]{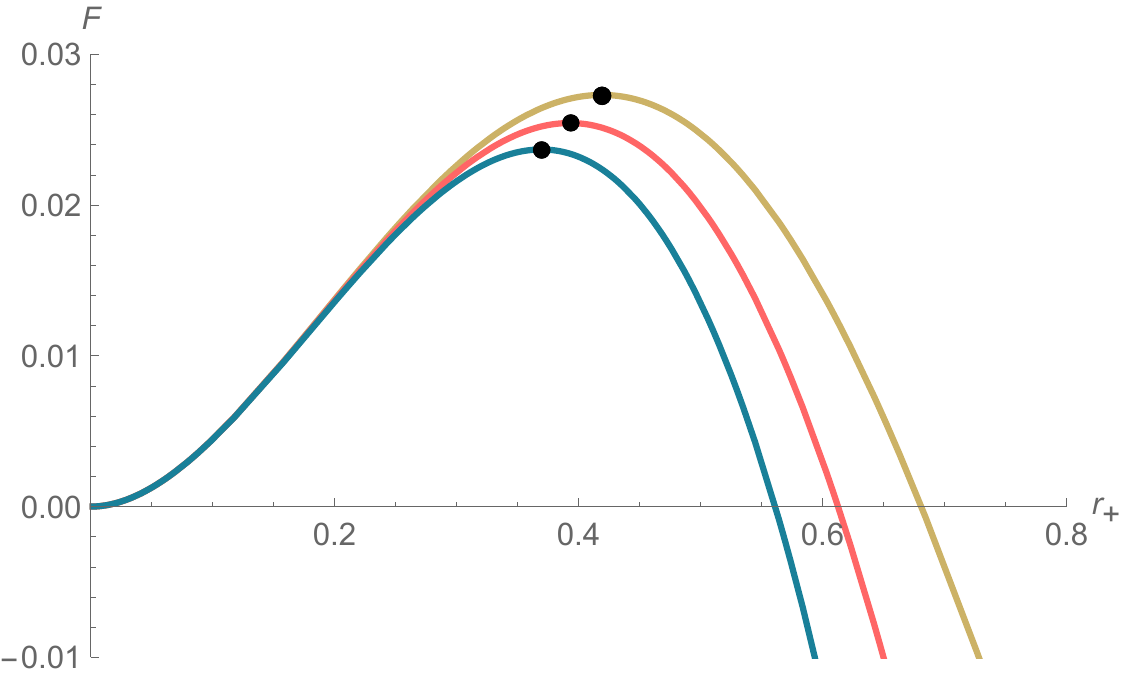}}
   \caption{Free energy as
a function of the order parameter $r_+$ for fixed ${\Phi} = \frac{\sqrt{3}}{2} $ and varying $\Omega$. The yellow, red, blue curve represents ${\Omega} = 0.5, 0.6, 0.7$ with common temperature $T = 0.39 >T_{\rm HP}$ of all three curves, for (a) relatively large range of $r_+$ (large black holes). (b) small range of $r_+$ (small black holes).}
   \label{fig4ab}
\end{figure}

\noindent
{\underline {Effect of angular velocity}} : Free energy in eqn. (\ref{2.21}) plotted for subcritical electric potential and for a sample of angular velocities at their respective HP temperatures is shown in figure-(\ref{fig4}), and for a fixed $T>T_{\rm HP}$ in figure-(\ref{fig4ab}a) and (\ref{fig4ab}b).
\noindent
The following conclusions can be drawn from figure-(\ref{fig4}):
\begin{itemize}
    \item In the sub critical electric potential limit, i.e., $0 < {\Phi} < \sqrt{3}$, and for a fixed value of $\Phi$, increasing the angular velocity decreases the Hawking-Page temperature ${T}_{\rm HP}$.
    \item With increasing angular velocity the Hawking-Page transition point shifts towards right. Size of both small and large black hole phases increases with increasing angular velocity.
    
\end{itemize}
\noindent
As $\Omega$ is varied (at fixed $\Phi$), following can be learnt from figure-(\ref{fig4ab}a) and (\ref{fig4ab}b) (which is also found to be true when variation w.r.t. to $\Phi$ is considered (at fixed $\Omega$): not plotted separately):
 \begin{itemize}
     \item For a given value of $\Phi$ in the sub critical limit, taking  $\Omega <1$ \& $T >T_{HP}$, with increasing angular velocity, the size of large black hole phase increases while size of the small black hole phase decreases.
      \item In the sub critical limit, for a given value of $\Phi$, with $\Omega <1$ \& $T >T_{HP}$, increasing angular velocity  lowers the free energy of both small and large black hole phases.
      \end{itemize}
\FloatBarrier
\noindent
Thus far, we studied the phase transitions keeping  $\Phi$ and $\Omega$ as fixed (below their respective critical limits), with temperature as the free parameter, with transition happening at a critical temperature $T_{\rm HP}$. However, unlike in the on-shell formalism, in the off-shell description, one can actually keep any of the three parameters as fixed, and vary the others to study changes in the phase structure straightforwardly. For instance, we can keep temperature as fixed, and look for changes in phase structure as one of $\Phi$ or $\Omega$ as a free parameter. As the expressions are involved, and the analysis is not very different from that presented above, we directly state the results. One can check that the phase transitions with respect to $\Phi$ at fixed $\Omega$ \&  temperature, and phase transitions with respect to $\Omega$ at fixed $\Phi$ \& temperature, do occur~\cite{Chamblin:1999tk,Caldarelli:1999xj}. The phase diagrams are qualitatively similar to the ones drawn in figure-(\ref{fig1} a), and hence not shown separately. In particular, phase transitions happen as a function of the order parameter $r_+$  at: (a) fixed temperature $T = 0.37$, fixed $\Omega = 0.5$ with varying $\Phi$, at the critical value $\Phi_c = \frac{\sqrt 3}{2}$ (b)  fixed temperature $T = 0.37$, fixed $\Phi = \frac{\sqrt 3}{2} $ with varying $\Omega$, with the critical point occurring at $\Omega = 0.5$.

\subsubsection{Vanishing electric potential limit: $\Phi = 0$} \label{2.2.3}

\begin{figure}[!htbp]
    \centering
     {  \includegraphics[width=0.6\linewidth]{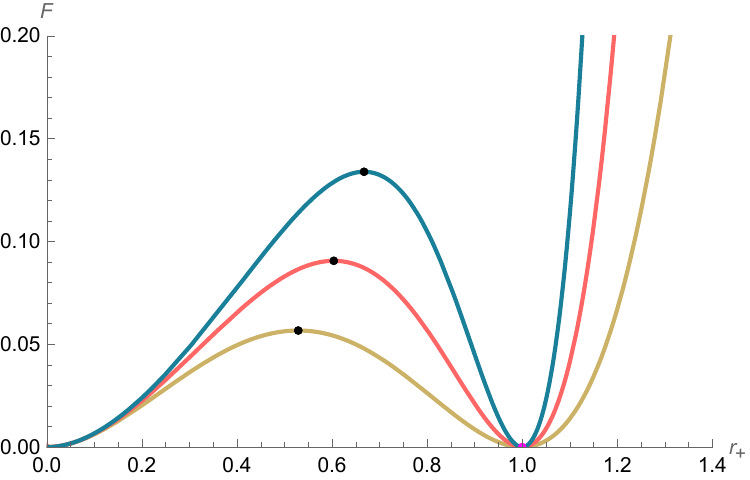}}~~~~
    \caption{ BW free energy as
a function of the order parameter $r_+$ for ${\Phi} = 0 $: The yellow, red, blue curves represent ${\Omega} = 0.5, 0.8, 0,9$ with respective Hawking-Page temperatures $ {T}_{HP} = 0.434, 0.350, 0.297 $.}
    \label{fig6}
\end{figure}

In the vanishing electric potential limit we set $\Phi = 0$ and take $0 < \Omega <1$. Equation (\ref{2.23}) simplifies to 
\be
a=\frac{1+r_+^2-f_\Omega}{2 \Omega} \, , \qquad {\rm where}~~ f_{\Omega} = \sqrt{\left(2-4 \Omega^2\right) r_+^2+r_+^4+1}\, .
\ee 
Hence the free energy in eqn. (\ref{2.21}) is
\begin{eqnarray}\label{2.25}
&&   F = \left(\frac{1}{8 r_+^2 \left(\Omega^2-1\right)^2}\right) \left(-4 \pi  r_+^3 T \left(f_{\Omega}-2 \Omega^2\right)+r_+^4 \left(3 f_{\Omega}-10
  \Omega^2+3\right) \right. \\ \nonumber
  &&  \hskip 2.0cm \left. +r_+^2 \left((6-4 f_{\Omega}) \Omega^2-1\right) +4 \pi 
   (f_{\Omega}-1) r_+ T +f_{\Omega}-4 \pi  r_+^5 T +3 r_+^6-1 \right) \, .
\end{eqnarray}
The on-shell temperature is
\be
T = \frac{ f_{\Omega}+r_+^2}{2 \pi  r_+} \, .
\ee
The off-shell phase diagram with vanishing electrical potential, at a given fixed value of $\Omega$ is qualitatively same as figure-(\ref{fig1}a). The effect of variable angular velocity on the phase diagram is shown in figure-(\ref{fig6}).
From fig \ref{fig6}, we see that for vanishing electric potential, all large black holes have same horizon radius $r_+ = 1$, independent of angular velocities, at
    \be \label{min2}
    T_{\rm HP} = \frac{2 \sqrt{1-\Omega^2}+1}{2 \pi } \, .
    \ee
 which should be contrasted with the case $\Omega=0, \Phi \neq 0$ in eqn. (\ref{min1}).  We note that the size of small black holes though vary.\\
    
\noindent
One can of course take the simplest case, where $\Phi = 0$ and $\Omega = 0$ giving the AdS Schwarzschild black hole. In this case, eqns. (\ref{offphi} ) and (\ref{Tonphi}) for free energy and the on shell temperature simplify, giving, 
\begin{equation}\label{gr3}
F = \frac{1}{4} r_+^2 \left(3 r_+^2-4 \pi  r_+ {T}+3\right) \, ,\qquad
    T = \frac{ \left(2 r_+^2+1\right)}{2 \pi  r_+}
\end{equation}
The phase diagram for AdS Schwarzschild black holes at different temperatures is qualitatively similar to sub critical electric potential limit shown in figure-(\ref{fig1}). A known fact recovered is that, both angular velocity and electric potential decrease the HP transition temperature. The AdS Schwarzschild black hole has the largest HP transition temperature, i.e.,  $T_{\rm HP} = 3/{2\pi}$ with $r_+=1$. 
\FloatBarrier

\subsubsection{Critical electric potential $\Phi = \Phi^* = \sqrt{3}$} \label{2.2.4}

\begin{figure}[!htbp]
    \centering
 \subfloat[]{\includegraphics[width=0.47\linewidth]{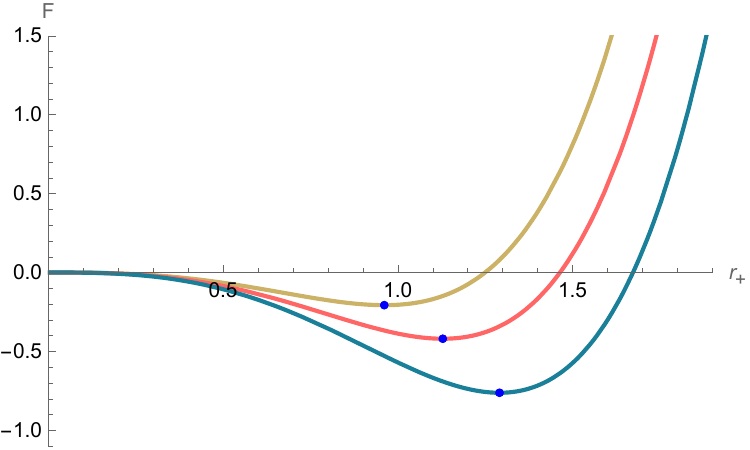}}~~~~~~
  \subfloat[]{ \includegraphics[width=0.47\linewidth]{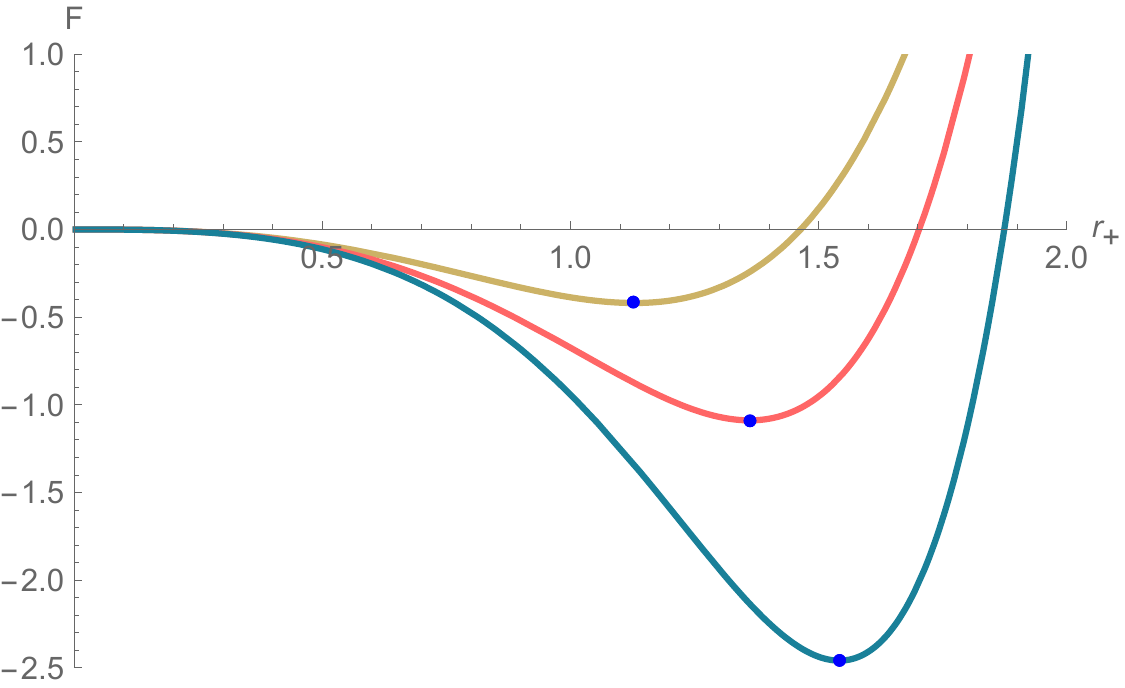}}
    \caption{(a) BW free energy as
a function of the order parameter $r_+$ for ${\Phi} = \Phi^* = \sqrt{3} $ and $\Omega = 0.5$ : The yellow, red, blue curves represents temperatures $ {T} = 0.25, 0.3, 0.35 $ respectively. (b) BW free energy as
a function of the order parameter $r_+$ for ${\Phi} = \Phi^* = \sqrt{3} $ and common temperature $T = 0.3$ : The yellow, red, blue curves represent $ \Omega = 0.5, 0.7,0.8 $ respectively.}
    \label{fig7}
\end{figure}

For the case of critical electric potential, $\Phi^* = \sqrt{3}$ with non-zero $\Omega$, the analysis is not analytically tractable, and hence we directly show the phase diagram in figure-(\ref{fig7}). 
\FloatBarrier
Following are the observations:
\begin{itemize}
    \item There is no locally unstable state, and hence no small black hole phase.
    \item  Locally stable large black holes always exist with negative free energy, for non-zero temperature. Large black hole phase is thus always preferred over pure AdS.
        \item At fixed $\Omega$, with increasing temperature, size (free energy) of the large black hole phase increases (decreases).
\item Similarly at fixed temperature, with increasing $\Omega$, size (free energy) of the large black hole phase increases (decreases).  

 \item Further, in the case of super critical electric potential limit $\Phi > \sqrt{3}$, the phase diagram and discussion is identical to the one given above from figure-(\ref{fig7}), with a qualitative difference that the free energy is lower.  
\end{itemize}

\subsubsection{Maximal rotational velocity $\Omega = 1$} \label{2.2.5}

\begin{figure}[!htbp]
    \centering
\subfloat[]{\includegraphics[width=0.47\linewidth]{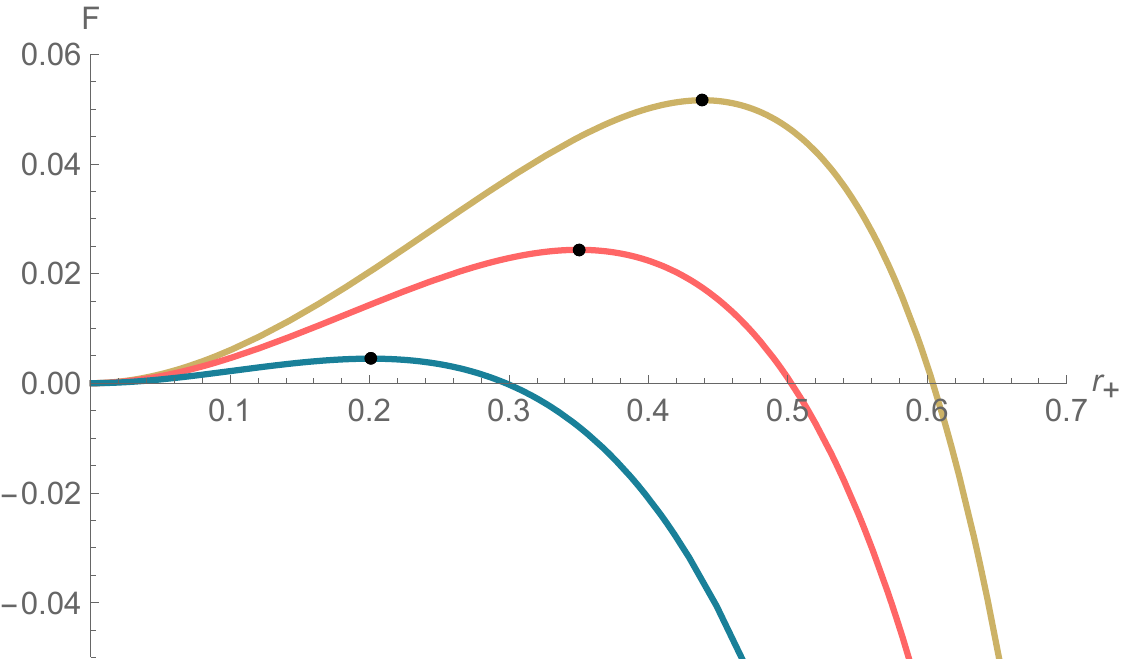}}~~~~
\subfloat[]{ \includegraphics[width=0.47\linewidth]{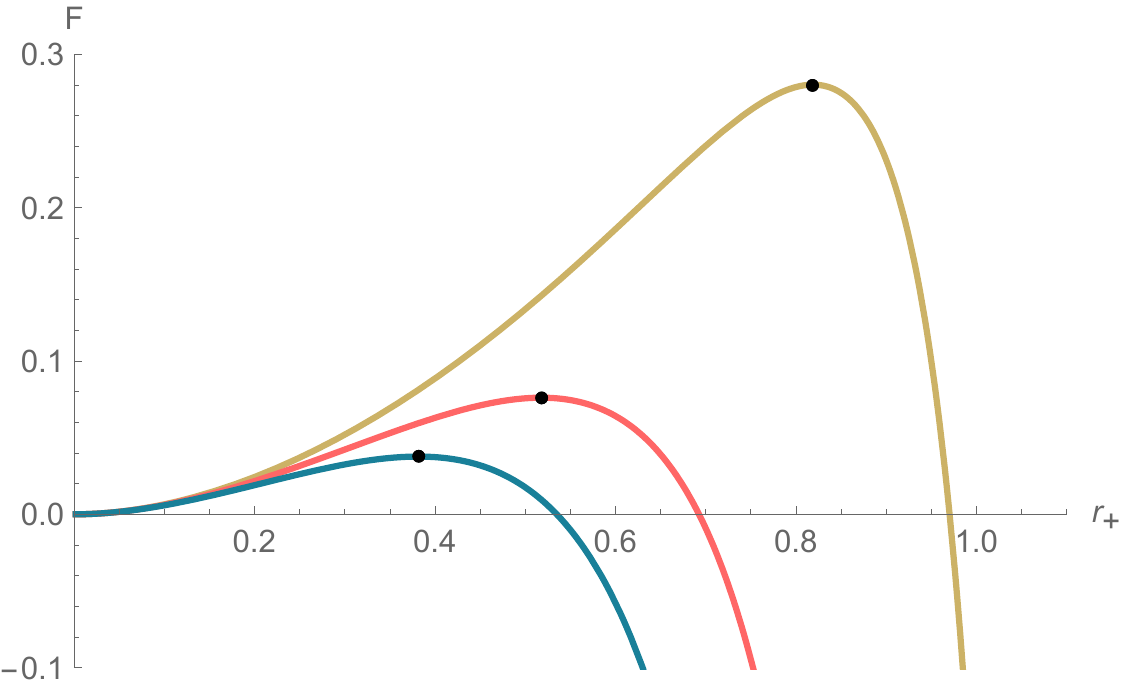}}
    \caption{ (a) BW free energy as
    	a function of the order parameter $r_+$ for $\Omega = 1,T = 0.35 $. The yellow, red, blue curves represent $ \Phi = \frac{\sqrt{3}}{4},  \frac{2\sqrt{3}}{4} ,  \frac{3\sqrt{3}}{4}  $ respectively (b) BW free energy as
a function of the order parameter $r_+$ for $\Omega = 1,{\Phi} = \frac{\sqrt{3}}{4} $. The yellow, red, blue curves represent temperatures $ T = 0.2, 0.3, 0.4 $ respectively.   }
    \label{fig9}
\end{figure}

In the maximal rotational velocity limit, the off-shell phase diagram for $\Omega = 1$ is shown in fig \ref{fig9}. We draw the following conclusions:
\begin{itemize}
    \item There is no locally stable state, i.e., no large black hole phase. Only locally unstable small black holes with positive free energy exist for non-zero temperature. 
    \item For a fixed value of $\Phi$ ($< \sqrt{3}$), increasing temperature decreases the size as well as free energy of the small black hole phases.  
    \item For  a fixed value of temperature T, with increasing electric potential size as well as free energy of the small black hole decreases.
\end{itemize}
In summary, the findings of this section from the off-shell description of AdS$_5$ black hole thermodynamics are found to be consistent with on-shell analysis, enriching the results in~\cite{Ezroura:2021vrt} with new information around the equilibrium phases. In addition, the off-shell description also paves way for construction of effective potentials in the dual holographic system, to be discussed later. 
\FloatBarrier

	\section{ Phase diagram of AdS$_5$ black holes with four-derivative corrections } \label{three}

 
Corrections to the entropy and other conserved charges following from the four-derivative corrected five-dimensional minimal gauged supergravity have been recently studied in~\cite{Cassani_2022,Bobev_2022}. In this section we briefly present the four derivative corrected thermodynamics and off-shell phase diagram of $AdS_5$ black holes. 
The two-derivative action supplemented with the four-derivative corrections is
\cite{Liu:2022sew,Cassani_2022}:
\begin{equation}\label{3.1}
\begin{aligned}
S\,=&\,\frac{1}{16\pi G}\int d^5x \,e \left\{{\tilde c}_0 R+12{\tilde c}_1g^2-\frac{{\tilde c}_2}{4}F^2-\frac{{\tilde c}_3}{12\sqrt{3}}\epsilon^{\mu\nu\rho\sigma\lambda}F_{\mu\nu}F_{\rho\sigma} A_{\lambda}\right.\\[2mm]
&\left.\,+\,\lambda_1 \alpha \left[{\cal X}_{\text{GB}}-\frac{1}{2}C_{\mu\nu\rho\sigma}F^{\mu\nu}F^{\rho\sigma}+\frac{1}{8}F^4-\frac{1}{2\sqrt{3}}\epsilon^{\mu\nu\rho\sigma\lambda}R_{\mu\nu\alpha\beta}R_{\rho\sigma}{}^{\alpha\beta} A_{\lambda}\right]\right\}\, .
\end{aligned}
\end{equation}
Here,  ${\cal X}_{\text{GB}}=R_{\mu\nu\rho\sigma}R^{\mu\nu\rho\sigma}-4 R_{\mu\nu}R^{\mu\nu}+R^2$ stands for the Gauss-Bonnet invariant,  $C_{\mu\nu\rho\sigma}=R_{\mu\nu\rho\sigma}-\frac{2}{3}\left(R_{\mu[\rho}g_{\sigma]\nu}+R_{\nu[\sigma}g_{\rho]\mu}\right)+\frac{1}{6}Rg_{\mu[\rho}g_{\sigma]\nu}$ denotes the Weyl tensor, and constants are relabeled as ${\tilde c}_{i}=1+\alpha g^2 {\delta \tilde c}_{i}$, where
\begin{equation}\label{3.2}
{\delta \tilde c}_{0}=4\lambda_2\, , \hspace{.5cm}{\delta\tilde c}_{1}=-10\lambda_1+4\lambda_2 \,, \hspace{.5cm} {\delta \tilde c}_{2}=4\lambda_1+4\lambda_2 \,, \hspace{.5cm}{\delta 
\tilde c}_{3}=-12\lambda_1+4\lambda_2 \,.
\end{equation}
Here, $\lambda_{1}$ and $\lambda_{2}$ are dimensionless couplings. The corrections which are controlled by the coupling $\lambda_2$ are mainly related to the two-derivative part of the Lagrangian, whereas, the four derivative corrections are controlled by the coupling $\lambda_{1}$. Let us note that the dimension of the parameter $\alpha$  is length$^2$ and it is further assumed that $\alpha {\cal R}\ll 1$. Here, the curvature scale of the solution is $\cal R$. Since $\alpha g^2$ is dimensionless too with possible corrections to two-derivative part of the Lagrangian, in our analysis we keep $\lambda_{1} << 1$ \& $\alpha << 1$.\\

\noindent
When setting up the computation of the on-shell action at linear order in $\alpha$, there are two crucial points that should be kept in mind. The first is that since we are working in the grand-canonical ensemble, the inverse temperature $\beta$ and the chemical potentials, $\Phi$, $\Omega_{1}$, $\Omega_{2}$, must be held fixed to their zeroth-order values given in Sec.~\ref{two}.  On the other hand, the action, the entropy and the conserved charges are allowed to receive corrections. The second point is that the corrections to the bulk metric and gauge field are not needed in order to compute the $\mathcal{O}(\alpha)$ corrections to the thermodynamics, as recently argued in \cite{Reall:2019sah}. Hence, the four derivative corrected on shell action can be evaluated on the uncorrected solution in eqn. (\ref{3.1}). The four derivative corrected quantities obey the same first law of thermodynamics as in  eqn. (\ref{2.9}), and quantum statistical relation in eqn. (\ref{2.10}). Now, using eqn. (\ref{2.12}) in the on-shell action, the corrected expressions of charges can be calculated~\cite{Cassani_2022,Bobev_2022}, and given as:
\begin{align} \label{QJScorrected}
Q &=\,\frac{\sqrt{3} \pi q}{4 G\left(1-a^2\right)^2 }\left(1+4  \lambda _2\alpha\right)+\lambda_1\alpha  \Delta Q(q,a,r_+)\,,\\
J&=\frac{a \pi \left(1+4 \lambda_2\alpha \right)}{4 \left(1-a^2\right)^3 G r_+^2}\left[\left(a^2+q\right)^2+\left(a^4+q+a^2 (2+q)\right) r_+^2+\left(1+2a^2\right) r_+^4+r_+^6\right] \nonumber \\ 
& \hskip 1.0cm +\lambda_1\alpha \Delta J(q,a,r_+)\,,\nonumber \\
{\cal S}&=\,\frac{\pi ^2 \left[r_+^4+a^4+a^2 \left(q+2 r_+^2\right)\right] }{2Gr_+\left(1-a^2\right)^2}\left(1+4 \lambda _2\alpha\right)+\lambda_1\alpha \Delta {\cal S}(q,a,r_+)\,\nonumber.
\end{align}
Detailed expressions for $\Delta Q(q,a,r_+),\Delta J(q,a,r_+),{\cal S}(q,a,r_+)$ were obtained in~\cite{Cassani_2022}, which for the special case of equal angular momenta are given in appendix \ref{A1}. We can now use these corrected charges to construct the Bragg-Williams free energy of $AdS_5$ black holes with four derivative corrections:
\begin{equation}\label{2.199}
    F (r_+, T, \Phi, \Omega, \alpha,\lambda_1,\lambda_2) = E - {T} {\cal S} - {\Phi} Q - 2{\Omega} J \, ,
\end{equation}
where the only quantities which receive the higher derivative corrections are the ones in eqn. (\ref{QJScorrected}) and the energy $E$ which can be obtained from the quantum statistical relation using the corrected on shell action. The full expression for the free energy can be computed straightforwardly from eqn. (\ref{2.199}), but is long and given in appendix \ref{A2}.  For the case of a non-rotating black holes, the higher derivative corrected free energy is:
\begin{eqnarray}\label{2.1999}
 && F (r_+, T, \Phi,\alpha,\lambda_1,\lambda_2) = F_0(r_+, T, \Phi) +   4 \pi  \alpha  r_+^3 T \left(-\lambda _2-\frac{3 \lambda _1
   \left(9 r_+^2+\Phi ^2+9\right)}{6 r_+^2+\Phi ^2-3}\right) \\
   &&+\frac{\alpha }{6 \left(6 r_+^2+\Phi ^2-3\right)} \left[
  6 \lambda _2 r_+^2 \left(-3 r_+^2 \left(\Phi ^2-3\right)+18
   r_+^4-\left(\Phi ^2-3\right)^2\right) \right. \nonumber \\
 && \left. +\lambda _1 \left(r_+^4
   \left(729-99 \Phi ^2\right)-12 r_+^2 \left(\Phi ^4+6 \Phi
   ^2-27\right)+486 r_+^6+\left(\Phi ^2-3\right)^2 \left(\Phi
   ^2+9\right)\right) \right] \nonumber  \, 
\end{eqnarray}
where the first term $F_0(r_+, T, \Phi) $ is the two-derivative result given in eqn. (\ref{offphi}). Although, it is not obvious, the on-shell temperature in the four-derivative theory, which can be obtained by solving $\partial_r F (r_+, T, \Phi,\alpha,\lambda_1,\lambda_2) =0$ turns out to be same as the one given in eqn. (\ref{Tonphi}), i.e., the expression one gets 
by solving  $\partial_r F_0(r_+, T, \Phi) =0 $. This is a consistent check, as in the current approach~\cite{Cassani_2022,Bobev_2022}, the temperature, as well as the potentials $\Phi, \Omega $, are not expected to be corrected by four-derivative terms. We checked that this feature continues to hold for the general off-shell free energy given in eqn. (\ref{2.199}). The phase diagram for the non-rotating case is qualitatively similar to the case with a non-zero angular velocity, although, one has to resort to numerical calculations in the latter case. We thus directly present the phase structure for the most general case with charges and rotation, following from the off-shell free energy in eqn. (\ref{2.199}). \\

\noindent

\noindent
The phase diagram following from the four-derivative corrected off-shell free energy in eqn. (\ref{2.199}) for the subcritical electric potential case, is shown in figure-(\ref{fig32}), and we infer the following:
\begin{figure}[!htbp]
    \centering
    \includegraphics[width=1.0\linewidth]{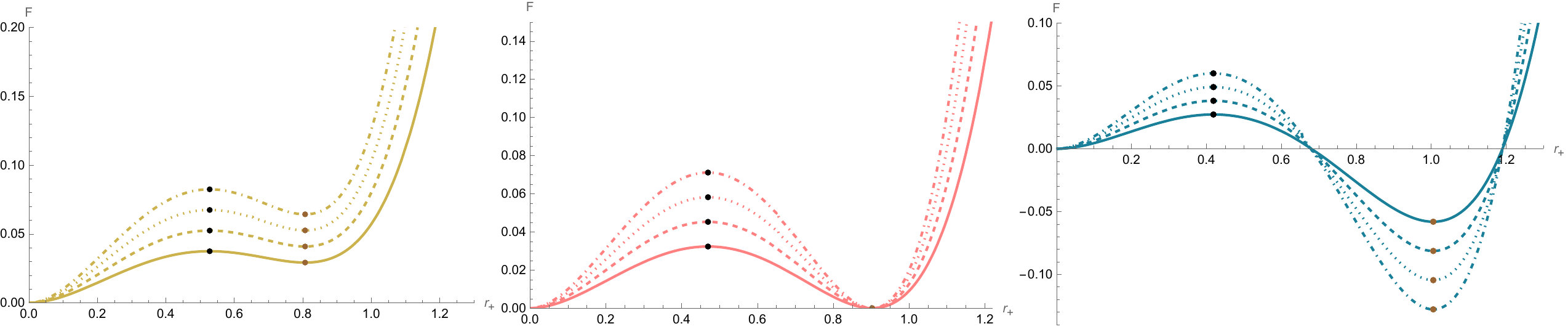}
    \caption{Off-shell free energy for the four-derivative case as
a function of the order parameter $r_+$. The left, middle and the right plots in the figure, correspond to the temperature values $T = .36 <T_{HP}, T = T_{HP} =  .372,  T = .39 > T_{HP}  $, respectively. In all the plots: the solid curves are two derivative plots with  ``$\alpha = \lambda_{1} = \lambda_{2} = 0$" while dashed, dotted and dot-dashed curves corresponds to four derivative correction with ``$\alpha = .001, \lambda_{1} = .001, \lambda_{2} = 100$",  ``$\alpha = .002, \lambda_{1} = .001, \lambda_{2} = 100$" and ``$\alpha = .003, \lambda_{1} = .001, \lambda_{2} = 100$" respectively. For all plots $\Omega = .5,{\Phi} = \frac{\sqrt{3}}{2}$.}
    \label{fig32}
\end{figure}
\FloatBarrier
\begin{itemize}
    \item  At $ T < T_{\rm HP}$ (left plot in figure-(\ref{fig32})), with increasing $\alpha, \lambda_{1}$, both small and large black hole further destabilizes.
     \item  At $ T = T_{\rm HP}$ (middle plot in figure-(\ref{fig32})), with increasing $\alpha, \lambda_{1}$, the small black hole further destabilizes.
      \item  At $ T > T_{\rm HP}$ (right plot in figure-(\ref{fig32})), with increasing $\alpha, \lambda_{1}$, the small black hole phase destabilizes, while the large black hole phase stabilizes.
      \item The four-derivative corrections do not modify the horizon radius $r_{+}$, for any temperature. 
\end{itemize}

\begin{figure}[!htbp]
    \centering
    \includegraphics[width=1.0\linewidth]{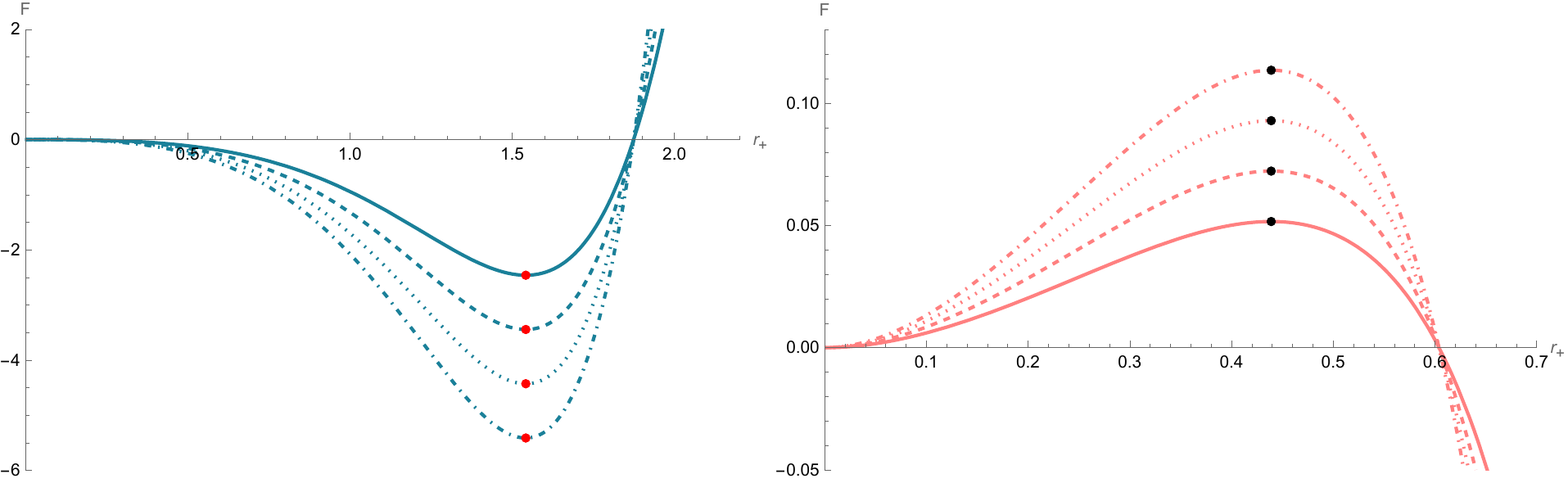}
    \caption{Off-shell free energy for the four-derivative case as
a function of the order parameter $r_+$. Left: $\Omega = 0.8,{\Phi} = \sqrt{3}, T = 0.3  $. Right: $\Omega = 1,{\Phi} = \frac{\sqrt{3}}{4}, T = 0.35  $.  In both plots: the solid curves are two derivative plots with  ``$\alpha = \lambda_{1} = \lambda_{2} = 0$" while dashed, dotted and dot-dashed curves corresponds to four derivative correction with ``$\alpha = .001, \lambda_{1} = .001, \lambda_{2} = 100$",  ``$\alpha = .002, \lambda_{1} = .001, \lambda_{2} = 100$" and ``$\alpha = .003, \lambda_{1} = .001, \lambda_{2} = 100$" respectively. }
    \label{fig33}
\end{figure}

\noindent
Phase diagram for the critical electric potential case is given in fig (\ref{fig33}(left)) and we observe that: in the presence of four derivative corrections, as expected, the large black hole phase further stabilizes.
We have checked that the effect of corrections in supercritical limit ($\Phi > \sqrt{3}$ ) is similar to critical limit.\\

\noindent
Effect of corrections for maximal velocity limit is given in fig (\ref{fig33}(right)). As expected, in the presence of four-derivative corrections, the small black hole phase further destabilizes.
\FloatBarrier

\section{Off-shell free energy of BPS black holes} \label{offshell}
In this section, we primarily focus on black holes in $\text{AdS}_5$ with a single charge and two equal angular momenta, which are supersymmetric and satisfy the BPS relation. In subsection-(\ref{onshell}) we briefly recall the construction of on-shell free energy of BPS black holes from\cite{Ezroura:2021vrt}, which will be used in the construction of our off-shell free energy for the two derivative and four derivative black hole cases in subsections-(\ref{2d}) and (\ref{4d}), respectively.  

\subsection{On-shell free energy} \label{onshell}

Following~\cite{Ezroura:2021vrt}, one starts by writing the BPS condition as \footnote{For section 4 we have scaled $Q$ by a factor of $\sqrt{3}$ to match with two derivative on shell results of \cite{Ezroura:2021vrt}} 
\begin{equation}
\label{eqn: bpscond}
    M = 3\, Q + 2J\, ,
\end{equation}
where the conjugate potentials satisfy $\Phi^*=3, \Omega^*=2$. The excess mass above the BPS bound is related to the parameters of the black hole as
\begin{equation}
\label{eqn:BPSm}
   M - 3Q - 2 J  = \frac{1}{2}N^2  \frac{3+2a-a^2}{(1-a)^2(1+a)^4} \Big( m-q(1+2 a)\Big) \, ,
\end{equation} 
which leads to the condition on the parameters as $ m = q(1+2a)$. Thus, the parameter $m$ can be eliminated in favor of $(q, a)$, which then characterize the BPS surface. We should note that due to the BPS condition, the horizon radius and charge parameter are constrained to be $ r_* ^2 = 2a+a^2$ and $q^* = 2a(1+a)^2$. We follow the convention in literature to call the starred variables as BPS quantities. Due to the above relations, the physical thermodynamical variables take special values in the BPS case and are denoted as $(M^*, Q^*, J^*,S^*)$, with $Q^*$ and $J^*$ satisfying a non-linear constraint.
Now, the key insight obtained in\cite{Ezroura:2021vrt} is that: a general grand canonical free energy $G(T,\Omega,\Phi)$, which is evaluated on the BPS mass relation $M = \Phi^* Q + \Omega^* J $, can be shown to be homogeneous of degree $1$ in terms of the variables $T, \Phi-\Phi^*$ and $\Omega-\Omega^*$. This allows one to define the BPS free energy which is of weight zero as~\cite{Ezroura:2021vrt}
\begin{equation}
W(\Phi',\Omega')= \frac{G}{T} = - S - \Phi'  Q - \Omega' J~.
\label{eqn:FreeEnergy}
\end{equation}
where the primed variables are defined as $\Phi' = (\Phi-\Phi^*)/T~, 
\Omega' =  (\Omega-\Omega^*)/T$. In the BPS limit where the temperature tends to zero, the primed variables can be identified as derivatives w.r.t. temperature computed at $T=0$. It was further noted in~\cite{Ezroura:2021vrt} that to obtain the correct phase diagram, it is imperative to change the basis and define new thermodynamic potentials  conjugate to the physical charges $Q, J$ as:
\begin{align}
\label{eqn:potstransform}
    \Phi' &= -\frac{2}{\tau} + \varphi' ~,\\
    \Omega' &= -\frac{1}{\tau} ~.
\end{align}
In the new basis mentioned above, the BPS free energy and the ``BPS temperature" become respectively~\cite{Choi:2021nnq,Ezroura:2021vrt}
\begin{align}\label{free2}
    W (a) &
    = \frac{2 a \left(\frac{\pi  \left(-2 a^2-5 a+1\right)}{\sqrt{\frac{2}{a}+1}}-\frac{1}{2} (3-a) a \varphi\prime \right)}{(1-a)^2 (5 a+1)}\, , \\
\tau_{\rm H} &= \frac{5 a+1}{(1-a) \left(\frac{3 \pi }{\sqrt{\frac{2}{a}+1}}+\frac{\varphi\prime }{2}\right)}  \label{tauH}.
\end{align}
The phase diagram emerging from the BPS free energy in eqn. (\ref{free2}) resembles closely the AdS black holes with a Hawking-Page type first order phase transition at $\tau_{\rm HP} = 0.863, a_{\rm HP} = 0.186$ (for $\varphi'=0$), though there are some differences as well, depending on the value chosen for $\varphi'$.  The BPS phase diagram can be studied for any real value of $\varphi'$~\cite{Dijkgraaf:2000fq,Larsen:2021wnu,Ezroura:2021vrt}, starting with the benchmark case $\varphi\prime = 0$ (where a  match between the BPS phase diagram and  the AdS Schwarzschild thermodynamics~\cite{Chamblin:1999tk, Witten:1998qj, Natsuume:2014sfa, Kubiznak:2016qmn} was found~\cite{Choi:2021nnq,Ezroura:2021vrt} ), $\varphi\prime \leq0$ and also $\varphi' \geq 0$. In particular, there exists
a small and large black hole branch as the BPS temperature is varied (which can be inferred from eqn. (\ref{tauH}) too), though there are some differences in all three ranges of $\varphi'$, particularly, around the large black hole branch.

\subsection{Off-shell BPS free energy from Landau's phenomenological approach} \label{alternative1}
We now try to understand the first order phase transition of BPS black holes at $\tau_{\rm H} = \tau_{\rm HP}$, using Landau's phenomenological theory~\cite{Landau:1980mil,Cappiello:2001tf}. This can be helpful in situations when the on-shell free energy is given, together with the on-shell data of various thermodynamic quantities, and one needs to study physics slightly away from equilibrium, without going into the construction from first principles.\\

\begin{figure}[!htbp]
    \centering
 {\includegraphics[width=0.55\linewidth]{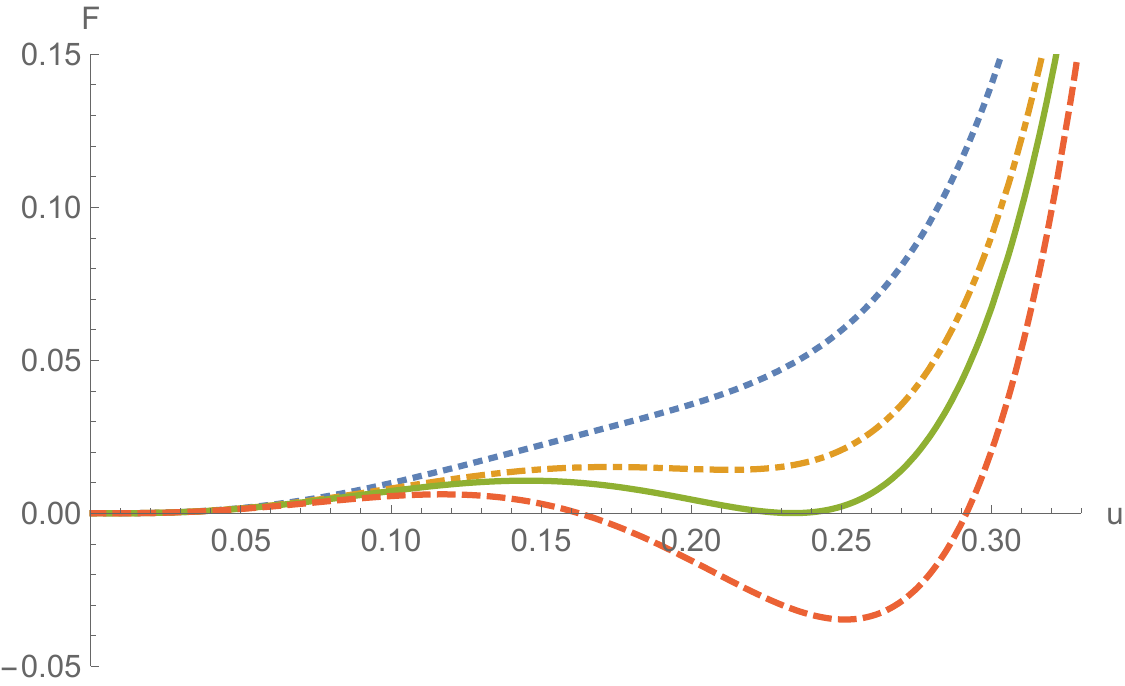}}
    \caption{BPS phase diagram following from the Landau free energy function defined in eqn. (\ref{Fseries3}). Temperature increases from top to bottom, taken respectively as $\tau = 0.9, 1.112, 1.212,1.412$.}
    \label{bps31}
\end{figure}

\noindent
We start with the on-shell expressions for BPS free energy $W(a)$ and the temperature $\tau_{\rm H}$ given in eqns. (\ref{free2}) and  (\ref{tauH}). 
Here, we limit ourselves to the case $\varphi'=0$ for brevity, though the discussion can be extended to general cases. We find it convenient to chose $u = \sqrt{a}$ as the order parameter, which makes a jump, having value $u=0$ for $\tau < \tau_{\rm HP}$ and taking the larger value of the solution of eqn. (\ref{tauH}) for $\tau > \tau_{\rm HP}$. We thus interpret eqn. (\ref{free2}) as the free energy at equilibrium and our aim is to consider a more general expression $F(u,\tau)$, which can describe the system away from the equilibrium. Then, eqn. (\ref{tauH}) can be thought of as the equilibrium condition which relates the BPS temperature to the order parameter. To proceed further, we take a small $u$ expansion of eqns. (\ref{free2}) and  (\ref{tauH}), giving
\begin{eqnarray}
&& W(u) \approx \sqrt{2} \pi  u^3-\frac{33 \pi  u^5}{2 \sqrt{2}} + {\mathcal O}(u^7) \, , \label {WS}\\
&& \tau_{\rm H}(u)  \approx \frac{25 u}{6 \sqrt{2} \pi }+\frac{\sqrt{2}}{3 \pi  u} + {\mathcal O}(u^3)\, . \label{tauS}
\end{eqnarray}
Let us make an ansatz based on Landau's theory as
\be \label{freeS}
F(u, \tau) =  \frac{\pi}{2 \sqrt{2}}  \left(c_3 u^3-c_4 \tau u^4+c_5
   u^5\right) \, ,
\ee
where $c_3, c_4$ and $c_5$ are three constants, which need to be determined. We first use the expression for the equilibrium temperature eqn. (\ref{tauS}) in eqn. (\ref{freeS}), giving us
\be \label{freeS1}
F(u) = \frac{1}{12} u^3 \left(3 \sqrt{2} \pi  c_3-2c_4\right) + \frac{1}{24} u^5 \left(6 \sqrt{2} \pi  c_5-25 c_4\right)\, .
\ee
We demand that eqn. (\ref{freeS1}) match with eqn. (\ref{WS}), which fixes two of the constants in terms of the third as:
\be \label{const2}
c_5= -\frac{198 \sqrt{2} \pi -25 c_4}{6 \sqrt{2} \pi } \, , \qquad c_3 = \frac{\sqrt{2} \left(c_4+6 \sqrt{2} \pi \right)}{3 \pi } \, .
\ee 
Using the above values of constants in eqn.  (\ref{freeS}), we get
\be \label{freeS2}
F(u, \tau)=-\frac{\pi  c_4 \tau u^4}{2 \sqrt{2}}+\frac{1}{24}
   \left(25c_4-198 \sqrt{2} \pi \right) u^5+\frac{1}{6}
   \left(c_4+6 \sqrt{2} \pi \right) u^3 \, .
\ee
To determine $c_4$, we now demand that the equilibrium temperature obtained by minimizing eqn. (\ref{freeS2}), i.e.,
\be \label{freeSm}
\partial_u F(u, \tau)=0 \, ,
\ee
match with eqn. (\ref{tauS}). This however does not fix $c_4$ uniquely, pointing to the possibility that it might be a $u$-dependent quantity. We thus now assume that $c_4= c_4(u)$ and refine our ansatz in eqn. (\ref{freeS}). With this assumption, we once again evaluate eqn. (\ref{freeSm}), and match it with eqn. (\ref{tauS}). This yields a differential equation for $c_4(u)$ as
\be
\frac{u \left(25 u^2+4\right) c_4'(u)+\left(125 u^2+12\right)
   c_4(u)+18 \sqrt{2} \pi  \left(4-55 u^2\right)}{6 \sqrt{2} \pi  u
   \left(u c_4'(u)+4 c_4(u)\right)} = \frac{25 u}{6 \sqrt{2} \pi }+\frac{\sqrt{2}}{3 \pi  u}\, ,
\ee
which can be solved exactly, but, the first few terms of the solution are enough in our case, and are
\be \label{const3}
c_4(u) = 18 \sqrt{2} \pi -135 \left(\sqrt{2} \pi \right)
   u^2+O\left(u^4\right)\, .
\ee
Using the solutions for the unknowns obtained in eqns. (\ref{const2}) and (\ref{const3}) in eqn. (\ref{freeS}), we get
the required Landau free energy function as
\be \label{Fseries3}
F(u, \tau) = 4 \sqrt{2} \pi  u^3-9 \left(\pi ^2\tau \right) u^4-12
   \left(\sqrt{2} \pi \right) u^5+\frac{135}{2} \pi ^2 \tau
   u^6+O\left(u^7\right)\, .
\ee
A plot of the above  eqn. (\ref{Fseries3}) is presented in figure-(\ref{bps31}), which shows the phase transition, though, not surprisingly, the value of transition temperature $\tau_{\rm HP}=1.112$ (at $u=0.233$) is different from the one obtained from the exact expression in eqn. (\ref{tauH}).
\FloatBarrier
\noindent
In summary, the Landau free energy in  eqn. (\ref{Fseries3}) correctly captures the first order phase transition, in terms of the order parameter $u$. $F$ assumes its absolute minimum for a given value of $u$ for $\tau > \tau_{\rm HP}$, for larger solution of eqn. (\ref{tauS}),  and a local maximum for the smaller solution of eqn. (\ref{tauS}) for $\tau < \tau_{\rm HP}$. 

\subsection{Exact Method for off-shell BPS free energy} \label{alternative}

The off-shell free energy constructed earlier in section-(\ref{2.2}) was exact, once the basic thermodynamic quantities are known.  Starting from the standard thermodynamic relation in eqn. (\ref{2.19}), the temperature  and potentials conjugate to charges were taken to be free parameters, acquiring their on-shell values only after a certain minimisation procedure (see eqn.(\ref{dFzero})). Once the on-shell values of thermodynamic potentials are used in eqn. (\ref{2.19}), the on-shell free energy of~\cite{Choi:2018vbz,Choi:2018hmj,Ezroura:2021vrt} is recovered.  A phenomenological construction valid close to phase transition point was considered in the previous subsection-(\ref{alternative1}), which is valid when the order parameter is small. We now look for a method which is valid in general, and will systematically yield off-shell BPS free energy. This construction will also allow us to propose BPS effective potentials in the boundary theory. This might lead to an easier off-shell construction in certain cases, where first principles derivation may not otherwise be possible or tedious to obtain. \\

\noindent
Consider a simple black hole system, where the on-shell free energy $G(r_+)$, and the Hawking temperature denoted as $T_H (r_+)$, are given. We call $T_H (r_+)$ as the on-shell value of temperature. The off-shell free energy is supposed to be a function of $r_+$ (order parameter), and an off-shell temperature $T$ (a free parameter at this stage), whose dependence on the order parameter is not fixed yet. We can thus propose an off-shell free energy as, 
\begin{equation} \label{gr}
F(r_+, T) = G(r_+) + g(r_+) \left(1- \frac{T}{T_H} \right) \, .
\end{equation}
where $T_H=T_H(r_+)$ is known, and $g(r_+)$ is at this stage an arbitrary function that needs to be determined. A consistency check for the off-shell function $F(r_+, T)$ is that it has to reduce to the on-shell free energy once the temperature is on-shell. This is ensured by our construction, since the second term in eqn. (\ref{gr}) vanishes at $T=T_H$.  Now, we recall eqn. (\ref{dFzero}), which says that minimizing $F(r_+, T)$ w.r.t $r_+$, should give the on-shell temperature $T_H$. If the on-shell temperature is already known (as in the present case), the same equation (eqn. (\ref{dFzero})) can be used as a constraint to determine the unknown function $g(r_+)$. That is, we impose
\begin{equation} \label{der}
 F'(r_+, T) = \left[ G'(r_+) + g(r_+) \left(1- \frac{T}{T_H(r_+)} \right)'  + g'(r_+)  \left(1- \frac{T}{T_H} \right) \right]  = 0  \, ,
\end{equation}
where the $'$ denotes derivative w.r.t. to $r_+$.
Now, after evaluating the derivatives, one can substitute $T=T_H$, which has the effect of setting the last term in eqn. (\ref{der}) to zero. Eqn. (\ref{der}) can now be solved straightforwardly to determine $g(r_+)$. The result can then be plugged back in eqn. (\ref{gr}) to get the required expression for off-shell free energy. The procedure outlined above is quite general, and is applicable to general thermodynamic systems, not necessarily black holes, even when there are multiple order parameters, and other thermodynamic potentials. Of course, it is clear that to obtain the off-shell free energy in this method, the on-shell quantities, such as, the temperature, potentials and free energy should be known beforehand. \\

\noindent
Let us now illustrate this procedure for the case of an AdS Schwarzschild black hole. The on-shell (Gibbs) free energy $G = M - TS$ can be constructed from the thermodynamic parameters given eqns.(\ref{temperature_CCLP}), (\ref{entropyCCLP}) and (\ref{CCLPcharges}) (in the limit $a=b=q=0$), which is given below together with the on-shell temperature as~\cite{Chamblin:1999tk,Banerjee:2010ve,Ezroura:2021vrt}:
\begin{align}
\label{eqn:sadsG}
    G &= \frac{r_+^2(1-r_+^2)}{4} ~,\\
\label{eqn:sadsT}
    T_H &= \frac{1+2 r_+ ^2}{2\pi r_+} ~.
\end{align}
Our trial off-shell free energy function is
\begin{equation} \label{gr1}
F(r_+, T) = \frac{r_+^2(1-r_+^2)}{4}  + g(r_+) \left[1- \left(\frac{2\pi r_+\,}{1+2 r_+ ^2}\right) T \right] \, .
\end{equation}
Implementing eqn. (\ref{der}), gives the unknown function  $g(r_+) = r^4+\frac{r^2}{2}$, which when used back in eqn. (\ref{gr1}) gives the off-shell free energy to be
\begin{equation} \label{gr2}
F(r_+, T) = \frac{1}{4} r^2 \left(3 r^2-4 \pi  r T+3\right) \, .
\end{equation}
This agrees with the expression obtained earlier in eqn. (\ref{gr3}). We checked that the method described above works for all the cases discussed in this section, including the cases where there are other variables, such as the electric potential and the angular velocities.

\subsubsection{Two derivative case} \label{2d}

One way to construct the off-shell free energy in the BPS case would be follow the procedure outlined in subsection-(\ref{onshell}) to systematically take the zero temperature limit, and compute eqn. (\ref{eqn:FreeEnergy}), while at the same time using the BPS relations in eqn. (\ref{eqn:BPSM}) and the constraints given in eqn. (\ref{eqn:chargeconstraint})~\cite{Choi:2018vbz,Choi:2018hmj,Ezroura:2021vrt}. Though this can be done, we find it more convenient to obtain the same using the alternative derivation of free energy presented in the previous subsection-(\ref{alternative}), as the on-shell free energy and temperature are already known.\\

\noindent
Our starting point is the trial off-shell BPS free energy function taken as:
\begin{equation}\label{g4}
 F(a, \tau,\varphi') = W(a) + g(a, \varphi\prime)\left(1- \frac{\tau}{\tau_H}\right) \, ,
\end{equation}
where the on-shell free energy $W(a)$ and  $\tau_H$ are the ones in eqn. (\ref{free2}) and eqn. (\ref{tauH}). Here, we choose the order parameter to be $a$, which might jump discretely at the phase transition point. The unknown function $g(a, \varphi\prime)$  is to be determined by minimising the trial function in eqn. (\ref{g4}) as per the procedure outlined in subsection-(\ref{alternative}). We skip the details, and note the result of the procedure as:
\begin{equation} \label{ga}
g(a, \varphi\prime) = \frac{a (a+1)^2 \left(\pi  \sqrt{\frac{a+2}{a}} \left(17 a^2+20
   a-1\right) \varphi +6 \pi ^2 \left(11 a^2+8
   a-1\right)+(a+2)^2 \varphi ^2\right)}{(a-1)^2 (5 a+1)
   \left(\pi  \sqrt{\frac{a+2}{a}} \left(11 a^2+8
   a-1\right)+(a+2)^2 \varphi \right)} \, .
\end{equation}
Using eqn. (\ref{ga}) in eqn. (\ref{g4}), the final form of the exact off-shell BPS free energy comes out to be
\begin{eqnarray} \label{Fbps}
 &&   F =  \frac{a}{(a-1)^2 (5 a+1)^2} \left(2 (5 a+1) \left(\frac{\pi  \left(-2 a^2-5 a+1\right)}{\sqrt{\frac{a+2}{a}}}+\frac{1}{2} (a-3) a \varphi\prime \right) \right. \nonumber \\
    && \hskip 1.0cm  \left. -\frac{(a+1)^2 \left((a+2) \varphi\prime +6 \pi  a \sqrt{\frac{a+2}{a}}\right) \left((1-a) \tau  \left(\frac{3 \pi }{\sqrt{\frac{a+2}{a}}}+\frac{\varphi\prime }{2}\right)-5 a-1\right)}{a+2}\right)\, .
\end{eqnarray}

\begin{figure}[!htbp]
    \centering
 \subfloat[]{\includegraphics[width=0.45\linewidth]{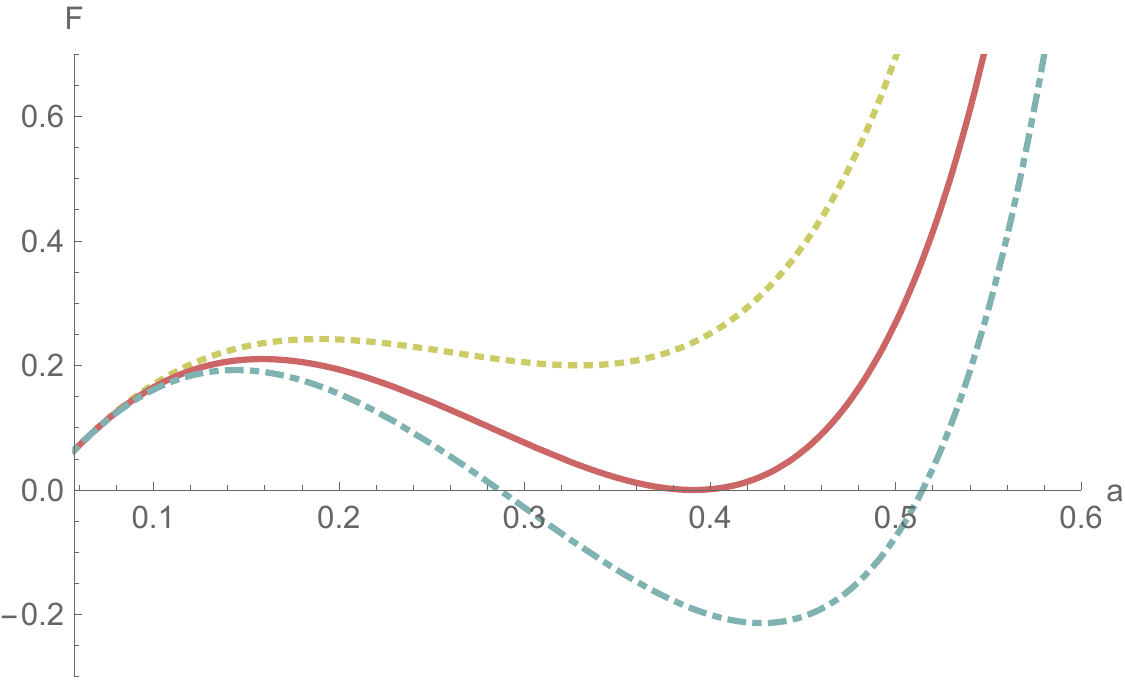}}~~~
 \subfloat[]{\includegraphics[width=0.45\linewidth]{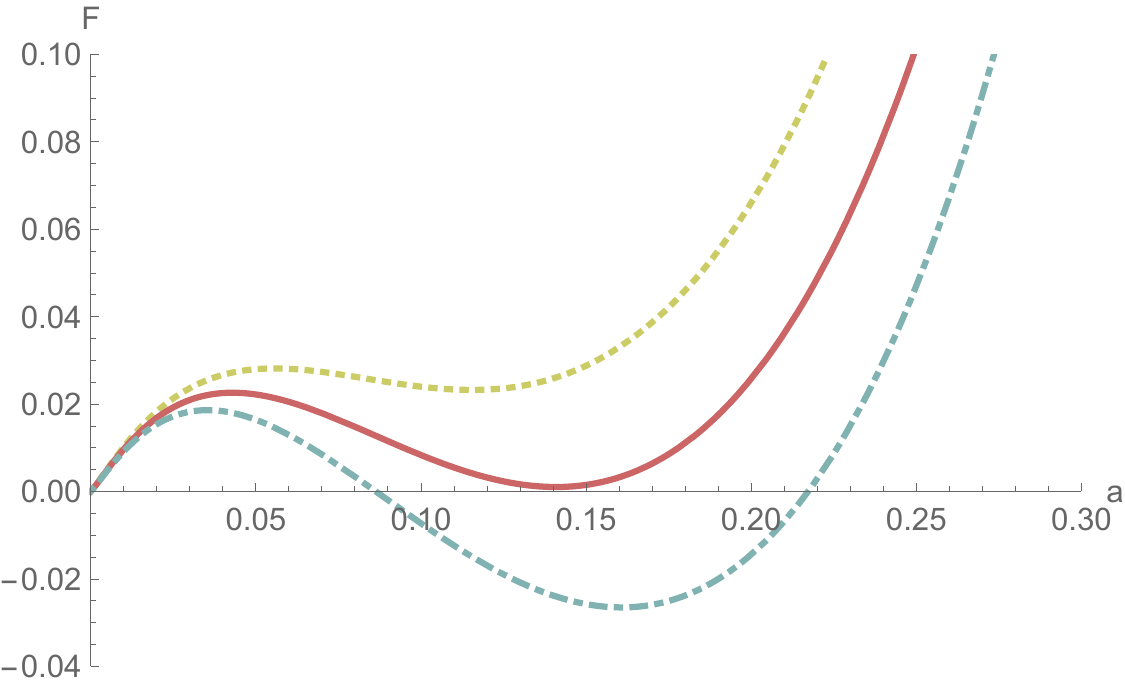}}
    \caption{Off-shell BPS free energy of eqn. (\ref{Fbps}) as
a function of the order parameter $a$ for: (a) $\varphi\prime = -\pi$ and values of $a$ and $F$ are limited at the lower end, starting from $a_{\rm min}=0.058$ and $F_{\rm min} = 0.064$ with $\tau_{\rm HP} = 2.166  $ (red curve)  (b) $\varphi\prime = 1$ with $\tau_{\rm HP} = 0.68$ (red curve). The temperature increases from top to bottom in both plots. }
    \label{bps3}
\end{figure}

\noindent
The off-shell BPS black hole phase diagram can now be studied in various cases, such as, the benchmark case ($\varphi\prime = 0$), $\varphi\prime \leq0$ and $\varphi' \geq 0$.
\noindent 
From eqn. (\ref{tauH}), one notes that for the large black hole case, $\tau$ diverges as $a \rightarrow 1$ and, as $a$ decreases, it attains a minimum value $\tau_{\rm cusp}$, which happens at $a_{\rm cusp}$. This also happens for the benchmark case $\varphi'=0$. However, when $\varphi'\leq0$, the parameter $a$ is limited to the range $1>a>a_{\min}$, where there turns out to be lower bound $\varphi'\geq \varphi'_{-}$~\cite{Ezroura:2021vrt}. Thus, the phase diagram in this case is expected to start from a finite positive value of free energy. The phase diagram following from the off shell BPS free energy in eqn. (\ref{Fbps}) for the  $\varphi' = -\pi$ and for the case $\varphi' = 1$ are shown respectively, in figures-(\ref{bps3})a and (\ref{bps3}b). The phase diagram for the benchmark case $\varphi\prime = 0$,  is similar to figure-(\ref{bps3}b), and discussed below. \\
\noindent
We make few observations before proceeding to discuss four derivative corrections to the BPS phase structure studied here.
First, we can take a closer look at  eqn. (\ref{Fbps}) by making a small $a$ expansion.  Assuming a real $\varphi'$, the first few terms of eqn. (\ref{Fbps}) go as
\be \label{Fseries1}
F(a,\tau)= 4 \sqrt{2} \pi  a^{3/2} -12 \sqrt{2} \pi  a^{5/2}-9 \pi ^2 a^2 \tau  +\varphi\prime  \left(\frac{87 \pi  a^{5/2} \tau }{2 \sqrt{2}}-3 \sqrt{2} \pi 
   a^{3/2} \tau -4 a^2+a\right)+ {\mathcal O}(a^{5/2}) \, .
\ee
We find that the above series in eqn. (\ref{Fseries1}) exactly matches with the one obtained from Landau's phenomenological approach in eqn. (\ref{Fseries3}), for $\varphi'=0$, and with the replacement $a = u^2$. This is a consistency check. \\

\noindent
{\underline{The benchmark case ~$\varphi' = 0$}}: \\

\noindent
For $\varphi' =0$, after some rearrangement, the free energy in eqn. (\ref{Fbps}) can actually be expressed more succinctly, directly in terms of the BPS charges as
\begin{equation} \label{fbps1}
	F(a,\tau) = W(a)  +  2\left(\frac{J^*+Q^*}{\tau_H}\right) \left(1- \frac{\tau}{\tau_H}\right) \, ,
\end{equation}
where
\begin{equation}
	W(a) =  - S^* + 2\left(\frac{J^*+Q^*}{\tau_{\rm H}}\right)  \, ,
	\label{on3}
\end{equation} \label{off2}
and the BPS charges are
\begin{align}
	\label{eqn:BPSM}
	M^* &= N^2 \frac{a(3 - a^2 -2a^3)}{(1-a)^4}~,\qquad 
	Q^* = N^2 \frac{ a}{(1-a)^2}~,
	\\ 
	\label{eqn:BPSJa}
	J^* &= N^2 \frac{a^2 (a+3)}{(1-a)^3} ~, \qquad
	\quad \qquad S^* = N^2\frac{2 \pi  a \sqrt{a (a+2)}}{(1-a)^2}~\,\nonumber ,
\end{align}
which are known to satisfy an additional constraint
\begin{equation} 
	\label{eqn:chargeconstraint}
	\left(Q^{*3} +\frac{1}{2}N^2J^{*2} \right) - \left(3Q^*+\frac{1}{2}N^2 \right) \left(3Q^{*2} - N^2 J^* \right)= 0~.
\end{equation} 
The on-shell BPS free energy of eqn. (\ref{free2})~\cite{Ezroura:2021vrt}  can be obtained from eqn. (\ref{on3}) upon using the values of BPS charges.\\

\noindent
Now, the off-shell BPS free energy in eqn. (\ref{fbps1}) can be written in a more traditional form as:

\be \label{fb1}
F \,=\,{\mathcal E} \, - \, \tau\, {\mathcal S}\, ,
\ee
which is the general structure one should expect, if the analogy with standard thermodynamics is correct. The quantities
${\mathcal E} $ and ${\mathcal S}$ appearing in eqn. (\ref{fb1}) can then be interpreted as the effective energy and entropy of the BPS system. In fact, from the usual thermodynamic relations,
\be \label{Seff}
{\mathcal S} = - \frac{\partial F (a,\tau)}{\partial \tau} =  2\left(\frac{J^*+Q^*}{\tau_H^2}\right) = \frac{18 \pi ^2 a^2 (a+1)^2}{(5 a+1)^2 \left(2-a-a^2\right)}\, ,
\ee
should be thought of as the conjugate of BPS temperature $\tau$, with the effective energy identified as
\be \label{Eeff}
{\mathcal E} = -S^* + 4\left(\frac{J^*+Q^*}{\tau_H}\right) = \frac{2 \pi  a^{3/2} \left(a^2+a+4\right)}{(a-1)^2 \sqrt{a+2} (5 a+1)}\, .
\ee
\noindent
Let us recollect that, the earlier proposal in~\cite{Choi:2018vbz,Choi:2018hmj} was to start from an on-shell free energy in the CFT\footnote{The CFT expression will appear in eqn. (\ref{fb2}), whose corresponding bulk result is given in eqn. (\ref{free2})}, where, after transcribing the expressions to bulk quantities, one takes $(Q^* + J^*)$ to be the conjugate to the BPS temperature $\tau$, with a Hawking-Page type transition,  beyond which, the large black hole phase is preferred over gas of  the `thermal' gravitons.  
However, there was a mismatch between the Hawking-Page (HP) type transition temperature and the confinement-deconfinement type temperature of the dual CFT~\cite{Choi:2018vbz}. This led to some speculations on the possible existence of new gravitational saddles, though, this was argued in~\cite{Copetti:2020dil} to be otherwise. Since, the free energy used in ~\cite{Choi:2018vbz,Choi:2018hmj,Ezroura:2021vrt} was on-shell, it was not straightforward to identify the conjugate to the BPS temperature. Considering the fact that there still exists a notion of thermodynamics on the BPS surface, despite the fact the original free energy and temperature vanish identically, it points towards the existence of non-zero energy, which competes with the second term in eqn. (\ref{fb1}), resulting in various equilibrium phases. Thus, the rearrangement of BPS quantities suggested by eqn. (\ref{fb1}), leading to novel effective entropy in eqn. (\ref{Seff}) as the thermodynamic conjugate to the BPS temperature $\tau$, and the non-zero effective energy 
in (\ref{Eeff}), seems plausible, and might be the correct way to identify thermodynamic variables, as they follow from standard relations. This of course is at best a better guess, and needs to be verified by other possible means. \\

\begin{figure}[!htbp]
    \centering
 \subfloat[]{\includegraphics[width=0.32\linewidth]{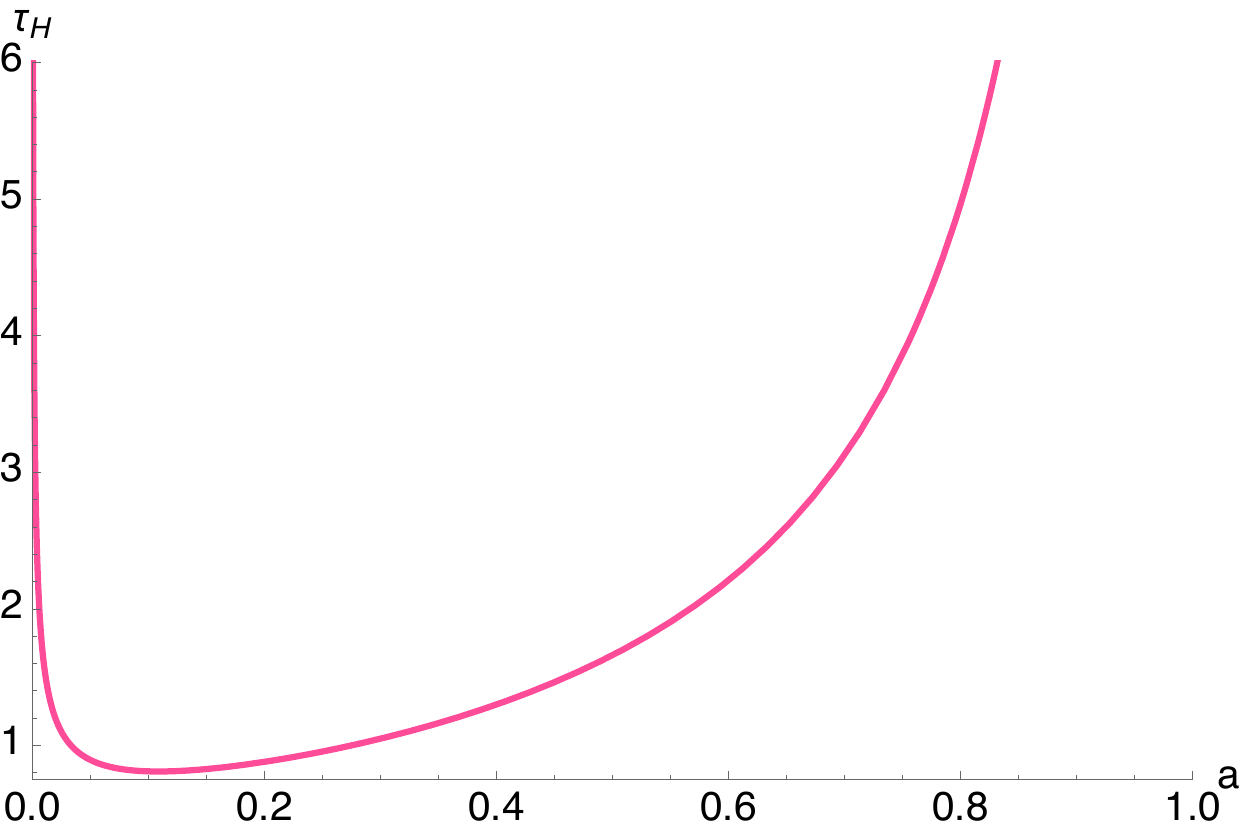}}
 \subfloat[]{\includegraphics[width=0.34\linewidth]{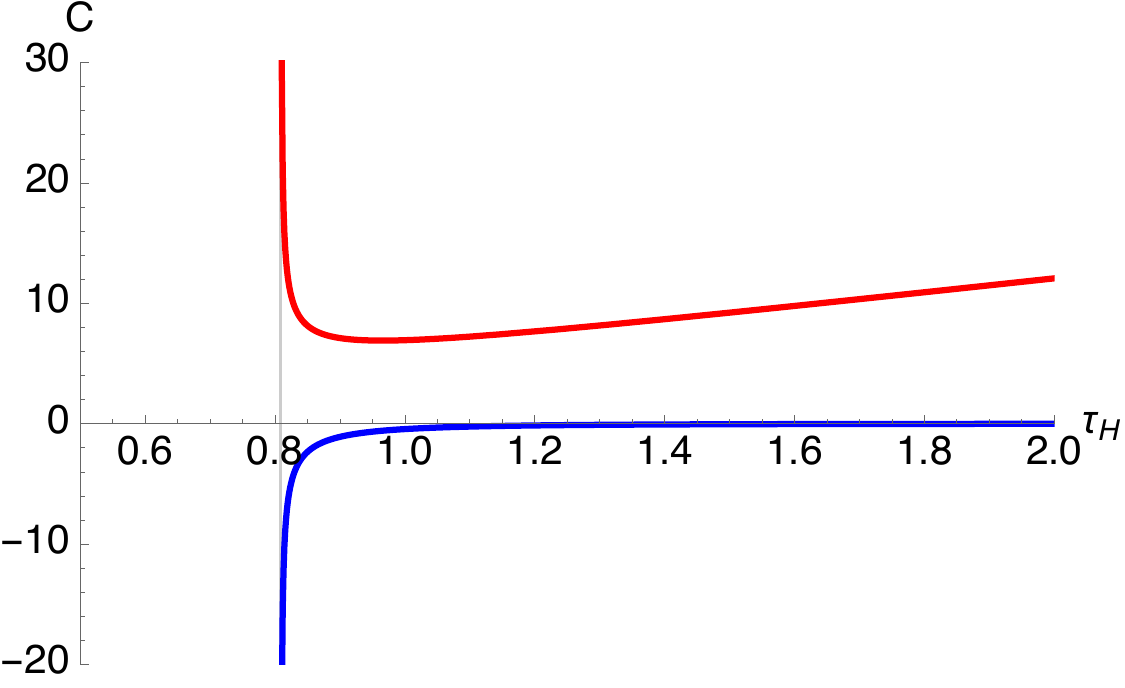}}
 \subfloat[]{\includegraphics[width=0.34\linewidth]{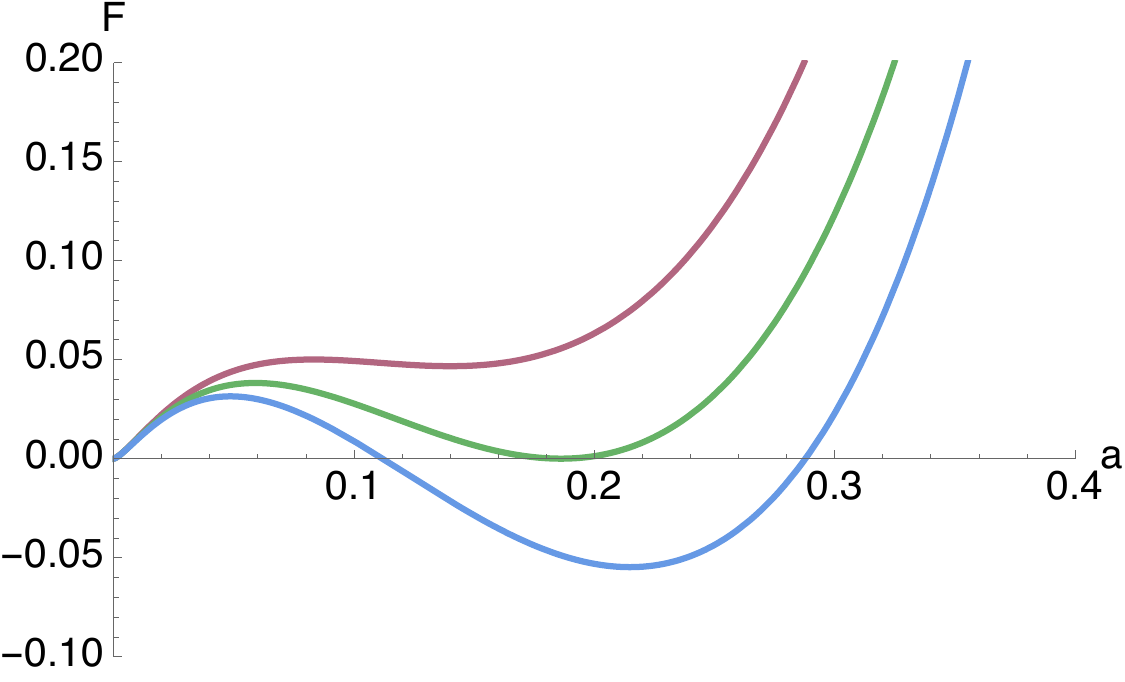}}
    \caption{(a) BPS temperature given in eqn.(\ref{tauH}), showing two black hole branches (b):   Specific heat $C=\tau_H \frac{d{\mathcal S}}{d\tau_H}$, calculated using eqn.(\ref{Seff}): positive for larger branch (red) and negative for smaller branch (blue) (c) Off-shell BPS free energy given in eqn.(\ref{fb1}) for three different temperatures:  $\tau_{\rm H} = 0.82 ({\rm red}), 0.8626 ({\rm green}, {\rm HP ~transition}), 0.9 ({\rm blue}) $  All the plots are for the case $\varphi'=0$.   }
    \label{fig40}
\end{figure}

\noindent 
In our off-shell formulation, the BPS HP transition temperature and the confinement-deconfinement temperature of the dual CFT are exactly the same (as we show in the sequel in section-(\ref{bef2d2})), unlike the case in~\cite{Choi:2018vbz,Choi:2018hmj}. The BPS temperature and the off-shell free energy plotted in figures-(\ref{fig40})(a) and (\ref{fig40})(c),  show the presence of small and large black hole branches, with the former being unstable and latter being stable. This is also seen from figure-(\ref{fig40})(b), where the specific heat of the smaller branch is negative and that of the larger branch is positive, separated by a divergence which appears at $a=0.108$. 
Exactly similar features will be seen while studying the boundary effective potential in section-(\ref{bef2d2})).
All these aspects are also pretty familiar from the thermodynamics of AdS Schwarzschild black holes, and it seems they continue to hold in the BPS limit, though there are subtle differences, as we discuss further below after the inclusion of four-derivative corrections.  The thermodynamics being discussed here on the BPS surface appears through a limiting procedure in the  grand canonical ensemble and characterised only by the BPS potentials $(\Phi',\Omega')$, which were traded in for the variables $(\tau,\varphi')$. The effective quantities ${\mathcal E} $ and ${\mathcal S} $ though are defined on the BPS surface and are quite different from the BPS mass $M^*$ (which has been eliminated using eqn. (\ref{eqn: bpscond})) and entropy $S^*$ of the black hole. The thermodynamics coming from equation-(\ref{fb1}) points towards a new BPS system, whose thermodynamic quantities are rearranged versions of the quantities given by the BPS black hole. This proposed new BPS thermodynamic system might also be attributed to new unknown gravitational saddles or an interacting mixture of multiple solutions. In fact, possible presence of new hairy black holes, solitons, and corresponding dual phases in the CFT have been mooted for long time, from consistent truncations of ${\mathcal N}=8$ supergravity with scalar hair~\cite{Basu:2010uz,Bhattacharyya:2010yg,Markeviciute:2016ivy,Markeviciute:2018yal,Markeviciute:2018cqs,Dias:2024edd}. Attractor mechanism ensures that in the BPS limit, the entropy of these black holes is still dependent only on a single parameter, and the weakly interacting mixture of black holes and solitons gives reasonable bounds on mass and entropy~\cite{Dias:2024edd}. However, these gravitational solutions have only been studied either perturbatively or numerically, especially, in the large charge limit. In addition, the energy of these hairy solutions are bounded by the BPS energy, unlike the situation here.  Understanding these discrepancies  might aid us in deciphering BPS black hole thermodynamics. It would also be nice to identify an interacting system of particles, whose microscopic formulation might give us the desired thermodynamic relations found here in eqns. (\ref{fb1}), (\ref{Seff}) and (\ref{Eeff}).  \\
\noindent
\FloatBarrier
\noindent 
Finally, the results in \cite{Choi:2018vbz,Choi:2018hmj,Ezroura:2021vrt} can be reproduced after taking appropriate on-shell limit of our off-shell expressions in this section. We emphasise that the off-shell description is more general, as it captures the various phases of  the system both at and around the equilibrium. The emergence of a hypothetical HP transition and the behaviour of the large black hole branch all work out cleanly for $\varphi'$ taken in the range $(\varphi'_-,\infty)$ or if $\varphi'=0$. Though, the BPS thermodynamics around the small black hole branch varies with the sign of $\varphi'$. Also, though we assumed $\varphi'=0$ for simplicity, it should be possible to generalise to non-zero $\varphi'$. We leave these issues for future.

\subsubsection{Four derivative case} \label{4d}

In this subsection, we obtain the off-shell free energy of BPS black holes, which follow from the analysis in section-(\ref{2d}), and in addition contain the four-derivative corrected BPS charges. The corrected charges and the entropy with higher derivative corrections and in the BPS limit, were obtained from the on-shell action in~\cite{Cassani_2022}. There are of course various ways to arrive at the BPS limit and we summarise the relations obtained in~\cite{Cassani_2022}.
\begin{eqnarray}
Q^{*}&=&\frac{ \pi a}{2G (1-a)^2}\left[1+4  \lambda _2\alpha +4\lambda_1\alpha \frac{1+8 a +36 a^2 +44 a^3 +19 a^4}{a\left(-1+8 a
   +11 a^2 \right)}\right]\,, \label{QfromI}\\[2mm]
J^{*}&=&\frac{\pi a^2 (3+a )}{2 G (1-a)^3}\left[1+4
   \lambda _2 \alpha +24 \lambda_1\alpha \frac{1+9 a +29 a^2 +25 a^3 +8 a^4 }{a (3+a ) \left(-1+8 a +11 a^2 \right)}\right], \label{QfromI1} \\
{\cal S}^{*}\,&=&\, \frac{\pi ^2 a \sqrt{a (a +2)}}{ G (1-a)^2}\left[1+4 \lambda _2\alpha   +48\lambda _1 \alpha   \frac{2 a^2 +5 a +2}{11 a^2 +8 a -1}\right]\,.\label{QfromI2}
\end{eqnarray}
In addition, the constraint in eqn. (\ref{eqn:chargeconstraint}) receives corrections depending on $\alpha,\lambda_1, \lambda_2$~\cite{Cassani_2022}. The zeroth-order terms in $\alpha$ of the above expressions agree with the BPS limit of entropy and the charges given in the case of two-derivative quantities in section-(\ref{2d}), as expected. The above quantities also satisfy the supersymmetry relation $E^* =  2 J^* + 3Q^* $~\cite{Cassani_2022}.\\

\begin{figure}[!htbp]
    \centering
  \includegraphics[width=1.0\linewidth]{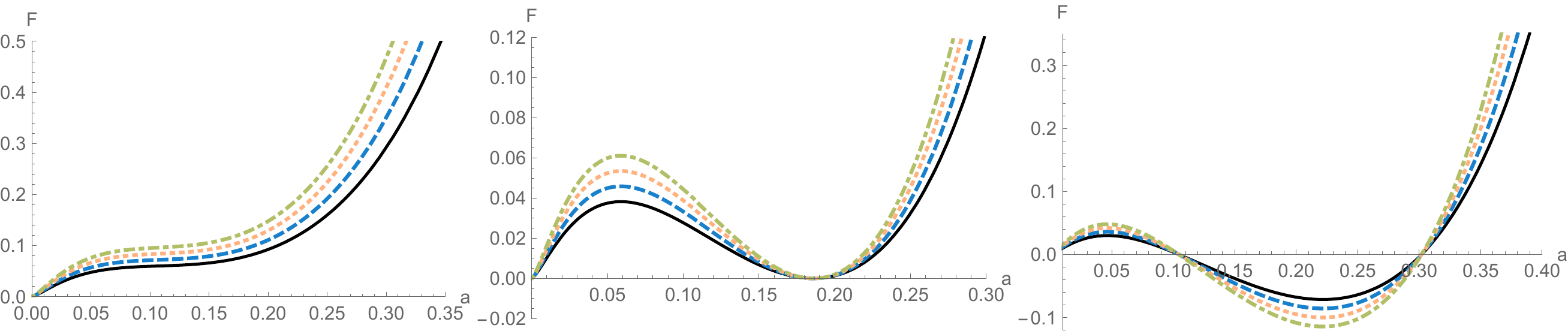}
    \caption{Off-shell BPS free energy for the four-derivative case as
a function of the order parameter $a$. The left, middle and the right plots in the figure, correspond to the BPS temperature values $\tau = 0.8,0.8626, 0.91$, respectively. In all the plots: for the solid curves we set  ``$\alpha = \lambda_{1} = \lambda_{2} = 0$", and for the dashed curves we set ``$ \lambda_{1} = .001, \lambda_{2} = 100$". For the dashed curves in each plot we take $\alpha = .0005$ (blue), $\alpha = .001$ (orange) and $\alpha = .0015$ (green).   }
    \label{fig31}
\end{figure}

\noindent  
In the present work, we once again concentrate on the benchmark case $\varphi'=0$, but the analysis can be done for more general values of $\varphi'$ also. We can now directly use the master formula for off-shell BPS free energy arrived at in terms of BPS charges in eqn. (\ref{fbps1}).
Thus, using the above corrected quantities of eqns. (\ref{QfromI}), (\ref{QfromI1}) and (\ref{QfromI2}) in eqn. (\ref{fbps1}), the off-shell free energy of BPS black holes in the four-derivative case can be computed to be
\begin{eqnarray} \label{offalpha}
&& F(a, \tau,\alpha,\lambda_1,\lambda_2) = F_{0}(a,\tau) + \frac{\left(8 \sqrt{a}\pi \alpha \left(f_{\lambda_1} + f_{\lambda_2}  \right) \right) }{(a-1)^2 (a+2)^{3/2} (5 a+1)^2 (a (11 a+8)-1)}\, , \\
&& f_{\lambda_1} = 3 \lambda_1 \left(3 \pi  \sqrt{a (a+2)} \left(29 a^6+96
   a^5+57 a^4-100 a^3-69 a^2-12 a-1\right) \tau \right. \\ \nonumber
   && \hskip 1.0cm \left. +90
   a^7+508 a^6+1158 a^5+1272 a^4+622 a^3+192 a^2+42 a+4\right) \, , \\
&& f_{\lambda_2} = \lambda_2 \left(9 \pi  \sqrt{a^3 (a+2)} \left(11 a^3-3
   a^2-9 a+1\right) (a+1)^2 \tau \right. \\ \nonumber
   && \hskip 1.0cm \left. +a (a+2) (5 a+1)
   \left(a^2+a+4\right) (a (11 a+8)-1)\right)\, ,
\end{eqnarray}
where $F_{0}(a,\tau)$ is the two-derivative off-shell BPS free energy given in eqn. (\ref{Fbps}), where we have to set $\varphi'=0$. The free energy is plotted in figure-(\ref{fig31}) for three different temperatures. Since the BPS temperature for two and four derivative cases is same, the effect of four derivative terms on the free energy curves can be traced. \\

\noindent
The four derivative terms modify the phase structure of BPS black holes (seen from figure-(\ref{fig31})), in a way similar to phase diagram of general black holes (see figure-(\ref{fig32})). For $ T < T_{\rm HP}$ (left plot in figure-(\ref{fig31})), both small and large black hole phases further destabilize. For $ T = T_{\rm HP}$ (middle plot in figure-(\ref{fig31})), the small black hole phase further destabilizes, and finally for $ T > T_{\rm HP}$ (right plot in figure-(\ref{fig31})), the small black hole phase further destabilizes, while the large black hole phase stabilizes.
\FloatBarrier

\noindent
{\underline{Asymptotic phase structure}}: 
\noindent
In~\cite{Ezroura:2021vrt}, a comparison was made between the phase structure of AdS Schwarzschild black holes and the BPS black holes, which appear similar but the asymptotic structures do not match, particularly for the large black hole branch. 
 In the present situation, we can obtain the corrected on-shell BPS free energy and study the effect of four-derivative corrections on the asymptotic structure, which can then be compared with behaviour of AdS Schwarzschild free energy.\\
 
 \noindent
The four-derivative corrected AdS Schwarzschild on shell free energy can be obtained from eqn. (\ref{2.1999}) by putting $\Phi = 0$, and setting temperature to the on-shell temperature in eqn. (\ref{gr3}):
\begin{equation}
 G(r_+,\alpha,\lambda_1,\lambda_2) = -\frac{1}{4} \left(r_+^2-1\right) r_+^2 \left(4 \alpha  \lambda _2+1\right)-\frac{9}{2}
   \alpha  \lambda _1 \left(r_+^2+1\right)^2   \, ,
\end{equation}
 whereas the rescaled free energy $W = G/T$ is
 \begin{equation}
 W = -\frac{\pi  r_+ \left(\left(r_+^2-1\right) r_+^2 \left(4 \alpha  \lambda _2+1\right)+18
   \alpha  \lambda _1 \left(r_+^2+1\right)^2\right)}{4 r_+^2+2}  \, .
\end{equation}
 For very small AdS-Schwarzschild black holes ($r_+ << 1$), $T \sim 1/r_+$ and $W$ has the expansion
 \be
 W \sim -9 r_+ \left(\pi  \alpha  \lambda _1\right)+\frac{1}{2} \pi  r_+^3 \left(4 \alpha  \lambda
   _2+1\right)+O\left(r_+^4\right) \, .
 \ee
Thus, $W \sim -r_+$ (four-derivative corrections modify the $r_+ << 1$ behaviour) and hence
\be
W \sim -T^{-1} \quad \mathrm{as} \quad T \rightarrow \infty \, \qquad {\rm (small~black~holes)}
\ee
For large black holes ($r_+ \rightarrow \infty$), we have $T \sim r_+$ and 
\be
W \sim -\frac{1}{4} \pi  r_+^3 \left(18 \alpha  \lambda _1+4 \alpha  \lambda
   _2+1\right)-\frac{3}{8} \pi  r_+ \left(18 \alpha  \lambda _1-4 \alpha  \lambda
   _2-1\right)\, + \cdots .
\ee
Thus, $W \sim -r_+^3$ (four-derivative corrections do not modify the $r_+ \rightarrow \infty$ behavior) and hence,
\be
W \sim  -T^{3} \quad \mathrm{as} \quad T \rightarrow \infty \, \qquad {\rm (large~black~holes)}
\ee
 
 \noindent
 We now check the asymptotic structure of four derivative corrected BPS black hole free energy.  To do this, we use the on-shell temperature of eqn. (\ref{tauH})(with $\varphi' = 0$) in eqn. (\ref{offalpha}) and obtain the four-derivative corrected on-shell BPS free energy as 
\begin{eqnarray} \label{4dcorrectedW}
&& W(a, \tau,\alpha) = W(a,\tau) -\frac{8 \pi  \sqrt{\frac{a}{a+2}} \alpha  \left(3 \left(a^3+5 a^2+5
   a+1\right) \lambda _1+a \left(2 a^2+5 a-1\right) \lambda
   _2\right)}{(a-1)^2 (5 a+1)} \, ,
\end{eqnarray}
where $W(a,\tau) $ is the BPS free energy in the two-derivative theory given in eqn. (\ref{free2})(with $\varphi' = 0$). From the above expression, we note the following asymptotic behaviors. For very small BPS black holes, as  $a \rightarrow 0^+$, $\tau$ diverges as $a^{-1/2}$ and where as 
\be
W(a,\alpha) = -12 \sqrt{2} \pi  \alpha  \sqrt{a} \lambda _1 - \sqrt{2} \pi  a^{3/2} \left(21 \alpha  \lambda _1-4 \alpha  \lambda
   _2-1\right) + O(a^{5/2}) \, ,
\ee
and thus,
\be
 W \sim -\tau^{-1} \quad \mathrm{as} \quad \tau \rightarrow \infty \, , \qquad {\rm (small~BPS~black~holes)}
 \ee
We note that the four-derivative corrections modify the small BPS black hole behaviour.
 For large BPS black hole $a \rightarrow 1^-$, $\tau$ diverges as $(1-a)^{-1}$ and $W$ goes as
 \be
 W(a,\alpha) =  -\frac{4 \pi  \left(4 \alpha  \left(3 \lambda _1+\lambda
   _2\right)+1\right)}{\sqrt{3} (1-a)}-\frac{2 \pi  \left(4 \alpha 
   \left(6 \lambda _1+\lambda _2\right)+1\right)}{\sqrt{3}
   (1-a)^2}-\frac{\pi  \left(4 \alpha  \left(6 \lambda _1+5 \lambda
   _2\right)+5\right)}{3 \sqrt{3}} + \cdots \, ,
 \ee
 and thus,
\be
 W \sim - \tau^{2} \quad \mathrm{as} \quad \tau \rightarrow \infty \, , \qquad {\rm (large~BPS~black~holes)}
 \ee 
Thus, the asymptotic structure of the small black hole branch is same for both the general Schwarzschild black holes in AdS, as well as BPS black holes (though the behaviour is modified by four-derivative corrections). On the other hand, the four-derivative corrections do not change the asymptotic behavior of free energy for large BPS black holes or Schwarzschild AdS black holes.
 In summary, in the presence of four-derivative corrections, the asymptotic behavior of free energy of BPS black holes w.r.t to the BPS temperature $\tau$, does not agree with the behavior of AdS Schwarzschild black holes with temperature $T$, for large black hole limit(as in two-derivative case~\cite{Ezroura:2021vrt}). Thus, the qualitative identification of $T$ and $\tau$ is not accurate even with the inclusion of four-derivative corrections.

\section{Effective potentials in the boundary theory} \label{2.4}

AdS/CFT correspondence gives a handle in trying to understand the microscopic description of supersymmetric black holes~\cite{Sen:1995in,Strominger:1996sh}, though the methods also address situations where supersymmetry is not preserved. A key insight in our understanding of microscopics of black holes is to view them as an ensemble of microstructures in a boundary gauge theory.  Last several years, a plethora of black holes in AdS spacetimes have been constructed, including charges and rotations and systematic methods have been developed to view them as microscopic constituents of a dual quantum field theory. A well appreciated limitation in this endeavor is the unavailability of reliable techniques to perform detailed checks at strong coupling in the dual gauge theories. Some computations at large $N$ work out nicely. For example, Bekenstein-Hawking entropy of a class of supersymmetric  black holes in AdS$_4$ was shown in~\cite{Benini:2015eyy}, to follow by preforming a  localisation computation in the ABJM theory~\cite{Benini:2015noa}. In other cases, some progress has been made recently~\cite{Pestun:2007rz,Nian:2019pxj,Cabo-Bizet:2018ehj} (see~\cite{Zaffaroni:2019dhb} and references therein), in particular, for the class of black holes of~\cite{Gutowski:2004yv,Chong:2005hr,Chong:2005da,Kunduri:2006ek}.\\

\noindent
 Since the class of  black holes considered here are solutions to the five-dimensional 
minimal gauged supergravity theory, one expects the holographically dual to exist in any four-dimensional $\mathcal{N}=1$ superconformal field theory, having a weakly coupled supergravity dual background. The charge and rotation parameters in the bulk map via AdS/CFT conjecture to the R-charges and rotations, with their respective thermodynamic potentials. 
Assuming such a set up, with a gauge theory dual to the $AdS_5$ black hole, we pursue the problem of  phenomenologically identifying an effective potential of the gauge theory at strong coupling, which describes its equilibrium properties. As the gauge theory is strongly coupled, a direct way to obtain the effective potential in terms of an order parameter (such as, the R charge) may in general be difficult.  In the last section, an off-shell construction of free energy was presented for this class of black holes, following the insights of~\cite{Ezroura:2021vrt}.  Now, it should be possible to make use of the AdS/CFT correspondence (as it has been done in certain non-supersymmetric situations in the past~\cite{Dey:2007vt,Dey:2006ds,nayak2008bragg,Banerjee:2010ve,Banerjee:2010ng}),  to propose an effective potential, the saddle points of which capture the various phases of the gauge theory. We should however draw attention to the fact that the effective potential constructed this way may not quite be unique, except probably close to the transition points.  In this section, we restrict ourselves to constructing effective potentials in gauge theories dual to general non-rotating black holes in AdS$_5$, with non-zero electric potential. In the following subsections, we will study the potentials constructed from off-shell free energy of BPS black holes where both the charge and rotation are present.

\subsection{Boundary effective potential: Two-derivative action} \label{bef2d}
\begin{figure}[!htbp]
    \centering
\subfloat[]{\includegraphics[width=0.5\linewidth]{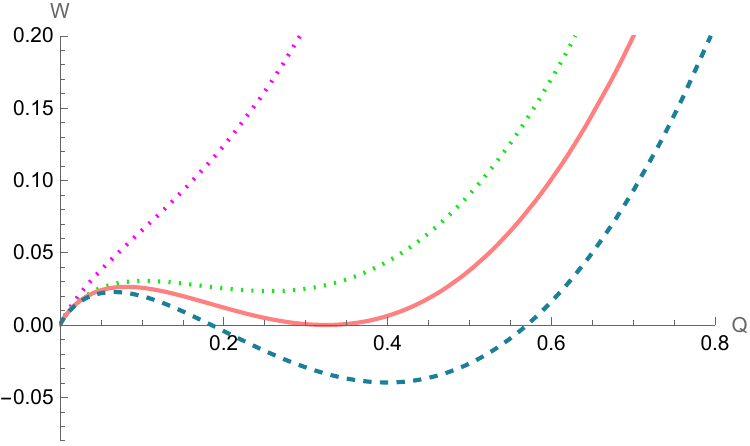}}~~~
   \subfloat[]{  \includegraphics[width=0.5\linewidth]{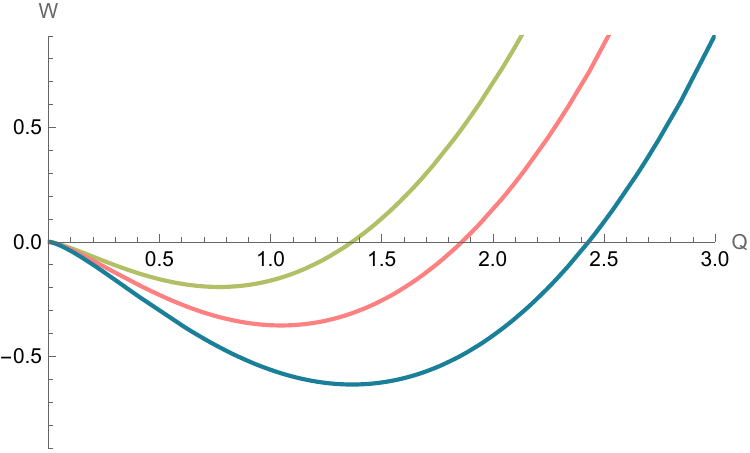}}
    \caption{(a) Behaviour of the boundary effective potential as
a function of the order parameter $Q$ for $\Phi = \frac{\sqrt{3}}{2}$. The temperature increases from top to bottom. The red solid curve corresponds to transition at $T = T_{c} = \frac{3 \sqrt{3}}{4 \pi } $ (b) Behaviour of the boundary effective potential as
a function of the order parameter $Q$ for $\Phi = \sqrt{3}$. The temperature increases from top to bottom. The green, red, blue curves corresponds to $T = 0.3, 0.35, 0.4 $ respectively}
    \label{figep1}
\end{figure}

\noindent
While in the bulk, the order parameter for the off-shell free energy was the horizon radius $r_+$, in the dual theory, one expects the corresponding
order parameter to be the physical charge, $Q$. To keep things analytically tractable, we consider non rotating $AdS_5$ black holes, and hence set $\Omega = 0$. The off-shell free energy for this situation was constructed in eqn. (\ref{offphi}).  We set $N=1$ from now on\footnote{One should, as such, scale $Q$ as $\frac{Q}{N^2}$ and F as $\frac{F}{N^2}$, as in the de-confined phase, both the free energy and charge are of order $N^2$. }. 
Since $a=0$, the charge and electric potential in eqn. (\ref{CCLPcharges}) and (\ref{electrostatic_pot_CCLP}) simplify to $
 Q = \frac{\sqrt{3}q}{2},~ \Phi = \frac{\sqrt{3}q}{r^2}   $. 
Using these expressions in eqn. (\ref{offphi}), the off-shell free energy can be expressed in the following form
\begin{equation} \label{bdyfree}
    W(Q, \Phi, T) = \frac{Q \left(6 Q-\Phi  \left(4 \pi  \sqrt{2} T \sqrt{\frac{Q}{\Phi }}+\Phi ^2-3\right)\right)}{2 \Phi ^2} \, .
\end{equation}
Eqn. (\ref{bdyfree}) is thus our proposed effective potential in the gauge theory, with the charge $Q$ as the order parameter. The phase diagram following from $W$ is shown in figure-(\ref{figep1}). We see a first order phase transition. This phase transition corresponds to the confinement-deconfinement transition
in the strongly coupled gauge theory as discussed in \cite{Witten:1998zw}.
\noindent
The temperature of the gauge theory can be found by extremising the effective potential $W$ with respect to the order
parameter $Q$ and this comes out to be
\begin{equation}\label{2.29}
    T = \frac{\sqrt{\frac{Q}{\Phi }} \left(12 Q-\Phi ^3+3 \Phi \right)}{6 \sqrt{2} \pi  Q} \, .
\end{equation}
To find the confinement-deconfinement
transition temperature, we solve $W = 0$ which gives
\begin{equation}\label{2.30}
    T = \frac{\sqrt{\frac{Q}{\Phi }} \left(6 Q-\Phi ^3+3 \Phi \right)}{4 \sqrt{2} \pi  Q}
\end{equation}
Now, equating equation \ref{2.29} and \ref{2.30}, the critical order parameter $Q\textsubscript{c}$ can be found to be
\begin{equation}
    Q\textsubscript{c} = -\frac{1}{6} \Phi  \left(\Phi ^2-3\right)
\end{equation}
Using $Q_c$ back in either eqn. (\ref{2.29}) or (\ref{2.30}) the critical temperature, $T_c$ for the confinement-deconfinement transition can be found to be
\begin{equation}
    T_c = \frac{\sqrt{9-3 \Phi ^2}}{2 \pi }
\end{equation}
Expression for $T_c$ is found to be same as $T_{HP}$ in the bulk. 
Phase diagram for critical electric potential limit is presented in figure-(\ref{figep1}b). 
\FloatBarrier

\noindent
To conclude this sub-section, for the supersymmetric black holes in AdS$_5$, we were able to propose a candidate off-shell effective potential in terms of an order parameter $Q$, which, on-shell, reproduces all the stable phases of the boundary CFT at finite temperatures and finite non-zero chemical potential. The results of this section can be generalised to the situation with non-zero angular velocity, which needs to be done numerically, but we do not pursue it here. Also, the four-derivative corrected boundary effective potential $W(Q, \Phi, T,\alpha)$ can be constructed analogous to the one done above, where the starting point is the bulk off-shell free energy in eqn. (\ref{2.199}) (expression given in appendix-(\ref{A2})). We checked that the phase diagram of the gauge theory obtained in this way from the corrected effective potential, captures the equilibrium features seen in the bulk.  

\subsection{Boundary effective potential for BPS black holes: Two-derivative} \label{bef2d2}
In section-(\ref{bef2d}), using the AdS/CFT correspondence, we proposed a phenomenological boundary effective potential in the gauge theory, using the off-shell free energy of gravitational thermodynamics of AdS$_5$ black holes. Having discussed the off-shell free energy of BPS black holes in a previous subsection-(\ref{2d}), it is natural to seek a boundary effective potential in this case too, particularly, because the off-shell phase diagram of BPS black holes closely resembles the AdS-Schwarzschild black  thermodynamics \cite{Chamblin:1999tk, Witten:1998qj, Natsuume:2014sfa, Kubiznak:2016qmn}. One should of course also be cautious in using the BPS black hole thermodynamics, while a more clear interpretation of equilibrium points of the free energy emerges (as the BPS temperature $\tau$ is varied), in analogy with the traditional black hole thermodynamics. However, supposing the existence of such a BPS phase structure,
it is expected that a boundary effective potential should also show an analogue confinement-deconfinement like transition w.r.t some order parameter defined in the BPS case. Then, such a holographic study of BPS thermodynamics may be amenable to detailed analysis while still preserving supersymmetry \cite{tHooft:1977nqb, Polyakov:1978vu, Witten:1998zw, Susskind:1979up,Aharony:2003sx}, uncovering deeper aspects of BPS surface. In this subsection, our aim is to propose such an effective potential at strong coupling of the boundary CFT, which captures the bulk BPS phases. \\

\noindent
Before proceeding,  one has to identify an order parameter in the gauge theory dual to the BPS black hole. The only available parameters on which the off-shell phase diagram was based were $(a,\varphi')$, though one keeps in mind that, these variables are thought of as functions of the potentials $(\Phi',\Omega')$. The off-shell BPS phase diagram of previous subsection was based on various possible real values of $\varphi'$, with $a$ as the order parameter.  As emphasised in~\cite{Ezroura:2021vrt}, the possible values of the new fugacity $\varphi'=\Phi' - 2 \Omega'$ can be inferred, by studying the connection of two quantities derived independently, namely, the  gravitational BPS free energy in eqn. (\ref{free2}) and the HHZ free energy (obtained from an extremisation principle in the BPS case). The HHZ free energy is genuinely complex~\cite{Hosseini:2017mds},
\begin{equation}
\label{hhz}
    H (\Delta, \omega) = -\frac{N^2}{2} \frac{\Delta^3}{\omega^2}~,
\end{equation}
with the complex potentials it depends on, satisfying a constraint: \cite{Hosseini:2017mds,Zaffaroni:2019dhb} $
    3\Delta - 2\omega = 2\pi i$. 
The potentials $\Delta$ and $\omega$ are conjugate to the charges $Q$ and $J$. After expressing $H (\Delta, \omega)$ of eqn. (\ref{hhz}) as a function of the parameter $a$ (with the potentials satisfying a complex constraint), a possible choice for the fugacity $\varphi'$ in the BPS free energy is $2\pi i$, whence, $- W(a) = H$. This value of $\varphi'$ is well motivated, as the BPS partition function and the superconformal index match in this case, and hence, the comparisons between the week and strong coupling relations in the CFT are justified. Our strategy on the other hand is to start from the expression in the off-shell BPS free energy in the bulk (eqn. (\ref{Fbps})), and use the AdS/CFT dictionary to obtain the boundary off-shell free energy. The expressions we constructed  for the off-shell BPS free energy in the bulk (eqn. (\ref{Fbps})) are valid for a general value of $\varphi'$ (taken to be real in black hole thermodynamics). Since, our bulk construction is in semi-classical gravity, we expect the boundary free energy we construct using AdS/CFT to be valid in general at strong coupling (for any real value of $\varphi'$). \\

\noindent
Further, as argued in \cite{Ezroura:2021vrt}, for the special value $\varphi'=0$, the bulk BPS free energy is exactly equal to the real part of HHZ free energy (see eqn. 3.75 of \cite{Ezroura:2021vrt}). So, it may be helpful to study this limit, in addition to other general values of $\varphi'$. With the above aspects in mind, we pick $\varphi'=0$, as the expressions for the free energy are slightly simpler and presentable, but generalisation to any other value is straightforward. Since in our analysis $\varphi'$ is real, a direct comparison of the strong coupling result we obtain for the free energy in the CFT, with either the HHZ result or the superconformal index may not be straightforward, as for that we need to be in the special point in the moduli space where $\varphi'$ is imaginary. \\

\noindent
{\underline{Two-derivative theory}}: 
For the case $\varphi'=0$ which we consider now, the on-shell BPS free energy is mapped to the real part of HHZ functional in eqn. (\ref{hhz})\cite{Choi:2018vbz,Choi:2018hmj,Ezroura:2021vrt}. The parameter ``$a$" can be replaced with ``$\omega_I$" where $\omega_I$ is the imaginary part of complex potential $\omega = \omega_R + i\omega_I$ and is given as (see section-(3.5.3) of \cite{Ezroura:2021vrt} for further details):
 \begin{equation}\label{2.57}
     a = \frac{\pi + \omega_I}{\pi-5 \omega_I} \, .
 \end{equation}
 We note that, due to the reality condition on the conserved charges, $\omega_R$ and $\omega_I$ are related.  
 We thus identify $\omega_I$ as the possible order parameter in the boundary gauge theory. Now,  using eqn. (\ref{2.57}), we can rewrite (\ref{Fbps})(with $\varphi' =0$) as
 \begin{eqnarray} \label{Fbbps}
&& F (\omega_I, \tau) = \frac{\pi  (\pi -5 \omega_I) (\omega_I +\pi )}{18 \omega_I^2} \left(-\sqrt{3} \sqrt{\frac{-3 \omega_I^2-2 \pi  \omega_I +\pi ^2}{(\pi -5 \omega_I)^2}}+\frac{2 \tau\omega_I (\omega_I +\pi ) (\pi -2 \omega_I)^2}{\pi  (\pi -5 \omega_I) (\pi -3 \omega_I)} \right. \nonumber \\
&& \hskip 6.0cm \left. +\frac{4 (\pi -2 \omega_I)^2}{\sqrt{3} \pi  (\pi -5 \omega_I) \sqrt{\frac{\pi -3 \omega_I}{\omega_I +\pi }}}\right)
 \end{eqnarray} 
 \begin{figure}[!htbp]
 \centering{
    {\includegraphics[width=0.6\linewidth]{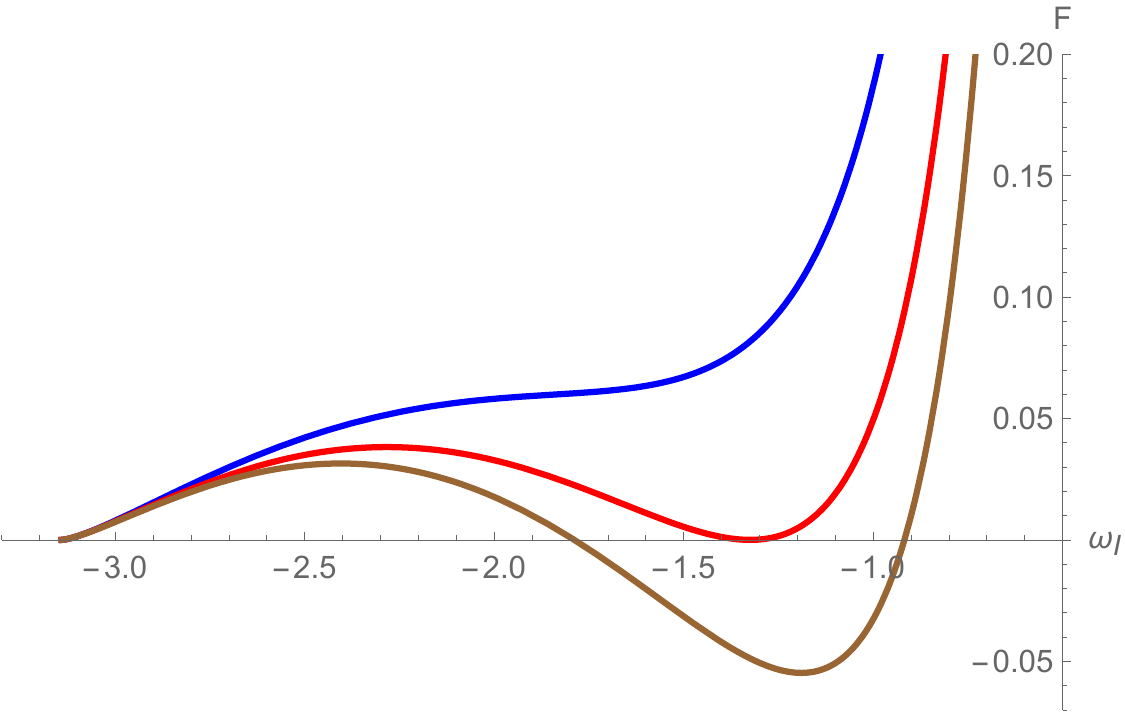}}
    }
    \caption{Effective potential of the boundary field theory as
a function of the order parameter $\omega_I$ for $\varphi\prime = 0$, for different BPS temperatures. The temperature increases from top to bottom. The ``BPS confinement-deconfinement transition temperature'' corresponds to red curve at $\tau_{\rm C} = .862$ and $\omega_{I,C} = -1.33$. }
\label{Fbbpsfig}
\end{figure} 
\FloatBarrier
Solving the minimisation condition on the free energy as $\partial_{\omega_I}\, F =0$, we get the BPS boundary ``CFT Temperature'' as
 \begin{equation}\label{2.59}
     \tau_{b} = -\frac{\sqrt{\frac{4 \pi }{\omega_I+\pi }-3}}{\sqrt{3} \omega_I} \, .
 \end{equation} 
In the bulk, the BPS HP transition temperature is found by simultaneously solving both the conditions in eqns. (\ref{2.24}) and (\ref{dFzero}), using the BPS off-shell free energy in eqn. (\ref{Fbps}), which gives (for $\varphi'=0$), $a_{\rm HP} = \frac{1}{4} \left(\sqrt{33}-5\right)$ at $\tau_{\rm HP-BPS} = 0.862 $ (which matches with~\cite{Choi:2018vbz,Choi:2018hmj,Ezroura:2021vrt}). The confinement-deconfinement transition temperature ``$\tau_C$", can analogously be found by first solving eqn. (\ref{dFzero}) to get the value of order parameter, where the free energy is the one in eqn. (\ref{Fbbps}). The critical value of the order parameter obtained in this way is,
 \be
 \omega_{I,C}=\frac{\left(\sqrt{33}-9\right) \pi }{5 \sqrt{33}-21}\, .
 \ee
 Using $\omega_{I,C}$ in eqn. (\ref{2.59}), we get $\tau_C=0.862$, precisely matching $\tau_{\rm HP-BPS}$. The phase diagram following from the potential in eqn. (\ref{Fbbps}) is shown in figure-(\ref{Fbbpsfig}). We note that the value of the order parameter  $\omega_I$ is restricted to vary from $-\pi$ to $0$.\\

\noindent
Earlier, in section-(\ref{2d}), we expressed the BPS free energy in the bulk, as in eqn. (\ref{fb1}), which allowed for speculations on the existence of a new BPS system, with lower energy and entropy, the HP transition temperature was same as the corresponding BPS black hole. The boundary free energy of the dual CFT in eqn. (\ref{Fbbps}) can in fact also be expressed as:
\be \label{fb2}
F(\omega_I, \tau) \,=\,{\mathcal E_b} \, - \, \tau\, {\mathcal S_b}\, ,
\ee
where the effective boundary entropy is
\be \label{Sbeff}
{\mathcal S_b} = - \frac{\partial F(\omega_I, \tau)}{\partial \tau} =  -\frac{(\pi -2 \omega_I)^2 (\omega_I+\pi )^2}{9 (\pi -3 \omega_I) \omega_I}\, ,
\ee
and the effective boundary energy is
\be
{\mathcal E_b} = \frac{(\omega_I+\pi )^{3/2} \left(16 \omega_I^2-7 \pi  \omega_I+\pi ^2\right)}{18
   \sqrt{3} \sqrt{\pi -3 \omega_I} \omega_I^2} \, .
\ee
These quantities could probably emerge from novel phases in the CFT, dual to the new BPS system supposed to exist in the bulk. As in the bulk, physical meaning of what such a phase with the above energy and entropy could be, and how to understand it, possibly from the index, needs further investigation. \\

\noindent
To conclude, for the BPS black holes in AdS$_5$, we were able to propose a candidate off-shell effective potential in terms of an order parameter $\omega_I$, which, on-shell, gives all the equilibrium phases in the dual CFT.  The boundary free energy we obtained in eqn. (\ref{Fbbps}), reduces on-shell (after substituting $\tau_b$ for $\tau$ and using eqn. (\ref{2.59})) to the real part of HHZ free energy and agrees with the results in ~\cite{Choi:2018vbz,Choi:2018hmj,Ezroura:2021vrt}. \\

\begin{figure}[!htbp]
    \centering
        \includegraphics[width=1.0\linewidth]{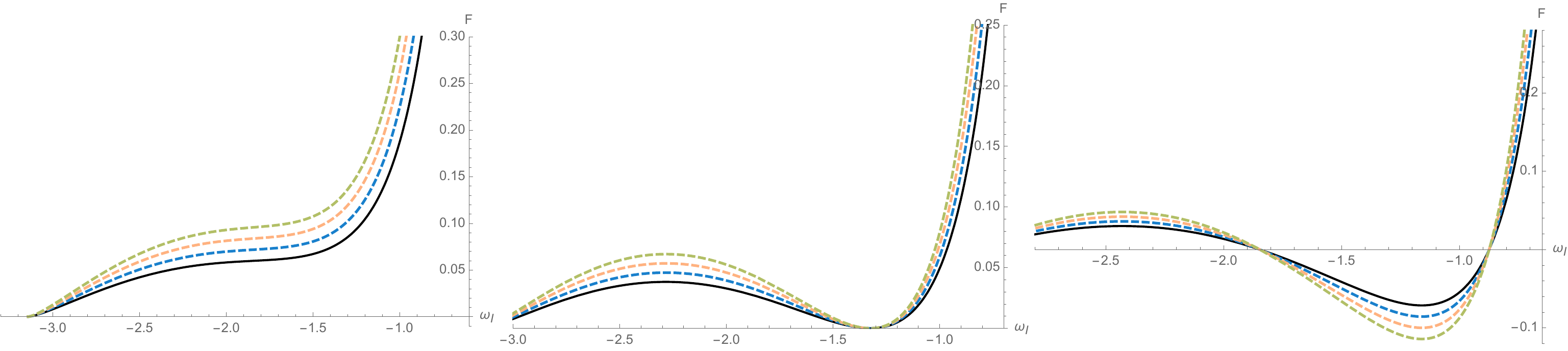}
    \caption{Comparative plot of the Effective potential of boundary gauge theory in the BPS case, with and without four-derivative corrections, plotted as
a function of the order parameter $\omega_I$. The left, middle and the right plots in the figure, correspond to the BPS temperature values $\tau = 0.8,0.8626, 0.91$, respectively. In all the plots: for the solid curves we set  ``$\alpha = \lambda_{1} = \lambda_{2} = 0$", and for the dashed curves we set ``$ \lambda_{1} = .001, \lambda_{2} = 100$". For the dashed curves in each plot we take $\alpha = .0005$ (blue), $\alpha = .001$ (orange) and $\alpha = .0015$ (green).  }
    \label{fig36}
\end{figure}

\noindent
{\underline{Four-derivative corrected theory}}: 
The off-shell BPS free energy with four derivative corrections given in eqn. (\ref{offalpha}) can be used to write a corrected effective potential in the boundary gauge theory. To do this, we replace $a$ with the order parameter $\omega_I$, analogous to the discussion in subsection-(\ref{bef2d}). In particular, we focus on the $\varphi'=0$ case as in subsection-(\ref{bef2d}). The parameter ``$a$" can be replaced with ``$\omega_I$" as in eqn. (\ref{2.57}), which allows us to express eqn. (\ref{offalpha}) as
\begin{eqnarray} \label{offalphab}
&& F(\omega_I, \tau,\alpha,\lambda_1,\lambda_2) =  F (\omega_I, \tau) + \frac{2 \alpha }{27 \left(\pi -3 \omega _I\right){}^{3/2} \omega _I^2
   \left(\pi ^2-3 \omega _I^2\right)}\left(f_{\lambda_3}\, + f_{\lambda_4} \right), \\
&& f_{\lambda_3} =   6 \lambda _1 \left(-18 \tau \sqrt{\pi -3 \omega _I} \omega
   _I^7+18 \omega _I^6 \left(2 \sqrt{3} \sqrt{\omega _I+\pi }-\pi 
   \tau \sqrt{\pi -3 \omega _I}\right) \right.  \nonumber \\
  && \hskip 1.0cm \left. +3 \pi  \omega _I^5  \left(2 \pi  \tau \sqrt{\pi -3 \omega _I}-31 \sqrt{3}
   \sqrt{\omega _I+\pi }\right)+3 \pi ^2 \omega _I^4 \left(20 \pi 
   \tau \sqrt{\pi -3 \omega _I}+23 \sqrt{3} \sqrt{\omega
   _I+\pi }\right) \right. \nonumber  \\
   && \hskip 1.0cm  \left. +2 \pi ^3 \omega _I^3 \left(9 \pi  \tau
   \sqrt{\pi -3 \omega _I}-25 \sqrt{3} \sqrt{\omega _I+\pi }\right)+30
   \pi ^4 \omega _I^2 \left(\sqrt{3} \sqrt{\omega _I+\pi }-\pi 
   \tau \sqrt{\pi -3 \omega _I}\right)  \right. \nonumber  \\
   && \hskip 1.0cm  \left. +3 \pi ^5 \omega _I
   \left(2 \pi  \tau \sqrt{\pi -3 \omega _I}-3 \sqrt{3}
   \sqrt{\omega _I+\pi }\right)+\sqrt{3} \pi ^6 \sqrt{\omega _I+\pi
   }\right) \, , \nonumber  \\
&& f_{\lambda_4}=  \lambda _2 \left(\omega _I+\pi \right) \left(\pi ^2-3 \omega _I^2\right)
   \left(24 \tau \sqrt{\pi -3 \omega _I} \omega _I^4+\pi 
   \omega _I^2 \left(37 \sqrt{3} \sqrt{\omega _I+\pi }-18 \pi 
   \tau \sqrt{\pi -3 \omega _I}\right) \right. \nonumber  \\ 
   && \hskip 1.0cm   \left. +2 \pi ^2 \omega _I
   \left(3 \pi  \tau \sqrt{\pi -3 \omega _I}-5 \sqrt{3}
   \sqrt{\omega _I+\pi }\right)-48 \sqrt{3} \sqrt{\omega _I+\pi } \omega
   _I^3+\sqrt{3} \pi ^3 \sqrt{\omega _I+\pi }\right) \, , \nonumber 
\end{eqnarray}
\noindent
The effective potential given in eqn. (\ref{offalphab}) gives the typical phase structure shown in figure-(\ref{fig36}), where the corrections coming from four derivative terms are shown as dashed curves. The modification to the phase diagram is found to be similar to the one in subsection-(\ref{4d}) from bulk analysis. 
\FloatBarrier
\noindent
To conclude this section, for the BPS black holes in AdS$_5$ with four derivative corrections, we  proposed an off-shell boundary effective potential in terms of an order parameter $\omega_I$, and the couplings $\alpha, \lambda_1, \lambda_2$ using AdS/CFT prescription, which corrects on-shell, the equilibrium phases in the dual CFT at strong coupling. 
\section{Conclusions} \label{untilnexttime}

In this work, we proposed novel connections between the gravitational thermodynamics of supersymmetric black holes in AdS$_5$ and boundary CFT using off-shell methods. We constructed the off-shell free energy of black holes and obtained the phenomenological effective potentials in the dual field theory. The effective potential for the boundary CFT obtained this way at strong coupling, and at finite temperature and chemical potential, yields the various equilibrium phases one expects from the AdS/CFT correspondence. The results were then extended with the inclusion of four-derivative corrections in the action and their effect on the phase structure was investigated in various limiting cases of the electric potential and angular velocity. Following the usual lore~\cite{Cassani_2022,Bobev_2022}, the changes to the thermodynamic quantities and the off-shell free energy were calculated by evaluating the effect of four derivative terms on the two-derivative solution. Owing to the presence of several higher derivative couplings, the modifications to the phase structure following from the four derivative terms was only studied qualitatively, in certain small ranges of the couplings. It would be nice to perform an exhaustive analysis. \\

\noindent
We extended the proposal of BPS phase structure in~\cite{Choi:2018vbz,Choi:2018hmj,Ezroura:2021vrt}, by constructing the off-shell BPS phase diagram, which shows remarkable resembles with the AdS Schwarzschild thermodynamics. The off-shell construction for the BPS case is new and allows us to extract certain thermodynamic quantities, which might give further insights into microscopic physics. The BPS phase diagram is given in terms of an off-shell BPS free energy which is a function of the parameter ``$a$" and ``BPS temperature $\tau$", which is constructed from a certain combination of potentials. We showed that, the four-derivative corrections in the action, shift the phase diagram of BPS black holes, and in particular, modify the asymptotic structure of the small black hole branch, which still agrees with the asymptotic structure of AdS Schwarzschild phase diagram (in the small black hole limit). The asymptotic structure of large black hole branch of AdS Schwarzschild black holes as well as BPS black holes remain unchanged by four-derivative corrections and continue to disagree, revealing some differences.
We followed this up by making a proposal for a phenomenological boundary effective potential dual to BPS black holes, which shows confinement-deconfinement like transition at a ``critical temperature" and captures the phases of gauge theory in the BPS limit. \\

\noindent
A possible avenue for future work follows earlier studies on Hawking-Page (HP) transition~\cite{Hawking:1982dh} in five space-time dimensions, which has an interpretation as a  
deconfining 
transition in 
the ${\cal{N}} = 4$  $SU(N)$ supersymmetric Yang-Mills (SYM) theory at large 
$N$~\cite{,Maldacena:1997re,Gubser:1998bc,Witten:1998qj}. This  transition is reflected in the boundary field theory as being a
jump in the corresponding free energy from order $N^0$ to $N^2$. This analogy was used in~\cite{Alvarez-Gaume:2005dvb}, to propose a phenomenological matrix model built out of the Wilson-Polyakov loop variable. It was also proposed to be in the same universality class belonging to the 
${\cal{N}} = 4$ $SU(N)$ SYM theory on $S^3$ at an infinite 't Hooft coupling ($\lambda$). This of course also involves the examples of $R$-charged black holes in AdS that undergo a continuous 
transitions~\cite{Chamblin:1999tk}, in addition to other interesting characteristics, shared with the black holes in studied in this work~\cite{Gross:1980he,Sundborg:1999ue,Cappiello:2001tf,Aharony:2003sx}.  
The boundary matrix model which takes into account these features, was  
extended in~\cite{Basu:2005pj}, which took care of the new dependence on charge, and also captures qualitatively the phase structure of bulk black holes. These models have undergone several consistency checks in recent years~\cite{Dey:2006ds,Dey:2007vt,Dey:2008bw,Chandrasekhar:2012vh,Yerra:2023wjo,Yerra_2024}. It would be interesting to develop such matrix models for the supersymmetric and BPS black holes which can phenomenologically reproduce the various equilibrium phases, leading to a better understanding of thermodynamics.\\

\noindent
There are other open questions. The identification of a BPS temperature ``$\tau$' in terms of certain scaled potentials was helpful in revealing a phase diagram of BPS black holes and its analogy with AdS Schwarzschild~\cite{Choi:2018vbz,Choi:2018hmj,Ezroura:2021vrt}. We took this analogy further by identifying an  entropy ${\mathcal S}$, which might be the thermodynamic conjugate of $\tau$. These quantities together with the effective  energy ${\mathcal E} $ put the BPS thermodynamics on a more traditional footing, though the interpretation of these quantities remains to be explored. 
Similarly, our identification of a boundary effective potential dual to BPS black holes shows a first order transition, mirroring the bulk phases, the physical meaning of which, including that of the equilibrium  phases it shows, is not clear, and deserves further study. Also, our off-shell thermodynamic analysis of AdS$_5$ black holes in this work, for both BPS and non-BPS cases, was restricted to the situation of equal angular momentum, mostly for clarity. There are interesting issues and certain novel features that arise in the phase diagram, in the case of unequal angular momentum and should be worked out. 

\section*{Acknowledgements}
We thank Sudipta Mukherji for helpful discussions, for sharing his insights during the development of section-(\ref{offshell}), and for comments on the draft. We thank the anonymous referee for helpful suggestions which improved the manuscript. The work of C.B. is supported by ARG-MATRICS grant no. ANRF/ARGM/2025/002280/MTR

\appendix
\section{Four derivative corrected expressions}\label{A}
\subsection{Four derivative corrected entropy, electric charge and angular momentum for $a=b$}\label{A1}
Here, we give the expressions for the four-derivative corrected entropy, electric charge and angular momentum which were obtained in detail in~\cite{Cassani_2022,Bobev_2022}, starting from the on-shell action. For the special case of equal rotational parameters $a=b$, these were used in section-(\ref{three}). 
Here we have also set $g=1$.
Entropy is
\begin{equation}
{\cal S}=\,\frac{\pi ^2 \left[r_+^4+a^4+a^2 \left(q+2 r_+^2\right)\right] }{2Gr_+\left(1-a^2\right)^2}\left(1+4 \lambda _2\alpha\right)+\lambda_1\alpha \Delta {\cal S}\,,
\end{equation}
with
{\small\begin{equation}
\begin{aligned}
\Delta {\cal S}\!&=\,F_{S}\left\{a^4 \left(a^2+q\right)^6 \left(a^2+2 q\right)+a^4 \left(a^2+q\right)^4 \left(2 a^6-5 a^2 q+3 a^4 q-3q^2\right) r_+^2\right.\\[1mm]
&+a^2\left(a^2+q\right)^3 \left[-7 a^8+a^{10}-38 a^2 q^2-4 q^3-2 a^4 q (37+8 q)-2 a^6 (14+13q)\right] r_+^4\\[1mm]
&\hspace{-3mm}-a^2 \left(a^2+q\right)^2 \left[14 a^{10}+13 q^3+a^2 q^2 (104+5 q)+3 a^8 (32+7 q)+16 a^6 (7+13q)+a^4 q (211+108 q)\right] r_+^6\\[1mm]
&-a^2 \left(a^2+q\right) \left[108 a^{10}+7 a^{12}+q^3 (23+2 q)+5 a^2 q^2 (31+4
   q)+4 a^8 (77+38 q)+6 a^4 q (55+53 q)\right.\\[1mm]
&\left.+a^6 \left(210+593q+34 q^2\right)\right] r_+^8-a^2 \left[40 a^{12}+2 (11-4 q) q^3+2a^{10} (140+9 q)+a^2 q^2 (155+48 q)\right.\\[1mm]
&\left.+a^8 (476+451 q)+a^4 q (347-2 (-260+q) q)+a^6 (224+q (920+153 q))\right]
   r_+^{10}-a^2 \left[84 a^{10}\right.\\[1mm]
&\left.+a^8 (322-15 q)+(14-23 q) q^2+14 a^6 (27+20 q)+a^4 (140+(409-55 q) q)\right.\\[1mm]
&\left.+a^2 q (110+(65-17
   q) q)\right] r_+^{12}+\left[-56 a^{10}+3 q^3+6 a^4 (-2+3 q) (4+7 q)+4 a^8 (-21+25 q)\right.\\[1mm]
&\left.+a^6 (-112+73 q)+a^2 q (-13+q(28+3 q))\right] r_+^{14}+\left[70 a^8+3 q^2+44 a^4 (1+4 q)+a^6 (196+135 q)\right.\\[1mm]
&\left.+a^2 \left(-7+45 q+48 q^2\right)\right]
   r_+^{16}+\left[168 a^6+3 q (3+q)+a^2 (42+75 q)+a^4 (232+78 q)\right] r_+^{18}\\[1mm]
&\left.+\left[140 a^4+9 (1+q)+a^2 (105+17
   q)\right] r_+^{20}+2 \left(9+28 a^2\right) r_+^{22}+9 r_+^{24}\right\}\,,
\end{aligned}
\end{equation}}
where 
\begin{equation}
{F}_{\cal S}=-\frac{2\pi^2}{Gr_+^3\left(1-a^2\right)^2\left(a^2+r_+^2\right){}^2{\cal D}}\,,
\end{equation}
and
\begin{equation}
\begin{aligned}
{\cal D}=\,&a^2 \left(a^2+q\right)^3\left(a^2+2 q\right)+a^2 \left(a^2+q\right)^2 \left(5 a^2+a^4+6 q\right) r_+^2\\[1mm]
&\hspace{-8mm}+\left(a^2+q\right) \left[3 a^6+8 a^2q-q^2+a^4 (10+q)\right] r_+^4-\left[-10 a^4-2 a^6+2 a^8+a^2 (-8+q) q+q^2\right] r_+^6\\[1mm]
& -\left[2 a^4+8 a^6+(-1+q)
   q+a^2 (-5+4 q)\right] r_+^8-\left(-1+3 a^2+12 a^4+2 q\right) r_+^{10}\\[1mm]
&-\left(1+8 a^2\right) r_+^{12}-2
   r_+^{14}\,.
\end{aligned}
\end{equation}

The electric charge is given by:
\begin{equation}
Q =\,\frac{\sqrt{3} \pi q}{4 G\left(1-a^2\right)^2 }\left(1+4  \lambda _2\alpha\right)+\lambda_1\alpha  \Delta Q\,,
\end{equation}
with 
{\small\begin{equation}
\begin{aligned}
\Delta Q\!&= F_{Q}\left\{a^4 q^8+2 a^2 \left(a^2-2 r_+^2\right) \left(1+r_+^2\right){}^3 \left(a^2+r_+^2\right){}^7 \left(a^2+r_+^2-2
   r_+^4\right)+a^2 q^7 \left(9 a^4+2 a^2 r_+^2+r_+^4\right)\right.\\[1mm]
&+q^4 \left(a^2+r_+^2\right){}^2 \left[3 r_+^{10}+5 a^8
   \left(21+14 r_+^2+r_+^4\right)-a^4 r_+^4 \left(87+193 r_+^2+3 r_+^4\right)\right.\\[1mm]
&\left.+a^2 r_+^6 \left(-22-3 r_+^2+10r_+^4\right)-a^6 r_+^2 \left(40+213 r_+^2+24 r_+^4\right)\right]+q^6 \left[-3 r_+^8+a^6 r_+^2 \left(19-22r_+^2\right)\right.\\[1mm]
&\hspace{-5mm}\left.-a^2 r_+^6 \left(9+2 r_+^2\right)+a^8 \left(35+4 r_+^2\right)-6 a^4 \left(r_+^4+2r_+^6\right)\right]+q^2 \left(a^2+r_+^2\right){}^4 \left[a^8 \left(1+r_+^2\right) \left(49+31 r_+^2+2r_+^4\right)\right.\\[1mm]
&\left.+3 r_+^8 \left(1+3 r_+^2+8 r_+^4\right)-a^4 r_+^4 \left(70+273 r_+^2+151 r_+^4+2 r_+^6\right)-a^6 r_+^2
   \left(29+125 r_+^2+84 r_+^4+6 r_+^6\right)\right.\\[1mm]
&\left.+a^2 r_+^6 \left(11-27 r_+^2+54 r_+^4+38 r_+^6\right)\right]+q^5
   \left(a^2+r_+^2\right) \left[-107 a^6 r_+^4+a^2 r_+^6 \left(-32+3 r_+^2\right)\right.\\[1mm]
&\left.-21 a^4 r_+^4 \left(2+3r_+^2\right)+a^8 \left(77+26 r_+^2\right)-3 \left(r_+^8+r_+^{10}\right)\right]\\[1mm]
&+q \left(1+r_+^2\right)
   \left(a^2+r_+^2\right){}^6 \left[a^6 \left(1+r_+^2\right) \left(15+4 r_+^2\right)+3 a^2 r_+^4 \left(-3-9 r_+^2+2r_+^4\right)\right.\\[1mm]
&\left.-a^4 r_+^2 \left(21+49 r_+^2+22 r_+^4\right)+3 \left(r_+^6+3 r_+^8+8 r_+^{10}\right)\right]+q^3\left(a^2+r_+^2\right){}^3 \left[a^8 \left(91+100 r_+^2+21 r_+^4\right)\right.\\[1mm]
&\left.+a^2 r_+^6 \left(13-15 r_+^2+44
   r_+^4\right)-a^4 r_+^4 \left(99+323 r_+^2+68 r_+^4\right)-a^6 r_+^2 \left(53+221 r_+^2+72 r_+^4\right)\right.\\[1mm]
&\left.\left.+3
   \left(r_+^{10}+r_+^{12}\right)\right]\right\}\, .
\end{aligned}
\end{equation}}
where
\begin{equation}
F_{Q}=-\frac{\pi }{\sqrt{3}G(1-a^2)^2\,r_{+}^4\left(a^2+r_{+}^2\right)^3{\cal D}}\, .
\end{equation}

The angular momentum reads:
\begin{equation}
J=\frac{a \pi \left(1+4 \lambda_2\alpha \right)}{4 \left(1-a^2\right)^3 G r_+^2}\left[\left(a^2+q\right)^2+\left(a^4+q+a^2 (2+q)\right) r_+^2+\left(1+2a^2\right) r_+^4+r_+^6\right] +\lambda_1\alpha \Delta J\,,
\end{equation}
with
{\small\begin{equation}
\begin{aligned}
\Delta J\!&=F_{J}\left\{-2 a^2 q^8+q^6 \left[-49 a^6+a^4 \left(33+7 a^2\right) r_+^2+51 a^2 \left(1+a^2\right) r_+^4+\left(9+7 a^2\right) r_+^6+3r_+^8\right]\right.\\[1mm]
&+q^7 \left[4 r_+^4+a^4 \left(-15+r_+^2\right)+a^2 r_+^2 \left(5+r_+^2\right)\right]-\left(1+r_+^2\right){}^2\left(a^2+r_+^2\right){}^7 \left[a^4 \left(1-3 r_+^2\right)\right.\\
&\hspace{-5mm}\left.-2 a^2 r_+^2 \left(4+19 r_+^2+7 r_+^4\right)+r_+^4\left(7+13 r_+^2+18 r_+^4\right)\right]+q^5 \left[-2 r_+^8 \left(-5+r_+^2\right)+a^4 r_+^4 \left(233+281 r_+^2-4 r_+^4\right)\right.\\
&\left.+a^2 r_+^6 \left(93+47 r_+^2-2
   r_+^4\right)+a^8 \left(-91+21 r_+^2+8 r_+^4\right)+a^6 r_+^2 \left(91+317 r_+^2+38 r_+^4\right)\right]\\
&+q^4 \left[r_+^{10} \left(-1+r_+^2\right)+a^2 r_+^8 \left(71+103 r_+^2-66 r_+^4\right)+2 a^4 r_+^6 \left(181+361 r_+^2-14r_+^4\right)\right.\\[1mm]
&\left.+a^{10} \left(-105+35 r_+^2+34 r_+^4\right)+5 a^8 r_+^2 \left(27+169 r_+^2+60 r_+^4\right)+2 a^6 r_+^4\left(265+715 r_+^2+152 r_+^4\right)\right]\\[1mm]
&+q \left(1+r_+^2\right) \left(a^2+r_+^2\right){}^5 \left[a^6 \left(-9+16 r_+^2+9 r_+^4\right)-2 r_+^6 \left(3+17 r_+^2+26r_+^4\right)\right.\\[1mm]
&\left.+2 a^4 r_+^2 \left(29+140 r_+^2+90 r_+^4+7 r_+^6\right)-a^2 r_+^4 \left(19-22 r_+^2+57 r_+^4+34r_+^6\right)\right]\\[1mm]
&+q^2 \left(a^2+r_+^2\right){}^3 \left[8 a^4 r_+^4 \left(13+82 r_+^2+58 r_+^4-2 r_+^6\right)+a^8 \left(-35+21 r_+^2+53r_+^4+9 r_+^6\right)\right.\\[1mm]
&\hspace{-10mm}\left.-r_+^8 \left(1+25 r_+^2+57 r_+^4+21 r_+^6\right)-4 a^2 r_+^6 \left(7+4 r_+^2+46 r_+^4+37r_+^6\right)+4 a^6 r_+^2 \left(40+215 r_+^2+209 r_+^4+46 r_+^6\right)\right]\\[1mm]
&+q^3 \left(a^2+r_+^2\right){}^2 \left[-2 r_+^8 \left(4+3 r_+^2+23 r_+^4\right)+a^4 r_+^4 \left(209+791 r_+^2+231 r_+^4-39r_+^6\right)\right.\\[1mm]
&\left.+5 a^2 r_+^6 \left(3+17 r_+^2-31 r_+^4-5 r_+^6\right)+a^8 \left(-77+35 r_+^2+59 r_+^4+3 r_+^6\right)\right.\\[1mm]
&\left.\left.+a^6r_+^2 \left(269+1119 r_+^2+687 r_+^4+53 r_+^6\right)\right]\right\}\,,
\end{aligned}
\end{equation}}
where
\begin{equation}
{F}_{J}=\frac{a\pi}{2G(1-a^2)^3(a^2+r_+^2)^2r_+^4 {\cal D}}\,.
\end{equation}
\FloatBarrier

\subsection{Four derivative corrected off-shell free energy}\label{A2}
\noindent
Using the above charges in eqn. (\ref{2.199}), we can compute the off-shell free energy of general black holes in AdS$_5$ with four derivative corrections given below as
{\small\begin{equation}
		F(a,q,r_{+},\Phi, T, \Omega, \lambda_{1},\lambda_{2},\alpha) = F_{0}(a,q,r_{+},\Phi, T, \Omega)+ \alpha(\lambda_{1} \Delta F_{1} + \lambda_{2} \Delta F_{2})   , 
\end{equation}}
where
{\small\begin{equation}
		\begin{aligned}
			F_{0}(a,q,r_{+},\Phi, T, \Omega)\!&=\frac{1}{4 \left(a^2-1\right)^3 r_+^2}\left( r_+ \left(r_+ \left(-2 \sqrt{3} \left(a^2-1\right) q \Phi -4 \pi  \left(a^2-1\right) r_+^3 T-8 \pi  a^2 \left(a^2-1\right) r_+ T\right.\right.\right.\\[1mm]
			&\left.\left.\left.-\left(r_+^4 \left(a^2-4 a \Omega +3\right)\right)-\left(2 a^2+1\right) r_+^2 \left(a^2-4 a \Omega +3\right)-a^2 \left(a^2+2\right) \left(a^2-4 a \Omega +3\right)\right.\right.\right.\\[1mm]
			&\left.\left.\left.+4 a q (a (a \Omega -2)+\Omega )\right)-4 \pi  a^2 \left(a^2-1\right) T \left(a^2+q\right)\right)-\left(a^2+q\right)^2 \left(a^2-4 a \Omega +3\right)\right) ,    
		\end{aligned}    
\end{equation}}
and
{\small\begin{equation}
		\Delta F_{1} = \frac{A_{1}+A_{2}+A_{3}+A_{4}+A_{5}+A_{6}}{B},
\end{equation}}
with
{\small\begin{equation}
		\begin{aligned}
			A_{1}\!&=12 \left(1-a^2\right) \left(a^2+r_+^2\right) \left(-2 r_+^6-\left(2 a^2+1\right) r_+^4+\left(a^2+q\right)^2\right) \left(9 r_+^{24}+2 \left(28 a^2+9\right) r_+^{22}+\left(140 a^4+(17 q+105) a^2\right.\right.\\[1mm]
			&\left.\left.+9 (q+1)\right) r_+^{20}+\left(168 a^6+(78 q+232) a^4+(75 q+42) a^2+3 q (q+3)\right) r_+^{18}+\left(70 a^8+(135 q+196) a^6\right.\right.\\[1mm]
			&\left.\left.+44 (4 q+1) a^4+\left(48 q^2+45 q-7\right) a^2+3 q^2\right) r_+^{16}+\left(-56 a^{10}+4 (25 q-21) a^8+(73 q-112) a^6\right.\right.\\[1mm]
			&\left.\left.+6 (3 q-2) (7 q+4) a^4+q (q (3 q+28)-13) a^2+3 q^3\right) r_+^{14}-a^2 \left(84 a^{10}+(322-15 q) a^8+14 (20 q+27) a^6\right.\right.\\[1mm]
			&\left.\left.+(q (409-55 q)+140) a^4+q (q (65-17 q)+110) a^2+q^2 (14-23 q)\right) r_+^{12}-a^2 \left(40 a^{12}+2 (9 q+140) a^{10}\right.\right.\\[1mm]
			&\left.\left.+(451 q+476) a^8+(q (153 q+920)+224) a^6+q (347-2 (q-260) q) a^4+q^2 (48 q+155) a^2\right.\right.\\[1mm]
			&\left.\left.+2 q^3 (11-4 q)\right) r_+^{10}-a^2 \left(a^2+q\right) \left(7 a^{12}+108 a^{10}+4 (38 q+77) a^8+(q (34 q+593)+210) a^6\right.\right.\\[1mm]
			&\left.\left.+6 q (53 q+55) a^4+5 q^2 (4 q+31) a^2+q^3 (2 q+23)\right) r_+^8-a^2 \left(a^2+q\right)^2 \left(14 a^{10}+3 (7 q+32) a^8\right.\right.\\[1mm]
			&\left.\left.+16 (13 q+7) a^6+q (108 q+211) a^4+q^2 (5 q+104) a^2+13 q^3\right) r_+^6+a^2 \left(a^2+q\right)^3 \left(a^{10}-7 a^8\right.\right.\\[1mm]
			&\left.\left.-2 (13 q+14) a^6-2 q (8 q+37) a^4-38 q^2 a^2-4 q^3\right) r_+^4+a^4 \left(a^2+q\right)^4 \left(2 a^6+3 q a^4-5 q a^2-3 q^2\right) r_+^2\right.\\[1mm]
			&\left.+a^4 \left(a^2+q\right)^6 \left(a^2+2 q\right)\right) \left(a^4+\left(2 r_+^2+q\right) a^2+r_+^4\right)-3 \left(1-a^2\right) \left(a^2+r_+^2\right) \left(-2 r_+^{14}-\left(8 a^2+1\right) r_+^{12}\right.\\[1mm]
			&\left.-\left(12 a^4+3 a^2+2 q-1\right) r_+^{10}-\left(8 a^6+2 a^4+(4 q-5) a^2+(q-1) q\right) r_+^8-\left(2 a^8-2 a^6-10 a^4+(q-8) q a^2\right.\right.\\[1mm]
			&\left.\left.+q^2\right) r_+^6+\left(a^2+q\right) \left(3 a^6+(q+10) a^4+8 q a^2-q^2\right) r_+^4+a^2 \left(a^2+q\right)^2 \left(a^4+5 a^2+6 q\right) r_+^2+a^2 \left(a^2+q\right)^3\right.\\[1mm]
			&\left.\left(a^2+2 q\right)\right) \left(9 r_+^{16}+2 \left(11 a^2+9\right) r_+^{14}-\left(a^4-(25 q+44) a^2-9\right) r_+^{12}-2 \left(22 a^6+(1-14 q) a^4-11 (q+1) a^2\right.\right.\\[1mm]
			&\left.\left.-3 q^2\right) r_+^{10}-\left(41 a^8+2 (9 q+44) a^6+(24 q+1) a^4+q (3-16 q) a^2+6 q^2\right) r_+^8-2 a^2 \left(5 a^8+(10 q+41) a^6\right.\right.\\[1mm]
			&\left.\left.+(54 q+22) a^4+2 q (11 q+13) a^2-(q-8) q^2\right) r_+^6+\left(a^2+q\right) \left(a^{10}-20 a^8-(36 q+41) a^6-q (12 q+49) a^4\right.\right.\\[1mm]
			&\left.\left.-19 q^2 a^2-3 q^3\right) r_+^4+2 a^2 \left(a^2+q\right)^3 \left(a^4-5 a^2-3 q\right) r_+^2+a^2 \left(a^2+q\right)^5\right) \left(q a^2+\left(a^2+r_+^2\right){}^2\right),  
		\end{aligned}    
\end{equation}}
\vspace{2mm}
{\small\begin{equation}
		\begin{aligned}
			A_{2}\!&=24 \pi  \left(1-a^2\right) r_+ T \left(a^2+r_+^2\right) \left(2 \left(28 a^2+9\right) r_+^{22}+r_+^{20} \left(140 a^4+a^2 (17 q+105)+9 (q+1)\right)+a^4 \left(a^2+q\right)^6\right.\\[1mm]
			&\left.\left(a^2+2q\right)+a^4 r_+^2 \left(a^2+q\right)^4 \left(2 a^6+3 a^4 q-5 a^2 q-3 q^2\right)+r_+^{18} \left(168 a^6+a^4 (78 q+232)+a^2 (75 q+42)\right.\right.\\[1mm]
			&\left.\left.+3 q (q+3)\right)+r_+^{16} \left(70 a^8+a^6 (135q+196)+44 a^4 (4 q+1)+a^2 \left(48 q^2+45 q-7\right)+3 q^2\right)\right.\\[1mm]
			&\left.+r_+^{14} \left(-56 a^{10}+4 a^8 (25 q-21)+a^6 (73 q-112)+6 a^4 (3 q-2) (7 q+4)+a^2 q (q (3 q+28)-13)+3q^3\right)\right.\\[1mm]
			&\left.-a^2 r_+^{12} \left(84 a^{10}+a^8 (322-15 q)+14 a^6 (20 q+27)+a^4 ((409-55 q) q+140)+a^2 q ((65-17 q) q+110)\right.\right.\\[1mm]
			&\left.\left.+(14-23 q) q^2\right)-a^2 r_+^6 \left(a^2+q\right)^2 \left(14a^{10}+3 a^8 (7 q+32)+16 a^6 (13 q+7)+a^4 q (108 q+211)\right.\right.\\[1mm]
			&\left.\left.+a^2 q^2 (5 q+104)+13 q^3\right)+a^2 r_+^4 \left(a^2+q\right)^3 \left(a^{10}-7 a^8-2 a^6 (13 q+14)-2 a^4 q (8 q+37)-38 a^2q^2-4 q^3\right)\right.\\[1mm]
			&\left.-a^2 r_+^{10} \left(40 a^{12}+2 a^{10} (9 q+140)+a^8 (451 q+476)+a^6 (q (153 q+920)+224)+a^4 q (347-2 (q-260) q)\right.\right.\\[1mm]
			&\left.\left.+a^2 q^2 (48 q+155)+2 (11-4 q) q^3\right)-a^2 r_+^8
			\left(a^2+q\right) \left(7 a^{12}+108 a^{10}+4 a^8 (38 q+77)+a^6 (q (34 q+593)\right.\right.\\[1mm]
			&\left.\left.+210)+6 a^4 q (53 q+55)+5 a^2 q^2 (4 q+31)+q^3 (2 q+23)\right)+9 r_+^{24}\right) \left(a^2
			q+\left(a^2+r_+^2\right){}^2\right) \left(a^4+a^2 \left(q+2 r_+^2\right)+r_+^4\right),    
		\end{aligned}    
\end{equation}}
\vspace{2mm}
{\small\begin{equation}
		\begin{aligned}
			A_{3}\!&=-12 a^2 \left(a^2+r_+^2\right) \left(q+\left(r_+^2+1\right) \left(a^2+r_+^2\right)\right) \left(a^4+\left(2 r_+^2+q\right) a^2+r_+^4\right) \left(2 a^2 q^8-\left(-15a^4+\left(a^2+5\right) r_+^2 a^2\right.\right.\\[1mm]
			&\left.\left.+\left(a^2+4\right) r_+^4\right) q^7-\left(3 r_+^8+\left(7 a^2+9\right) r_+^6+51 \left(a^4+a^2\right) r_+^4+a^4 \left(7 a^2+33\right) r_+^2-49a^6\right) q^6-\left(-2 \left(a^2+1\right) r_+^{10}\right.\right.\\[1mm]
			&\left.\left.+\left(-4 a^4+47 a^2+10\right) r_+^8+a^2 \left(38 a^4+281 a^2+93\right) r_+^6+a^4 \left(8 a^4+317 a^2+233\right) r_+^4+7 a^6\left(3 a^2+13\right) r_+^2\right.\right.\\[1mm]
			&\left.\left.-91 a^8\right) q^5-\left(\left(1-66 a^2\right) r_+^{12}+\left(-28 a^4+103 a^2-1\right) r_+^{10}+a^2 \left(304 a^4+722 a^2+71\right) r_+^8\right.\right.\\[1mm]
			&\left.\left.+2 a^4 \left(150a^4+715 a^2+181\right) r_+^6+a^6 \left(34 a^4+845 a^2+530\right) r_+^4+5 a^8 \left(7 a^2+27\right) r_+^2-105 a^{10}\right) q^4\right.\\[1mm]
			&\left.-\left(a^2+r_+^2\right){}^2 \left(-\left(\left(25a^2+46\right) r_+^{12}\right)-\left(39 a^4+155 a^2+6\right) r_+^{10}+\left(53 a^6+231 a^4+85 a^2-8\right) r_+^8\right.\right.\\[1mm]
			&\left.\left.+a^2 \left(3 a^6+687 a^4+791 a^2+15\right) r_+^6+a^4 \left(59 a^4+1119a^2+209\right) r_+^4+a^6 \left(35 a^2+269\right) r_+^2-77 a^8\right) q^3\right.\\[1mm]
			&\left.-\left(a^2+r_+^2\right){}^3 \left(\left(9 r_+^6+53 r_+^4+21 r_+^2-35\right) a^8+4 r_+^2 \left(46 r_+^6+209r_+^4+215 r_+^2+40\right) a^6\right.\right.\\[1mm]
			&\left.\left.+8 r_+^4 \left(-2 r_+^6+58 r_+^4+82 r_+^2+13\right) a^4-4 r_+^6 \left(37 r_+^6+46 r_+^4+4 r_+^2+7\right) a^2-r_+^8 \left(21 r_+^6+57 r_+^4+25r_+^2+1\right)\right) q^2\right.\\[1mm]
			&\left.-\left(r_+^2+1\right) \left(a^2+r_+^2\right){}^5 \left(-2 \left(17 a^2+26\right) r_+^{10}+\left(14 a^4-57 a^2-34\right) r_+^8+2 \left(90 a^4+11 a^2-3\right)r_+^6\right.\right.\\[1mm]
			&\left.\left.+a^2 \left(9 a^4+280 a^2-19\right) r_+^4+2 a^4 \left(8 a^2+29\right) r_+^2-9 a^6\right) q+\left(r_+^2+1\right){}^2 \left(a^2+r_+^2\right){}^7 \left(18 r_+^8+\left(13-14a^2\right) r_+^6\right.\right.\\[1mm]
			&\left.\left.+\left(7-38 a^2\right) r_+^4-a^2 \left(3 a^2+8\right) r_+^2+a^4\right)\right),    
		\end{aligned}    
\end{equation}}
\vspace{2mm}
{\small\begin{equation}
		\begin{aligned}
			A_{4}\!&=12 a \Omega  \left(a^2+r_+^2\right) \left(q a^2+\left(a^2+r_+^2\right){}^2\right) \left(a^4+\left(2 r_+^2+q\right) a^2+r_+^4\right) \left(2 a^2 q^8-\left(-15 a^4+\left(a^2+5\right)r_+^2 a^2\right.\right.\\[1mm]
			&\left.\left.+\left(a^2+4\right) r_+^4\right) q^7-\left(3 r_+^8+\left(7 a^2+9\right) r_+^6+51 \left(a^4+a^2\right) r_+^4+a^4 \left(7 a^2+33\right) r_+^2-49 a^6\right) q^6\right.\\[1mm]
			&\left.-\left(-2\left(a^2+1\right) r_+^{10}+\left(-4 a^4+47 a^2+10\right) r_+^8+a^2 \left(38 a^4+281 a^2+93\right) r_+^6+a^4 \left(8 a^4+317 a^2+233\right) r_+^4\right.\right.\\[1mm]
			&\left.\left.+7 a^6 \left(3 a^2+13\right)r_+^2-91 a^8\right) q^5-\left(\left(1-66 a^2\right) r_+^{12}+\left(-28 a^4+103 a^2-1\right) r_+^{10}+a^2 \left(304 a^4+722 a^2+71\right) r_+^8\right.\right.\\[1mm]
			&\left.\left.+2 a^4 \left(150 a^4+715 a^2+181\right)r_+^6+a^6 \left(34 a^4+845 a^2+530\right) r_+^4+5 a^8 \left(7 a^2+27\right) r_+^2-105 a^{10}\right) q^4\right.\\[1mm]
			&\left.-\left(a^2+r_+^2\right){}^2 \left(-\left(\left(25 a^2+46\right)r_+^{12}\right)-\left(39 a^4+155 a^2+6\right) r_+^{10}+\left(53 a^6+231 a^4+85 a^2-8\right) r_+^8\right.\right.\\[1mm]
			&\left.\left.+a^2 \left(3 a^6+687 a^4+791 a^2+15\right) r_+^6+a^4 \left(59 a^4+1119
			a^2+209\right) r_+^4+a^6 \left(35 a^2+269\right) r_+^2-77 a^8\right) q^3\right.\\[1mm]
			&\left.-\left(a^2+r_+^2\right){}^3 \left(\left(9 r_+^6+53 r_+^4+21 r_+^2-35\right) a^8+4 r_+^2 \left(46 r_+^6+209r_+^4+215 r_+^2+40\right) a^6\right.\right.\\[1mm]
			&\left.\left.+8 r_+^4 \left(-2 r_+^6+58 r_+^4+82 r_+^2+13\right) a^4-4 r_+^6 \left(37 r_+^6+46 r_+^4+4 r_+^2+7\right) a^2-r_+^8 \left(21 r_+^6+57 r_+^4+25r_+^2+1\right)\right) q^2\right.\\[1mm]
			&\left.-\left(r_+^2+1\right) \left(a^2+r_+^2\right){}^5 \left(-2 \left(17 a^2+26\right) r_+^{10}+\left(14 a^4-57 a^2-34\right) r_+^8+2 \left(90 a^4+11 a^2-3\right)r_+^6\right.\right.\\[1mm]
			&\left.\left.+a^2 \left(9 a^4+280 a^2-19\right) r_+^4+2 a^4 \left(8 a^2+29\right) r_+^2-9 a^6\right) q+\left(r_+^2+1\right){}^2 \left(a^2+r_+^2\right){}^7 \left(18 r_+^8+\left(13-14
			a^2\right) r_+^6\right.\right.\\[1mm]
			&\left.\left.+\left(7-38 a^2\right) r_+^4-a^2 \left(3 a^2+8\right) r_+^2+a^4\right)\right),    
		\end{aligned}    
\end{equation}}
\vspace{2mm}
{\small\begin{equation}
		\begin{aligned}
			A{5}\!&=-12 \left(1-a^2\right) q r_+^2 \left(a^4+a^2 \left(q+2 r_+^2\right)+r_+^4\right) \left(a^4 q^8+2 a^2 \left(r_+^2+1\right){}^3 \left(a^2+r_+^2\right){}^7 \left(a^2-2 r_+^2\right)\right.\\[1mm]
			&\left.\left(a^2-2 r_+^4+r_+^2\right)+a^2 q^7 \left(9 a^4+2 a^2 r_+^2+r_+^4\right)+q^5 \left(a^2+r_+^2\right) \left(26 a^8 r_+^2+77 a^8+3 \left(a^2-1\right) r_+^8\right.\right.\\[1mm]
			&\left.\left.-a^2 \left(63 a^2+32\right) r_+^6-a^4 \left(107 a^2+42\right) r_+^4-3 r_+^{10}\right)+q \left(r_+^2+1\right) \left(a^2+r_+^2\right){}^6 \left(15 a^6+\left(6 a^2+9\right) r_+^8\right.\right.\\[1mm]
			&\left.\left.+\left(-22 a^4-27 a^2+3\right) r_+^6+a^2 \left(4 a^4-49 a^2-9\right) r_+^4+a^4 \left(19 a^2-21\right) r_+^2+24 r_+^{10}\right)\right.\\[1mm]
			&\left.+q^6 \left(35 a^8-3 \left(4 a^2+3\right) a^2 r_+^6-\left(2 a^2+3\right) r_+^8+\left(4 a^2+19\right) a^6 r_+^2-2 \left(11 a^2+3\right) a^4 r_+^4\right)+q^4 \left(a^2+r_+^2\right){}^2\right.\\[1mm]
			&\left.\left(105 a^8+\left(10 a^2+3\right) r_+^{10}+10 a^6 \left(7 a^2-4\right) r_+^2-3 \left(a^4+a^2\right) r_+^8-a^2 \left(24 a^4+193 a^2+22\right) r_+^6\right.\right.\\[1mm]
			&\left.\left.+a^4 \left(5 a^4-213 a^2-87\right) r_+^4\right)+q^3 \left(a^2+r_+^2\right){}^3 \left(a^8 \left(21 r_+^4+100 r_+^2+91\right)-a^6 r_+^2 \left(72 r_+^4+221 r_+^2+53\right)\right.\right.\\[1mm]
			&\left.\left.-a^4 r_+^4 \left(68 r_+^4+323 r_+^2+99\right)+a^2 r_+^6 \left(44 r_+^4-15 r_+^2+13\right)+3 \left(r_+^{12}+r_+^{10}\right)\right)+q^2 \left(a^2+r_+^2\right){}^4\right.\\[1mm]
			&\left.\left(49 a^8+\left(38 a^2+24\right) r_+^{12}+a^6 \left(80 a^2-29\right) r_+^2+\left(-2 a^4+54 a^2+9\right) r_+^{10}+a^4 \left(33 a^4-125 a^2-70\right) r_+^4\right.\right.\\[1mm]
			&\left.\left.-\left(6 a^6+151 a^4+27 a^2-3\right) r_+^8+a^2 \left(2 a^6-84 a^4-273 a^2+11\right) r_+^6\right)\right)     
		\end{aligned}    
\end{equation}}
\vspace{2mm}

{\small\begin{equation}
		\begin{aligned}
			A_{6}\!&=4 \sqrt{3} \left(1-a^2\right) \Phi  \left(a^2 q+\left(a^2+r_+^2\right){}^2\right) \left(a^4+a^2 \left(q+2 r_+^2\right)+r_+^4\right) \left(a^4 q^8+2 a^2 \left(r_+^2+1\right){}^3 \left(a^2+r_+^2\right){}^7\right.\\[1mm]
			&\left.\left(a^2-2 r_+^2\right) \left(a^2-2 r_+^4+r_+^2\right)+a^2 q^7 \left(9 a^4+2 a^2 r_+^2+r_+^4\right)+q^5 \left(a^2+r_+^2\right) \left(26 a^8 r_+^2+77 a^8+3 \left(a^2-1\right) r_+^8\right.\right.\\[1mm]
			&\left.\left.-a^2 \left(63 a^2+32\right) r_+^6-a^4 \left(107 a^2+42\right) r_+^4-3 r_+^{10}\right)+q \left(r_+^2+1\right) \left(a^2+r_+^2\right){}^6 \left(15 a^6+\left(6 a^2+9\right) r_+^8\right.\right.\\[1mm]
			&\left.\left.+\left(-22 a^4-27 a^2+3\right) r_+^6+a^2 \left(4 a^4-49 a^2-9\right) r_+^4+a^4 \left(19 a^2-21\right) r_+^2+24 r_+^{10}\right)\right.\\[1mm]
			&\left.+q^6 \left(35 a^8-3 \left(4 a^2+3\right) a^2 r_+^6-\left(2 a^2+3\right) r_+^8+\left(4 a^2+19\right) a^6 r_+^2-2 \left(11 a^2+3\right) a^4 r_+^4\right)+q^4 \left(a^2+r_+^2\right){}^2\right.\\[1mm]
			&\left.\left(105 a^8+\left(10 a^2+3\right) r_+^{10}+10 a^6 \left(7 a^2-4\right) r_+^2-3 \left(a^4+a^2\right) r_+^8-a^2 \left(24 a^4+193 a^2+22\right) r_+^6\right.\right.\\[1mm]
			&\left.\left.+a^4 \left(5 a^4-213 a^2-87\right) r_+^4\right)+q^3 \left(a^2+r_+^2\right){}^3 \left(a^8 \left(21 r_+^4+100 r_+^2+91\right)-a^6 r_+^2 \left(72 r_+^4+221 r_+^2+53\right)\right.\right.\\[1mm]
			&\left.\left.-a^4 r_+^4 \left(68 r_+^4+323 r_+^2+99\right)+a^2 r_+^6 \left(44 r_+^4-15 r_+^2+13\right)+3 \left(r_+^{12}+r_+^{10}\right)\right)+q^2 \left(a^2+r_+^2\right){}^4\right.\\[1mm]
			&\left.\left(49 a^8+\left(38 a^2+24\right) r_+^{12}+a^6 \left(80 a^2-29\right) r_+^2+\left(-2 a^4+54 a^2+9\right) r_+^{10}+a^4 \left(33 a^4-125 a^2-70\right) r_+^4\right.\right.\\[1mm]
			&\left.\left.-\left(6 a^6+151 a^4+27 a^2-3\right) r_+^8+a^2 \left(2 a^6-84 a^4-273 a^2+11\right) r_+^6\right)\right)  ,  
		\end{aligned}    
\end{equation}}
\vspace{2mm}
{\small\begin{equation}
		\begin{aligned}
			B\!&=6 \left(a^2-1\right)^3 r_+^4 \left(a^2+r_+^2\right){}^3 \left(a^2 \left(a^2+q\right)^3 \left(a^2+2 q\right)-\left(8 a^2+1\right) r_+^{12}-r_+^{10} \left(12 a^4+3 a^2+2 q-1\right)\right.\\[1mm]
			&\left.+a^2 r_+^2 \left(a^2+q\right)^2 \left(a^4+5 a^2+6 q\right)+r_+^4 \left(a^2+q\right) \left(3 a^6+a^4 (q+10)+8 a^2 q-q^2\right)\right.\\[1mm]
			&\left.-r_+^8 \left(8 a^6+2 a^4+a^2 (4 q-5)+(q-1) q\right)-r_+^6 \left(2 a^8-2 a^6-10 a^4+a^2 (q-8) q+q^2\right)-2 r_+^{14}\right)\\[1mm]
			&\left(a^2 q+\left(a^2+r_+^2\right){}^2\right) \left(a^4+a^2 \left(q+2 r_+^2\right)+r_+^4\right)  ,  
		\end{aligned}    
\end{equation}}
and
{\small\begin{equation}
		\begin{aligned}
			\Delta F_{2}\!&=\frac{1}{\left(a^2-1\right)^3 r_+^2}\left( r_+ \left(r_+ \left(-2 \sqrt{3} \left(a^2-1\right) q \Phi -4 \pi  \left(a^2-1\right) r_+^3 T-8 \pi  a^2 \left(a^2-1\right) r_+ T\right.\right.\right.\\[1mm]
			&\left.\left.\left.-\left(r_+^4 \left(a^2-4 a \Omega +3\right)\right)-\left(2 a^2+1\right) r_+^2 \left(a^2-4 a \Omega +3\right)-a^2 \left(a^2+2\right) \left(a^2-4 a \Omega +3\right)\right.\right.\right.\\[1mm]
			&\left.\left.\left.+4 a q (a (a \Omega -2)+\Omega )\right)-4 \pi  a^2 \left(a^2-1\right) T \left(a^2+q\right)\right)-\left(a^2+q\right)^2 \left(a^2-4 a \Omega +3\right)\right) .
		\end{aligned}   
\end{equation}}
where the parameters ``$a$" and ``$q$" can be eliminated in favor of $r_+$ and the potentials $\Phi$ and $\Omega$, using eq. (\ref{2.23}). 

\section{On-shell phase diagram with four derivative corrections} \label{B}

\subsection{General black holes in AdS$_5$}

The on-shell phase diagram of general black holes in AdS$_5$ was studied in detail in various limits in~\cite{Ezroura:2021vrt}, considering the two derivative action. We presented the off-shell phase diagram in two derivative and four derivative theories in sections-(\ref{two}) and (\ref{three}). Here, just for consistency check, our aim is to present the on-shell phase diagram of general black holes in AdS$_5$ and also in the BPS limit, with four derivative corrections, which will also correct the phase diagram in~\cite{Ezroura:2021vrt}. 
The four-derivative corrected on-shell free energy can be computed from eqn. (\ref{2.199}), by substituting in it, the value of equilibrium  temperature, electric potential and angular velocity from eqn. (\ref{temperature_CCLP},\ref{electrostatic_pot_CCLP},\ref{angular_velocities_CCLP}), which we call as $G(T,\Phi,\Omega,\alpha,\lambda_1,\lambda_2)$  given as
{\small{\begin{equation}
			G(a,\Phi,\Omega,\lambda_1, \lambda_2,\alpha) = G_{0}(a,\Phi, \Omega)+ \alpha(\lambda_{1}\Delta G_{1}+\lambda_{2}\Delta G_{2})
\end{equation}}}
where
{\small\begin{equation}
		\begin{aligned}
			G_{0}(a,\Phi, \Omega)\!&=\frac{a}{12 \sqrt{3} \left(\Phi +\sqrt{3}\right)^4 (a-\Omega )^2 (a \Omega -1)^2}\left(\Phi ^7 (a-\Omega ) \left(a^2-2 a \Omega +1\right)+\sqrt{3} \Phi ^6 \left(a \left(-\left(a^2 \left(\Omega ^2-6\right)\right)\right.\right.\right.\\[1mm]
			&\left.\left.\left.-15 a \Omega +11 \Omega ^2+4\right)-5 \Omega \right)-9 \Phi ^5 \left(a \left(2 \left(a^2-4\right) \Omega ^2-5 a^2+9 a \Omega -1\right)+3 \Omega \right)\right.\\[1mm]
			&\left.-15 \sqrt{3} \Phi ^4 \left(a \left(a^2 \left(3 \Omega ^2-4\right)+3 a \Omega -5 \Omega ^2+2\right)+\Omega \right)-45 \Phi ^3 \left(a \left(a^2 \left(4 \Omega ^2-3\right)-3 a \Omega -2 \Omega ^2+5\right)-\Omega \right)\right.\\[1mm]
			&\left.-27 \sqrt{3} \Phi ^2 \left(a \left(a^2 \left(5 \Omega ^2-2\right)-9 a \Omega +\Omega ^2+8\right)-3 \Omega \right)-27 \Phi  \left(a \left(a^2 \left(6 \Omega ^2-1\right)-15 a \Omega +4 \Omega ^2+11\right)-5 \Omega \right)\right.\\[1mm]
			&\left.-27 \sqrt{3} (a \Omega -1) (a (a \Omega -2)+\Omega )\right), 
		\end{aligned}
\end{equation}}
{\small\begin{equation}
		\begin{aligned}
			\Delta G_{1} \! &= -\frac{1}{18 \sqrt{3} \left(\Phi +\sqrt{3}\right)^4 (a-\Omega )^2 (a \Omega -1)^4}\left(\sqrt{3} \left(a^2-1\right) \Phi ^8 (a-\Omega )^2 \left(a^2-4 a \Omega +3\right)+162 \Phi  \Omega  (a \Omega -1)\right.\\[1mm]
			&\left.\left(a^2+2 a \Omega -3\right) \left(a \left(-\left(9 a^2+23\right) \Omega ^2+a^2+18 a \Omega ^3+4 a \Omega +3\right)+6 \Omega \right)+81 \sqrt{3} \Omega ^2 (a \Omega -1)^2 \right.\\[1mm]
			&\left.\left(a \left(-16 \left(a^2+2\right) \Omega +a \left(a^2+14\right)+24 a \Omega ^2\right)+9\right)+6 \Phi ^7 (a-\Omega ) \left(a \left(3 a^4-20 a^3 \Omega +2 \left(a^2-3\right) a \Omega ^3\right.\right.\right.\\[1mm]
			&\left.\left.\left.+10 a^2+\left(a^4+14 a^2-3\right) \Omega ^2+2 a \Omega -9\right)+6 \Omega \right)-3 \sqrt{3} \Phi ^6 \left(a^6 \left(-25 \Omega ^4+4 \Omega ^2-15\right)\right.\right.\\[1mm]
			&\left.\left.+2 a^5 \Omega  \left(8 \Omega ^4+23 \Omega ^2+55\right)-2 a^4 \left(7 \Omega ^4+118 \Omega ^2+27\right)+8 a^3 \Omega  \left(17 \Omega ^2+10\right)+a^2 \left(45-49 \Omega ^4\right)\right.\right.\\[1mm]
			&\left.\left.+2 a \Omega  \left(5 \Omega ^2-39\right)+24 \Omega ^2\right)-18 \Phi ^5 \left(a^6 \left(-39 \Omega ^4+17 \Omega ^2-10\right)+a^5 \Omega  \left(83 \Omega ^2+51\right)\right.\right.\\[1mm]
			&\left.\left.+2 a^4 \left(6 \Omega ^6+16 \Omega ^4-107 \Omega ^2-14\right)+4 a^3 \Omega  \left(-6 \Omega ^4+9 \Omega ^2+20\right)+a^2 \left(-25 \Omega ^4+47 \Omega ^2+30\right)\right.\right.\\[1mm]
			&\left.\left.+a \left(9 \Omega ^3-75 \Omega \right)+18 \Omega ^2\right)+9 \sqrt{3} \Phi ^4 \left(a^6 \left(63 \Omega ^4-56 \Omega ^2+15\right)+2 a^5 \Omega  \left(72 \Omega ^4-83 \Omega ^2-9\right)\right.\right.\\[1mm]
			&\left.\left.+a^4 \left(-72 \Omega ^6-430 \Omega ^4+398 \Omega ^2+14\right)+16 a^3 \Omega  \left(7 \Omega ^4+23 \Omega ^2-10\right)+a^2 \left(47 \Omega ^4-352 \Omega ^2-45\right)\right.\right.\\[1mm]
			&\left.\left.-42 a \Omega  \left(\Omega ^2-5\right)-30 \Omega ^2\right)-54 \Phi ^3 \left(a^6 \left(14 \Omega ^4+17 \Omega ^2-3\right)-a^5 \Omega  \left(128 \Omega ^4+3 \Omega ^2+11\right)\right.\right.\\[1mm]
			&\left.\left.+6 a^4 \left(4 \Omega ^6+55 \Omega ^4-13 \Omega ^2+1\right)+4 a^3 \Omega  \left(4 \Omega ^4-81 \Omega ^2+10\right)+a^2 \left(-124 \Omega ^4+227 \Omega ^2+9\right)\right.\right.\\[1mm]
			&\left.\left.+a \left(87 \Omega ^3-93 \Omega \right)-6 \Omega ^2\right)-27 \sqrt{3} \Phi ^2 \left(a^6 \left(57 \Omega ^4+4 \Omega ^2-1\right)-2 a^5 \Omega  \left(72 \Omega ^4+43 \Omega ^2+3\right)\right.\right.\\[1mm]
			&\left.\left.+a^4 \left(-48 \Omega ^6+334 \Omega ^4+4 \Omega ^2+6\right)+8 a^3 \Omega  \left(36 \Omega ^4-41 \Omega ^2+2\right)+a^2 \left(-479 \Omega ^4+232 \Omega ^2+3\right)\right.\right.\\[1mm]
			&\left.\left.+a \left(286 \Omega ^3-90 \Omega \right)-48 \Omega ^2\right) \right),
		\end{aligned}    
\end{equation}}
{\small\begin{equation}
		\begin{aligned}
			\Delta G_{2}\!&=-\frac{1}{18 \sqrt{3} \left(\Phi +\sqrt{3}\right)^4 (a-\Omega )^2 (a \Omega -1)^4}\left(6 a (a \Omega -1)^2 \left(-\Phi ^7 (a-\Omega ) \left(a^2-2 a \Omega +1\right)\right.\right.\\[1mm]
			&\left.\left.+\sqrt{3} \Phi ^6 \left(a \left(a^2 \left(\Omega ^2-6\right)+15 a \Omega -11 \Omega ^2-4\right)+5 \Omega \right)+9 \Phi ^5 \left(a \left(2 \left(a^2-4\right) \Omega ^2-5 a^2+9 a \Omega -1\right)+3 \Omega \right)\right.\right.\\[1mm]
			&\left.\left.+15 \sqrt{3} \Phi ^4 \left(a \left(a^2 \left(3 \Omega ^2-4\right)+3 a \Omega -5 \Omega ^2+2\right)+\Omega \right)+45 \Phi ^3 \left(a \left(a^2 \left(4 \Omega ^2-3\right)-3 a \Omega -2 \Omega ^2+5\right)-\Omega \right)\right.\right.\\[1mm]
			&\left.\left.+27 \sqrt{3} \Phi ^2 \left(a \left(a^2 \left(5 \Omega ^2-2\right)-9 a \Omega +\Omega ^2+8\right)-3 \Omega \right)+27 \Phi  \left(a \left(a^2 \left(6 \Omega ^2-1\right)-15 a \Omega +4 \Omega ^2+11\right)-5 \Omega \right)\right.\right.\\[1mm]
			&\left.\left.+27 \sqrt{3} (a \Omega -1) (a (a \Omega -2)+\Omega )\right) \right) .   
		\end{aligned}    
\end{equation}}
\noindent
Similarly the on-shell temperature can be shown to be

{\small\begin{equation}
		\begin{aligned}
			T(a,\Phi,\Omega)\!&= \frac{\sqrt{a \left(\sqrt{3} \left(a^2-1\right) \Phi +3 a \Omega -3\right)}}{6 \pi  a \left(\Phi +\sqrt{3}\right) \left(\sqrt{3} \Phi +3\right) \sqrt{a-\Omega } (a \Omega -1)}\left( \left(\sqrt{3} \Phi +3\right) \left(\Phi ^2-3\right) \Omega -a \left(\sqrt{3} \Phi ^3+\Phi ^2 \left(9-6 \Omega ^2\right)\right.\right.\\[1mm]
			&\left.\left.+3 \sqrt{3} \Phi  \left(3-4 \Omega ^2\right)-18 \Omega ^2+9\right)\right).
			\end{aligned}   
\end{equation}}\\
\noindent
We present the phase diagram following from $G$ for various cases, such as,  subcritical electric potential limit, critical electric potential limit and the maximal rotation velocity limit. These are shown respectively in figures-(\ref{G1}), (\ref{G2}) and (\ref{G3}) below.
\begin{figure}[!htbp]
	\centering
	\includegraphics[width=0.6\linewidth]{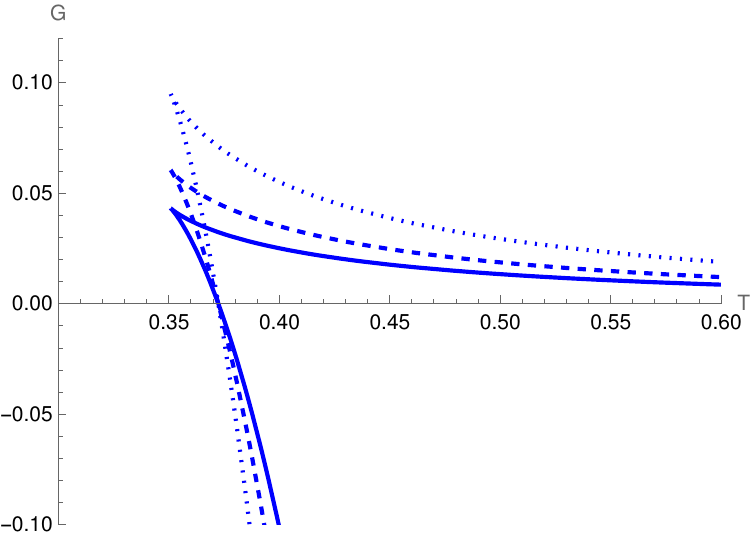}
	\caption{On-shell phase diagram of general black holes in AdS$_5$ with four-derivative corrections: $\Omega = 0.5,{\Phi} = \frac{\sqrt{3}}{2},  $: The solid, dashed, dotted curve represents  ``$\alpha = \lambda_{1} = \lambda_{2} = 0$", ``$\alpha = 0.001, \lambda_{1} = 0.001, \lambda_{2} = 100$", ``$\alpha = 0.002, \lambda_{1} = 0.003,\lambda_{2} = 100$" respectively. }
	\label{G1}
\end{figure}
\FloatBarrier
\noindent
 From the on-shell phase diagrams in the sub critical electric potential limit fig (\ref{G1}), we find that:
\begin{itemize}
	\item  For $ T < T_{\rm HP}$, with increasing $\alpha, \lambda_{1}$, both small and large black holes further destabilize.
	\item  For $ T > T_{\rm HP}$, with increasing $\alpha, \lambda_{1}$, the small black hole phase destabilizes, while the large black hole phase stabilizes.
	
\end{itemize}
\begin{figure}[!htbp]
	\centering
	\includegraphics[width=0.6\linewidth]{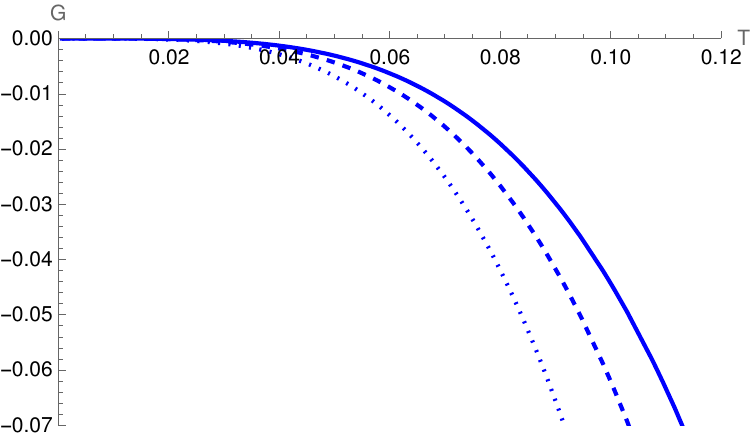}
	\caption{On-shell phase diagram of general black holes in AdS$_5$ with four-derivative corrections:  $\Omega = 0.8,{\Phi} = \sqrt{3},  $: The solid, dashed, dotted curve represents  ``$\alpha = \lambda_{1} = \lambda_{2} = 0$", ``$\alpha = 0.001, \lambda_{1} = 0.001, \lambda_{2} = 100$", ``$\alpha = 0.002, \lambda_{1} = 0.003,\lambda_{2} = 100$" respectively. }
	\label{G2}
\end{figure}
\FloatBarrier

\noindent
For the critical electric potential case in fig (\ref{G2}), we observe that, in the presence of four derivative corrections, the large black hole phase further stabilizes.


\begin{figure}[!htbp]
	\centering
	\includegraphics[width=0.6\linewidth]{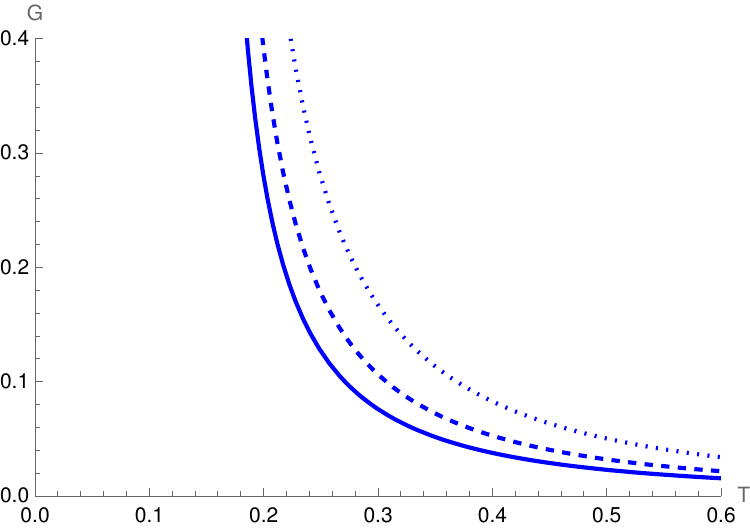}
	\caption{On-shell phase diagram of general black holes in AdS$_5$ with four-derivative corrections: $\Omega = 1,{\Phi} = \frac{\sqrt{3}}{4},  $: The solid, dashed, dotted curve represents  ``$\alpha = \lambda_{1} = \lambda_{2} = 0$", ``$\alpha = 0.001, \lambda_{1} = 0.001, \lambda_{2} = 100$", ``$\alpha = 0.002, \lambda_{1} = 0.003,\lambda_{2} = 100$" respectively. }
	\label{G3}
\end{figure}
\FloatBarrier
\noindent
For the maximum angular velocity limit with the phase diagram shown in fig (\ref{G3}), in the presence of four-derivative corrections, the small black hole phase further destabilizes.

\subsection{BPS black holes in AdS$_5$}
\begin{figure}[!htbp]
	\centering
	\includegraphics[width=0.6\linewidth]{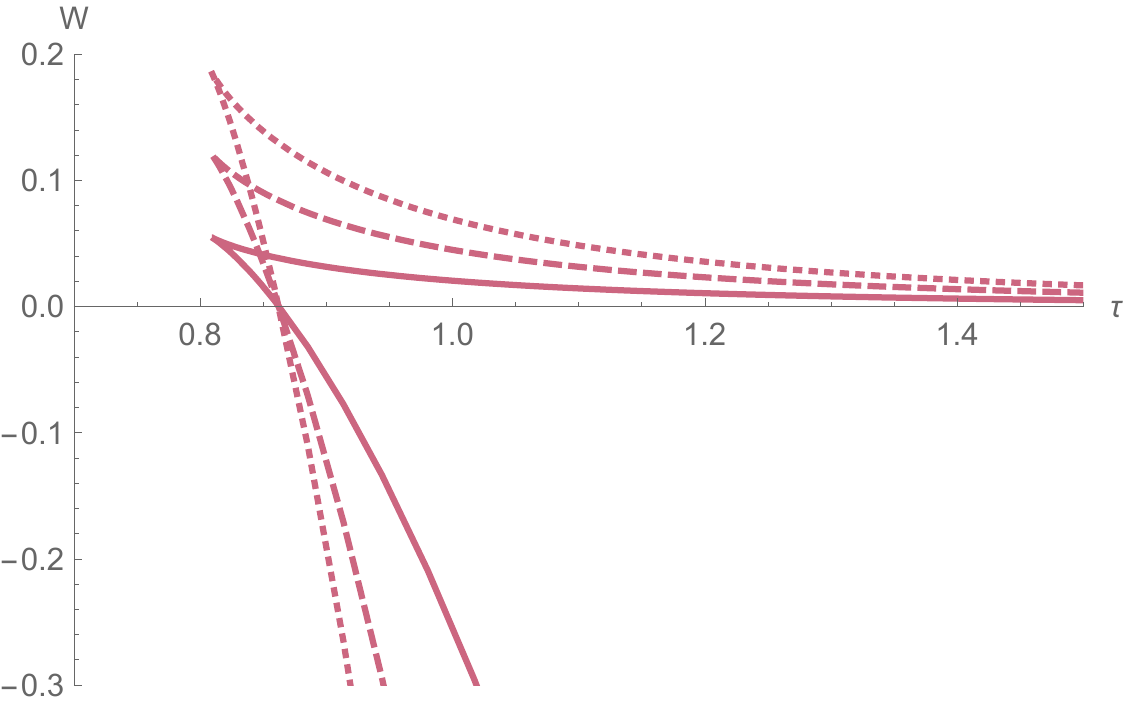}
	\caption{On-shell Gibbs free energy for BPS black holes with four derivative corrections: The solid, dashed and dotted curves represent the cases  ``$\alpha = \lambda_{1} = \lambda_{2} = 0$", ``$\alpha = 0.003, \lambda_{1} = 0.003, \lambda_{2} = 100$", ``$\alpha = 0.006, \lambda_{1} = 0.007,\lambda_{2} = 100$" respectively. }
	\label{4dWalpha}
\end{figure}

The on-shell phase diagram of BPS black holes with four derivative corrections, based on the free energy in eqn. (\ref{4dcorrectedW}) is given in figure-(\ref{4dWalpha}).
\FloatBarrier
\noindent
We can clearly see that our off shell conclusions for effect of four derivative corrections are consistent with on shell conclusions in all limiting cases considered in section-(\ref{4d}).

\bibliographystyle{apsrev4-1}
\bibliography{bwbib} 

\begin{thebibliography}{163}%
\makeatletter
\providecommand \@ifxundefined [1]{%
 \@ifx{#1\undefined}
}%
\providecommand \@ifnum [1]{%
 \ifnum #1\expandafter \@firstoftwo
 \else \expandafter \@secondoftwo
 \fi
}%
\providecommand \@ifx [1]{%
 \ifx #1\expandafter \@firstoftwo
 \else \expandafter \@secondoftwo
 \fi
}%
\providecommand \natexlab [1]{#1}%
\providecommand \enquote  [1]{``#1''}%
\providecommand \bibnamefont  [1]{#1}%
\providecommand \bibfnamefont [1]{#1}%
\providecommand \citenamefont [1]{#1}%
\providecommand \href@noop [0]{\@secondoftwo}%
\providecommand \href [0]{\begingroup \@sanitize@url \@href}%
\providecommand \@href[1]{\@@startlink{#1}\@@href}%
\providecommand \@@href[1]{\endgroup#1\@@endlink}%
\providecommand \@sanitize@url [0]{\catcode `\\12\catcode `\$12\catcode
  `\&12\catcode `\#12\catcode `\^12\catcode `\_12\catcode `\%12\relax}%
\providecommand \@@startlink[1]{}%
\providecommand \@@endlink[0]{}%
\providecommand \url  [0]{\begingroup\@sanitize@url \@url }%
\providecommand \@url [1]{\endgroup\@href {#1}{\urlprefix }}%
\providecommand \urlprefix  [0]{URL }%
\providecommand \Eprint [0]{\href }%
\providecommand \doibase [0]{http://dx.doi.org/}%
\providecommand \selectlanguage [0]{\@gobble}%
\providecommand \bibinfo  [0]{\@secondoftwo}%
\providecommand \bibfield  [0]{\@secondoftwo}%
\providecommand \translation [1]{[#1]}%
\providecommand \BibitemOpen [0]{}%
\providecommand \bibitemStop [0]{}%
\providecommand \bibitemNoStop [0]{.\EOS\space}%
\providecommand \EOS [0]{\spacefactor3000\relax}%
\providecommand \BibitemShut  [1]{\csname bibitem#1\endcsname}%
\let\auto@bib@innerbib\@empty
\bibitem [{\citenamefont {Bekenstein}(1973)}]{Bekenstein:1973ur}%
  \BibitemOpen
  \bibfield  {author} {\bibinfo {author} {\bibfnamefont {J.~D.}\ \bibnamefont
  {Bekenstein}},\ }\href {\doibase 10.1103/PhysRevD.7.2333} {\bibfield
  {journal} {\bibinfo  {journal} {Phys. Rev. D}\ }\textbf {\bibinfo {volume}
  {7}},\ \bibinfo {pages} {2333} (\bibinfo {year} {1973})}\BibitemShut
  {NoStop}%
\bibitem [{\citenamefont {Hawking}(1975)}]{Hawking:1975vcx}%
  \BibitemOpen
  \bibfield  {author} {\bibinfo {author} {\bibfnamefont {S.~W.}\ \bibnamefont
  {Hawking}},\ }\href {\doibase 10.1007/BF02345020} {\bibfield  {journal}
  {\bibinfo  {journal} {Commun. Math. Phys.}\ }\textbf {\bibinfo {volume}
  {43}},\ \bibinfo {pages} {199} (\bibinfo {year} {1975})},\ \bibinfo {note}
  {[Erratum: Commun.Math.Phys. 46, 206 (1976)]}\BibitemShut {NoStop}%
\bibitem [{\citenamefont {Hawking}\ and\ \citenamefont
  {Page}(1983)}]{Hawking:1982dh}%
  \BibitemOpen
  \bibfield  {author} {\bibinfo {author} {\bibfnamefont {S.~W.}\ \bibnamefont
  {Hawking}}\ and\ \bibinfo {author} {\bibfnamefont {D.~N.}\ \bibnamefont
  {Page}},\ }\href {\doibase 10.1007/BF01208266} {\bibfield  {journal}
  {\bibinfo  {journal} {Commun. Math. Phys.}\ }\textbf {\bibinfo {volume}
  {87}},\ \bibinfo {pages} {577} (\bibinfo {year} {1983})}\BibitemShut
  {NoStop}%
\bibitem [{\citenamefont {Strominger}\ and\ \citenamefont
  {Vafa}(1996)}]{Strominger:1996sh}%
  \BibitemOpen
  \bibfield  {author} {\bibinfo {author} {\bibfnamefont {A.}~\bibnamefont
  {Strominger}}\ and\ \bibinfo {author} {\bibfnamefont {C.}~\bibnamefont
  {Vafa}},\ }\href {\doibase 10.1016/0370-2693(96)00345-0} {\bibfield
  {journal} {\bibinfo  {journal} {Phys. Lett. B}\ }\textbf {\bibinfo {volume}
  {379}},\ \bibinfo {pages} {99} (\bibinfo {year} {1996})},\ \Eprint
  {http://arxiv.org/abs/hep-th/9601029} {arXiv:hep-th/9601029} \BibitemShut
  {NoStop}%
\bibitem [{\citenamefont {Chamblin}\ \emph {et~al.}(1999)\citenamefont
  {Chamblin}, \citenamefont {Emparan}, \citenamefont {Johnson},\ and\
  \citenamefont {Myers}}]{Chamblin:1999tk}%
  \BibitemOpen
  \bibfield  {author} {\bibinfo {author} {\bibfnamefont {A.}~\bibnamefont
  {Chamblin}}, \bibinfo {author} {\bibfnamefont {R.}~\bibnamefont {Emparan}},
  \bibinfo {author} {\bibfnamefont {C.~V.}\ \bibnamefont {Johnson}}, \ and\
  \bibinfo {author} {\bibfnamefont {R.~C.}\ \bibnamefont {Myers}},\ }\href
  {\doibase 10.1103/PhysRevD.60.064018} {\bibfield  {journal} {\bibinfo
  {journal} {Phys. Rev. D}\ }\textbf {\bibinfo {volume} {60}},\ \bibinfo
  {pages} {064018} (\bibinfo {year} {1999})},\ \Eprint
  {http://arxiv.org/abs/hep-th/9902170} {arXiv:hep-th/9902170} \BibitemShut
  {NoStop}%
\bibitem [{\citenamefont {Caldarelli}\ \emph {et~al.}(2000)\citenamefont
  {Caldarelli}, \citenamefont {Cognola},\ and\ \citenamefont
  {Klemm}}]{Caldarelli:1999xj}%
  \BibitemOpen
  \bibfield  {author} {\bibinfo {author} {\bibfnamefont {M.~M.}\ \bibnamefont
  {Caldarelli}}, \bibinfo {author} {\bibfnamefont {G.}~\bibnamefont {Cognola}},
  \ and\ \bibinfo {author} {\bibfnamefont {D.}~\bibnamefont {Klemm}},\ }\href
  {\doibase 10.1088/0264-9381/17/2/310} {\bibfield  {journal} {\bibinfo
  {journal} {Class. Quant. Grav.}\ }\textbf {\bibinfo {volume} {17}},\ \bibinfo
  {pages} {399} (\bibinfo {year} {2000})},\ \Eprint
  {http://arxiv.org/abs/hep-th/9908022} {arXiv:hep-th/9908022} \BibitemShut
  {NoStop}%
\bibitem [{\citenamefont {Silva}(2006)}]{Silva:2006xv}%
  \BibitemOpen
  \bibfield  {author} {\bibinfo {author} {\bibfnamefont {P.~J.}\ \bibnamefont
  {Silva}},\ }\href {\doibase 10.1088/1126-6708/2006/10/022} {\bibfield
  {journal} {\bibinfo  {journal} {JHEP}\ }\textbf {\bibinfo {volume} {10}},\
  \bibinfo {pages} {022} (\bibinfo {year} {2006})},\ \Eprint
  {http://arxiv.org/abs/hep-th/0607056} {arXiv:hep-th/0607056} \BibitemShut
  {NoStop}%
\bibitem [{\citenamefont {Sen}(2008{\natexlab{a}})}]{Sen:2007qy}%
  \BibitemOpen
  \bibfield  {author} {\bibinfo {author} {\bibfnamefont {A.}~\bibnamefont
  {Sen}},\ }\href {\doibase 10.1007/s10714-008-0626-4} {\bibfield  {journal}
  {\bibinfo  {journal} {Gen. Rel. Grav.}\ }\textbf {\bibinfo {volume} {40}},\
  \bibinfo {pages} {2249} (\bibinfo {year} {2008}{\natexlab{a}})},\ \Eprint
  {http://arxiv.org/abs/0708.1270} {arXiv:0708.1270 [hep-th]} \BibitemShut
  {NoStop}%
\bibitem [{\citenamefont {Benini}\ and\ \citenamefont
  {Zaffaroni}(2015)}]{Benini:2015noa}%
  \BibitemOpen
  \bibfield  {author} {\bibinfo {author} {\bibfnamefont {F.}~\bibnamefont
  {Benini}}\ and\ \bibinfo {author} {\bibfnamefont {A.}~\bibnamefont
  {Zaffaroni}},\ }\href {\doibase 10.1007/JHEP07(2015)127} {\bibfield
  {journal} {\bibinfo  {journal} {JHEP}\ }\textbf {\bibinfo {volume} {07}},\
  \bibinfo {pages} {127} (\bibinfo {year} {2015})},\ \Eprint
  {http://arxiv.org/abs/1504.03698} {arXiv:1504.03698 [hep-th]} \BibitemShut
  {NoStop}%
\bibitem [{\citenamefont {Almheiri}\ and\ \citenamefont
  {Kang}(2016)}]{Almheiri:2016fws}%
  \BibitemOpen
  \bibfield  {author} {\bibinfo {author} {\bibfnamefont {A.}~\bibnamefont
  {Almheiri}}\ and\ \bibinfo {author} {\bibfnamefont {B.}~\bibnamefont
  {Kang}},\ }\href {\doibase 10.1007/JHEP10(2016)052} {\bibfield  {journal}
  {\bibinfo  {journal} {JHEP}\ }\textbf {\bibinfo {volume} {10}},\ \bibinfo
  {pages} {052} (\bibinfo {year} {2016})},\ \Eprint
  {http://arxiv.org/abs/1606.04108} {arXiv:1606.04108 [hep-th]} \BibitemShut
  {NoStop}%
\bibitem [{\citenamefont {Hosseini}\ and\ \citenamefont
  {Zaffaroni}(2016)}]{Hosseini:2016tor}%
  \BibitemOpen
  \bibfield  {author} {\bibinfo {author} {\bibfnamefont {S.~M.}\ \bibnamefont
  {Hosseini}}\ and\ \bibinfo {author} {\bibfnamefont {A.}~\bibnamefont
  {Zaffaroni}},\ }\href {\doibase 10.1007/JHEP08(2016)064} {\bibfield
  {journal} {\bibinfo  {journal} {JHEP}\ }\textbf {\bibinfo {volume} {08}},\
  \bibinfo {pages} {064} (\bibinfo {year} {2016})},\ \Eprint
  {http://arxiv.org/abs/1604.03122} {arXiv:1604.03122 [hep-th]} \BibitemShut
  {NoStop}%
\bibitem [{\citenamefont {Benini}\ \emph {et~al.}(2016)\citenamefont {Benini},
  \citenamefont {Hristov},\ and\ \citenamefont {Zaffaroni}}]{Benini:2015eyy}%
  \BibitemOpen
  \bibfield  {author} {\bibinfo {author} {\bibfnamefont {F.}~\bibnamefont
  {Benini}}, \bibinfo {author} {\bibfnamefont {K.}~\bibnamefont {Hristov}}, \
  and\ \bibinfo {author} {\bibfnamefont {A.}~\bibnamefont {Zaffaroni}},\ }\href
  {\doibase 10.1007/JHEP05(2016)054} {\bibfield  {journal} {\bibinfo  {journal}
  {JHEP}\ }\textbf {\bibinfo {volume} {05}},\ \bibinfo {pages} {054} (\bibinfo
  {year} {2016})},\ \Eprint {http://arxiv.org/abs/1511.04085} {arXiv:1511.04085
  [hep-th]} \BibitemShut {NoStop}%
\bibitem [{\citenamefont {Hosseini}\ \emph {et~al.}(2017)\citenamefont
  {Hosseini}, \citenamefont {Hristov},\ and\ \citenamefont
  {Zaffaroni}}]{Hosseini:2017mds}%
  \BibitemOpen
  \bibfield  {author} {\bibinfo {author} {\bibfnamefont {S.~M.}\ \bibnamefont
  {Hosseini}}, \bibinfo {author} {\bibfnamefont {K.}~\bibnamefont {Hristov}}, \
  and\ \bibinfo {author} {\bibfnamefont {A.}~\bibnamefont {Zaffaroni}},\ }\href
  {\doibase 10.1007/JHEP07(2017)106} {\bibfield  {journal} {\bibinfo  {journal}
  {JHEP}\ }\textbf {\bibinfo {volume} {07}},\ \bibinfo {pages} {106} (\bibinfo
  {year} {2017})},\ \Eprint {http://arxiv.org/abs/1705.05383} {arXiv:1705.05383
  [hep-th]} \BibitemShut {NoStop}%
\bibitem [{\citenamefont {Choi}\ \emph
  {et~al.}(2018{\natexlab{a}})\citenamefont {Choi}, \citenamefont {Kim},
  \citenamefont {Kim},\ and\ \citenamefont {Nahmgoong}}]{Choi:2018vbz}%
  \BibitemOpen
  \bibfield  {author} {\bibinfo {author} {\bibfnamefont {S.}~\bibnamefont
  {Choi}}, \bibinfo {author} {\bibfnamefont {J.}~\bibnamefont {Kim}}, \bibinfo
  {author} {\bibfnamefont {S.}~\bibnamefont {Kim}}, \ and\ \bibinfo {author}
  {\bibfnamefont {J.}~\bibnamefont {Nahmgoong}},\ }\href@noop {} {\  (\bibinfo
  {year} {2018}{\natexlab{a}})},\ \Eprint {http://arxiv.org/abs/1811.08646}
  {arXiv:1811.08646 [hep-th]} \BibitemShut {NoStop}%
\bibitem [{\citenamefont {Choi}\ and\ \citenamefont
  {Kim}(2024)}]{Choi:2019miv}%
  \BibitemOpen
  \bibfield  {author} {\bibinfo {author} {\bibfnamefont {S.}~\bibnamefont
  {Choi}}\ and\ \bibinfo {author} {\bibfnamefont {S.}~\bibnamefont {Kim}},\
  }\href {\doibase 10.1007/JHEP08(2024)228} {\bibfield  {journal} {\bibinfo
  {journal} {JHEP}\ }\textbf {\bibinfo {volume} {08}},\ \bibinfo {pages} {228}
  (\bibinfo {year} {2024})},\ \Eprint {http://arxiv.org/abs/1904.01164}
  {arXiv:1904.01164 [hep-th]} \BibitemShut {NoStop}%
\bibitem [{\citenamefont {Larsen}\ \emph {et~al.}(2020)\citenamefont {Larsen},
  \citenamefont {Nian},\ and\ \citenamefont {Zeng}}]{Larsen:2019oll}%
  \BibitemOpen
  \bibfield  {author} {\bibinfo {author} {\bibfnamefont {F.}~\bibnamefont
  {Larsen}}, \bibinfo {author} {\bibfnamefont {J.}~\bibnamefont {Nian}}, \ and\
  \bibinfo {author} {\bibfnamefont {Y.}~\bibnamefont {Zeng}},\ }\href {\doibase
  10.1007/JHEP06(2020)001} {\bibfield  {journal} {\bibinfo  {journal} {JHEP}\
  }\textbf {\bibinfo {volume} {06}},\ \bibinfo {pages} {001} (\bibinfo {year}
  {2020})},\ \Eprint {http://arxiv.org/abs/1907.02505} {arXiv:1907.02505
  [hep-th]} \BibitemShut {NoStop}%
\bibitem [{\citenamefont {Copetti}\ \emph {et~al.}(2022)\citenamefont
  {Copetti}, \citenamefont {Grassi}, \citenamefont {Komargodski},\ and\
  \citenamefont {Tizzano}}]{Copetti:2020dil}%
  \BibitemOpen
  \bibfield  {author} {\bibinfo {author} {\bibfnamefont {C.}~\bibnamefont
  {Copetti}}, \bibinfo {author} {\bibfnamefont {A.}~\bibnamefont {Grassi}},
  \bibinfo {author} {\bibfnamefont {Z.}~\bibnamefont {Komargodski}}, \ and\
  \bibinfo {author} {\bibfnamefont {L.}~\bibnamefont {Tizzano}},\ }\href
  {\doibase 10.1007/JHEP04(2022)132} {\bibfield  {journal} {\bibinfo  {journal}
  {JHEP}\ }\textbf {\bibinfo {volume} {04}},\ \bibinfo {pages} {132} (\bibinfo
  {year} {2022})},\ \Eprint {http://arxiv.org/abs/2008.04950} {arXiv:2008.04950
  [hep-th]} \BibitemShut {NoStop}%
\bibitem [{\citenamefont {Hosseini}\ \emph
  {et~al.}(2018{\natexlab{a}})\citenamefont {Hosseini}, \citenamefont
  {Hristov},\ and\ \citenamefont {Zaffaroni}}]{Hosseini:2018dob}%
  \BibitemOpen
  \bibfield  {author} {\bibinfo {author} {\bibfnamefont {S.~M.}\ \bibnamefont
  {Hosseini}}, \bibinfo {author} {\bibfnamefont {K.}~\bibnamefont {Hristov}}, \
  and\ \bibinfo {author} {\bibfnamefont {A.}~\bibnamefont {Zaffaroni}},\ }\href
  {\doibase 10.1007/JHEP05(2018)121} {\bibfield  {journal} {\bibinfo  {journal}
  {JHEP}\ }\textbf {\bibinfo {volume} {05}},\ \bibinfo {pages} {121} (\bibinfo
  {year} {2018}{\natexlab{a}})},\ \Eprint {http://arxiv.org/abs/1803.07568}
  {arXiv:1803.07568 [hep-th]} \BibitemShut {NoStop}%
\bibitem [{\citenamefont {Hosseini}\ \emph
  {et~al.}(2018{\natexlab{b}})\citenamefont {Hosseini}, \citenamefont
  {Hristov}, \citenamefont {Passias},\ and\ \citenamefont
  {Zaffaroni}}]{Hosseini:2018usu}%
  \BibitemOpen
  \bibfield  {author} {\bibinfo {author} {\bibfnamefont {S.~M.}\ \bibnamefont
  {Hosseini}}, \bibinfo {author} {\bibfnamefont {K.}~\bibnamefont {Hristov}},
  \bibinfo {author} {\bibfnamefont {A.}~\bibnamefont {Passias}}, \ and\
  \bibinfo {author} {\bibfnamefont {A.}~\bibnamefont {Zaffaroni}},\ }\href
  {\doibase 10.1007/JHEP12(2018)001} {\bibfield  {journal} {\bibinfo  {journal}
  {JHEP}\ }\textbf {\bibinfo {volume} {12}},\ \bibinfo {pages} {001} (\bibinfo
  {year} {2018}{\natexlab{b}})},\ \Eprint {http://arxiv.org/abs/1809.10685}
  {arXiv:1809.10685 [hep-th]} \BibitemShut {NoStop}%
\bibitem [{\citenamefont {Crichigno}\ and\ \citenamefont
  {Jain}(2020)}]{Crichigno:2020ouj}%
  \BibitemOpen
  \bibfield  {author} {\bibinfo {author} {\bibfnamefont {P.~M.}\ \bibnamefont
  {Crichigno}}\ and\ \bibinfo {author} {\bibfnamefont {D.}~\bibnamefont
  {Jain}},\ }\href {\doibase 10.1007/JHEP09(2020)124} {\bibfield  {journal}
  {\bibinfo  {journal} {JHEP}\ }\textbf {\bibinfo {volume} {09}},\ \bibinfo
  {pages} {124} (\bibinfo {year} {2020})},\ \Eprint
  {http://arxiv.org/abs/2005.00550} {arXiv:2005.00550 [hep-th]} \BibitemShut
  {NoStop}%
\bibitem [{\citenamefont {Zaffaroni}(2020)}]{Zaffaroni:2019dhb}%
  \BibitemOpen
  \bibfield  {author} {\bibinfo {author} {\bibfnamefont {A.}~\bibnamefont
  {Zaffaroni}},\ }\href {\doibase 10.1007/s41114-020-00027-8} {\bibfield
  {journal} {\bibinfo  {journal} {Living Rev. Rel.}\ }\textbf {\bibinfo
  {volume} {23}},\ \bibinfo {pages} {2} (\bibinfo {year} {2020})},\ \Eprint
  {http://arxiv.org/abs/1902.07176} {arXiv:1902.07176 [hep-th]} \BibitemShut
  {NoStop}%
\bibitem [{\citenamefont {Fu}\ \emph {et~al.}(2017)\citenamefont {Fu},
  \citenamefont {Gaiotto}, \citenamefont {Maldacena},\ and\ \citenamefont
  {Sachdev}}]{Fu:2016vas}%
  \BibitemOpen
  \bibfield  {author} {\bibinfo {author} {\bibfnamefont {W.}~\bibnamefont
  {Fu}}, \bibinfo {author} {\bibfnamefont {D.}~\bibnamefont {Gaiotto}},
  \bibinfo {author} {\bibfnamefont {J.}~\bibnamefont {Maldacena}}, \ and\
  \bibinfo {author} {\bibfnamefont {S.}~\bibnamefont {Sachdev}},\ }\href
  {\doibase 10.1103/PhysRevD.95.026009} {\bibfield  {journal} {\bibinfo
  {journal} {Phys. Rev. D}\ }\textbf {\bibinfo {volume} {95}},\ \bibinfo
  {pages} {026009} (\bibinfo {year} {2017})},\ \bibinfo {note} {[Addendum:
  Phys.Rev.D 95, 069904 (2017)]},\ \Eprint {http://arxiv.org/abs/1610.08917}
  {arXiv:1610.08917 [hep-th]} \BibitemShut {NoStop}%
\bibitem [{\citenamefont {Iliesiu}\ and\ \citenamefont
  {Turiaci}(2021)}]{Iliesiu:2020qvm}%
  \BibitemOpen
  \bibfield  {author} {\bibinfo {author} {\bibfnamefont {L.~V.}\ \bibnamefont
  {Iliesiu}}\ and\ \bibinfo {author} {\bibfnamefont {G.~J.}\ \bibnamefont
  {Turiaci}},\ }\href {\doibase 10.1007/JHEP05(2021)145} {\bibfield  {journal}
  {\bibinfo  {journal} {JHEP}\ }\textbf {\bibinfo {volume} {05}},\ \bibinfo
  {pages} {145} (\bibinfo {year} {2021})},\ \Eprint
  {http://arxiv.org/abs/2003.02860} {arXiv:2003.02860 [hep-th]} \BibitemShut
  {NoStop}%
\bibitem [{\citenamefont {Heydeman}\ \emph {et~al.}(2022)\citenamefont
  {Heydeman}, \citenamefont {Iliesiu}, \citenamefont {Turiaci},\ and\
  \citenamefont {Zhao}}]{Heydeman:2020hhw}%
  \BibitemOpen
  \bibfield  {author} {\bibinfo {author} {\bibfnamefont {M.}~\bibnamefont
  {Heydeman}}, \bibinfo {author} {\bibfnamefont {L.~V.}\ \bibnamefont
  {Iliesiu}}, \bibinfo {author} {\bibfnamefont {G.~J.}\ \bibnamefont
  {Turiaci}}, \ and\ \bibinfo {author} {\bibfnamefont {W.}~\bibnamefont
  {Zhao}},\ }\href {\doibase 10.1088/1751-8121/ac3be9} {\bibfield  {journal}
  {\bibinfo  {journal} {J. Phys. A}\ }\textbf {\bibinfo {volume} {55}},\
  \bibinfo {pages} {014004} (\bibinfo {year} {2022})},\ \Eprint
  {http://arxiv.org/abs/2011.01953} {arXiv:2011.01953 [hep-th]} \BibitemShut
  {NoStop}%
\bibitem [{\citenamefont {Choi}\ and\ \citenamefont
  {Larsen}(2021)}]{Choi:2021nnq}%
  \BibitemOpen
  \bibfield  {author} {\bibinfo {author} {\bibfnamefont {S.}~\bibnamefont
  {Choi}}\ and\ \bibinfo {author} {\bibfnamefont {F.}~\bibnamefont {Larsen}},\
  }\enquote {\bibinfo {title} {{Effective Field Theory of Quantum Black
  Holes}},}\ \ (\bibinfo {year} {2021})\ \Eprint
  {http://arxiv.org/abs/2108.04028} {arXiv:2108.04028 [hep-th]} \BibitemShut
  {NoStop}%
\bibitem [{\citenamefont {Gutowski}\ and\ \citenamefont
  {Reall}(2004{\natexlab{a}})}]{Gutowski:2004ez}%
  \BibitemOpen
  \bibfield  {author} {\bibinfo {author} {\bibfnamefont {J.~B.}\ \bibnamefont
  {Gutowski}}\ and\ \bibinfo {author} {\bibfnamefont {H.~S.}\ \bibnamefont
  {Reall}},\ }\href {\doibase 10.1088/1126-6708/2004/02/006} {\bibfield
  {journal} {\bibinfo  {journal} {JHEP}\ }\textbf {\bibinfo {volume} {02}},\
  \bibinfo {pages} {006} (\bibinfo {year} {2004}{\natexlab{a}})},\ \Eprint
  {http://arxiv.org/abs/hep-th/0401042} {arXiv:hep-th/0401042} \BibitemShut
  {NoStop}%
\bibitem [{\citenamefont {Gutowski}\ and\ \citenamefont
  {Reall}(2004{\natexlab{b}})}]{Gutowski:2004yv}%
  \BibitemOpen
  \bibfield  {author} {\bibinfo {author} {\bibfnamefont {J.~B.}\ \bibnamefont
  {Gutowski}}\ and\ \bibinfo {author} {\bibfnamefont {H.~S.}\ \bibnamefont
  {Reall}},\ }\href {\doibase 10.1088/1126-6708/2004/04/048} {\bibfield
  {journal} {\bibinfo  {journal} {JHEP}\ }\textbf {\bibinfo {volume} {04}},\
  \bibinfo {pages} {048} (\bibinfo {year} {2004}{\natexlab{b}})},\ \Eprint
  {http://arxiv.org/abs/hep-th/0401129} {arXiv:hep-th/0401129} \BibitemShut
  {NoStop}%
\bibitem [{\citenamefont {Sen}(2008{\natexlab{b}})}]{Sen:2008yk}%
  \BibitemOpen
  \bibfield  {author} {\bibinfo {author} {\bibfnamefont {A.}~\bibnamefont
  {Sen}},\ }\href {\doibase 10.1088/1126-6708/2008/11/075} {\bibfield
  {journal} {\bibinfo  {journal} {JHEP}\ }\textbf {\bibinfo {volume} {11}},\
  \bibinfo {pages} {075} (\bibinfo {year} {2008}{\natexlab{b}})},\ \Eprint
  {http://arxiv.org/abs/0805.0095} {arXiv:0805.0095 [hep-th]} \BibitemShut
  {NoStop}%
\bibitem [{\citenamefont {Sen}(2009)}]{Sen:2008vm}%
  \BibitemOpen
  \bibfield  {author} {\bibinfo {author} {\bibfnamefont {A.}~\bibnamefont
  {Sen}},\ }\href {\doibase 10.1142/S0217751X09045893} {\bibfield  {journal}
  {\bibinfo  {journal} {Int. J. Mod. Phys. A}\ }\textbf {\bibinfo {volume}
  {24}},\ \bibinfo {pages} {4225} (\bibinfo {year} {2009})},\ \Eprint
  {http://arxiv.org/abs/0809.3304} {arXiv:0809.3304 [hep-th]} \BibitemShut
  {NoStop}%
\bibitem [{\citenamefont {Dabholkar}\ \emph {et~al.}(2011)\citenamefont
  {Dabholkar}, \citenamefont {Gomes}, \citenamefont {Murthy},\ and\
  \citenamefont {Sen}}]{Dabholkar:2010rm}%
  \BibitemOpen
  \bibfield  {author} {\bibinfo {author} {\bibfnamefont {A.}~\bibnamefont
  {Dabholkar}}, \bibinfo {author} {\bibfnamefont {J.}~\bibnamefont {Gomes}},
  \bibinfo {author} {\bibfnamefont {S.}~\bibnamefont {Murthy}}, \ and\ \bibinfo
  {author} {\bibfnamefont {A.}~\bibnamefont {Sen}},\ }\href {\doibase
  10.1007/JHEP04(2011)034} {\bibfield  {journal} {\bibinfo  {journal} {JHEP}\
  }\textbf {\bibinfo {volume} {04}},\ \bibinfo {pages} {034} (\bibinfo {year}
  {2011})},\ \Eprint {http://arxiv.org/abs/1009.3226} {arXiv:1009.3226
  [hep-th]} \BibitemShut {NoStop}%
\bibitem [{\citenamefont {Chowdhury}\ \emph {et~al.}(2024)\citenamefont
  {Chowdhury}, \citenamefont {Sen}, \citenamefont {Shanmugapriya},\ and\
  \citenamefont {Virmani}}]{Chowdhury:2024ngg}%
  \BibitemOpen
  \bibfield  {author} {\bibinfo {author} {\bibfnamefont {C.}~\bibnamefont
  {Chowdhury}}, \bibinfo {author} {\bibfnamefont {A.}~\bibnamefont {Sen}},
  \bibinfo {author} {\bibfnamefont {P.}~\bibnamefont {Shanmugapriya}}, \ and\
  \bibinfo {author} {\bibfnamefont {A.}~\bibnamefont {Virmani}},\ }\href
  {\doibase 10.1007/JHEP04(2024)136} {\bibfield  {journal} {\bibinfo  {journal}
  {JHEP}\ }\textbf {\bibinfo {volume} {04}},\ \bibinfo {pages} {136} (\bibinfo
  {year} {2024})},\ \Eprint {http://arxiv.org/abs/2401.13730} {arXiv:2401.13730
  [hep-th]} \BibitemShut {NoStop}%
\bibitem [{\citenamefont {Benini}\ and\ \citenamefont
  {Zaffaroni}(2017)}]{Benini:2016hjo}%
  \BibitemOpen
  \bibfield  {author} {\bibinfo {author} {\bibfnamefont {F.}~\bibnamefont
  {Benini}}\ and\ \bibinfo {author} {\bibfnamefont {A.}~\bibnamefont
  {Zaffaroni}},\ }\href@noop {} {\bibfield  {journal} {\bibinfo  {journal}
  {Proc. Symp. Pure Math.}\ }\textbf {\bibinfo {volume} {96}},\ \bibinfo
  {pages} {13} (\bibinfo {year} {2017})},\ \Eprint
  {http://arxiv.org/abs/1605.06120} {arXiv:1605.06120 [hep-th]} \BibitemShut
  {NoStop}%
\bibitem [{\citenamefont {Benini}\ \emph {et~al.}(2017)\citenamefont {Benini},
  \citenamefont {Hristov},\ and\ \citenamefont {Zaffaroni}}]{Benini:2016rke}%
  \BibitemOpen
  \bibfield  {author} {\bibinfo {author} {\bibfnamefont {F.}~\bibnamefont
  {Benini}}, \bibinfo {author} {\bibfnamefont {K.}~\bibnamefont {Hristov}}, \
  and\ \bibinfo {author} {\bibfnamefont {A.}~\bibnamefont {Zaffaroni}},\ }\href
  {\doibase 10.1016/j.physletb.2017.05.076} {\bibfield  {journal} {\bibinfo
  {journal} {Phys. Lett. B}\ }\textbf {\bibinfo {volume} {771}},\ \bibinfo
  {pages} {462} (\bibinfo {year} {2017})},\ \Eprint
  {http://arxiv.org/abs/1608.07294} {arXiv:1608.07294 [hep-th]} \BibitemShut
  {NoStop}%
\bibitem [{\citenamefont {Choi}\ \emph
  {et~al.}(2018{\natexlab{b}})\citenamefont {Choi}, \citenamefont {Kim},
  \citenamefont {Kim},\ and\ \citenamefont {Nahmgoong}}]{Choi:2018hmj}%
  \BibitemOpen
  \bibfield  {author} {\bibinfo {author} {\bibfnamefont {S.}~\bibnamefont
  {Choi}}, \bibinfo {author} {\bibfnamefont {J.}~\bibnamefont {Kim}}, \bibinfo
  {author} {\bibfnamefont {S.}~\bibnamefont {Kim}}, \ and\ \bibinfo {author}
  {\bibfnamefont {J.}~\bibnamefont {Nahmgoong}},\ }\href@noop {} {\  (\bibinfo
  {year} {2018}{\natexlab{b}})},\ \Eprint {http://arxiv.org/abs/1810.12067}
  {arXiv:1810.12067 [hep-th]} \BibitemShut {NoStop}%
\bibitem [{\citenamefont {Cabo-Bizet}\ \emph
  {et~al.}(2019{\natexlab{a}})\citenamefont {Cabo-Bizet}, \citenamefont
  {Cassani}, \citenamefont {Martelli},\ and\ \citenamefont
  {Murthy}}]{Cabo-Bizet:2018ehj}%
  \BibitemOpen
  \bibfield  {author} {\bibinfo {author} {\bibfnamefont {A.}~\bibnamefont
  {Cabo-Bizet}}, \bibinfo {author} {\bibfnamefont {D.}~\bibnamefont {Cassani}},
  \bibinfo {author} {\bibfnamefont {D.}~\bibnamefont {Martelli}}, \ and\
  \bibinfo {author} {\bibfnamefont {S.}~\bibnamefont {Murthy}},\ }\href
  {\doibase 10.1007/JHEP10(2019)062} {\bibfield  {journal} {\bibinfo  {journal}
  {JHEP}\ }\textbf {\bibinfo {volume} {10}},\ \bibinfo {pages} {062} (\bibinfo
  {year} {2019}{\natexlab{a}})},\ \Eprint {http://arxiv.org/abs/1810.11442}
  {arXiv:1810.11442 [hep-th]} \BibitemShut {NoStop}%
\bibitem [{\citenamefont {Benini}\ and\ \citenamefont
  {Milan}(2020)}]{Benini:2018ywd}%
  \BibitemOpen
  \bibfield  {author} {\bibinfo {author} {\bibfnamefont {F.}~\bibnamefont
  {Benini}}\ and\ \bibinfo {author} {\bibfnamefont {E.}~\bibnamefont {Milan}},\
  }\href {\doibase 10.1103/PhysRevX.10.021037} {\bibfield  {journal} {\bibinfo
  {journal} {Phys. Rev. X}\ }\textbf {\bibinfo {volume} {10}},\ \bibinfo
  {pages} {021037} (\bibinfo {year} {2020})},\ \Eprint
  {http://arxiv.org/abs/1812.09613} {arXiv:1812.09613 [hep-th]} \BibitemShut
  {NoStop}%
\bibitem [{\citenamefont {Arabi~Ardehali}(2019)}]{ArabiArdehali:2019tdm}%
  \BibitemOpen
  \bibfield  {author} {\bibinfo {author} {\bibfnamefont {A.}~\bibnamefont
  {Arabi~Ardehali}},\ }\href {\doibase 10.1007/JHEP06(2019)134} {\bibfield
  {journal} {\bibinfo  {journal} {JHEP}\ }\textbf {\bibinfo {volume} {06}},\
  \bibinfo {pages} {134} (\bibinfo {year} {2019})},\ \Eprint
  {http://arxiv.org/abs/1902.06619} {arXiv:1902.06619 [hep-th]} \BibitemShut
  {NoStop}%
\bibitem [{\citenamefont {Honda}(2019)}]{Honda:2019cio}%
  \BibitemOpen
  \bibfield  {author} {\bibinfo {author} {\bibfnamefont {M.}~\bibnamefont
  {Honda}},\ }\href {\doibase 10.1103/PhysRevD.100.026008} {\bibfield
  {journal} {\bibinfo  {journal} {Phys. Rev. D}\ }\textbf {\bibinfo {volume}
  {100}},\ \bibinfo {pages} {026008} (\bibinfo {year} {2019})},\ \Eprint
  {http://arxiv.org/abs/1901.08091} {arXiv:1901.08091 [hep-th]} \BibitemShut
  {NoStop}%
\bibitem [{\citenamefont {Cabo-Bizet}\ \emph
  {et~al.}(2019{\natexlab{b}})\citenamefont {Cabo-Bizet}, \citenamefont
  {Cassani}, \citenamefont {Martelli},\ and\ \citenamefont
  {Murthy}}]{Cabo-Bizet:2019osg}%
  \BibitemOpen
  \bibfield  {author} {\bibinfo {author} {\bibfnamefont {A.}~\bibnamefont
  {Cabo-Bizet}}, \bibinfo {author} {\bibfnamefont {D.}~\bibnamefont {Cassani}},
  \bibinfo {author} {\bibfnamefont {D.}~\bibnamefont {Martelli}}, \ and\
  \bibinfo {author} {\bibfnamefont {S.}~\bibnamefont {Murthy}},\ }\href
  {\doibase 10.1007/JHEP08(2019)120} {\bibfield  {journal} {\bibinfo  {journal}
  {JHEP}\ }\textbf {\bibinfo {volume} {08}},\ \bibinfo {pages} {120} (\bibinfo
  {year} {2019}{\natexlab{b}})},\ \Eprint {http://arxiv.org/abs/1904.05865}
  {arXiv:1904.05865 [hep-th]} \BibitemShut {NoStop}%
\bibitem [{\citenamefont {Kim}\ \emph {et~al.}(2021)\citenamefont {Kim},
  \citenamefont {Kim},\ and\ \citenamefont {Song}}]{Kim:2019yrz}%
  \BibitemOpen
  \bibfield  {author} {\bibinfo {author} {\bibfnamefont {J.}~\bibnamefont
  {Kim}}, \bibinfo {author} {\bibfnamefont {S.}~\bibnamefont {Kim}}, \ and\
  \bibinfo {author} {\bibfnamefont {J.}~\bibnamefont {Song}},\ }\href {\doibase
  10.1007/JHEP01(2021)025} {\bibfield  {journal} {\bibinfo  {journal} {JHEP}\
  }\textbf {\bibinfo {volume} {01}},\ \bibinfo {pages} {025} (\bibinfo {year}
  {2021})},\ \Eprint {http://arxiv.org/abs/1904.03455} {arXiv:1904.03455
  [hep-th]} \BibitemShut {NoStop}%
\bibitem [{\citenamefont {Cabo-Bizet}\ and\ \citenamefont
  {Murthy}(2020)}]{Cabo-Bizet:2019eaf}%
  \BibitemOpen
  \bibfield  {author} {\bibinfo {author} {\bibfnamefont {A.}~\bibnamefont
  {Cabo-Bizet}}\ and\ \bibinfo {author} {\bibfnamefont {S.}~\bibnamefont
  {Murthy}},\ }\href {\doibase 10.1007/JHEP09(2020)184} {\bibfield  {journal}
  {\bibinfo  {journal} {JHEP}\ }\textbf {\bibinfo {volume} {09}},\ \bibinfo
  {pages} {184} (\bibinfo {year} {2020})},\ \Eprint
  {http://arxiv.org/abs/1909.09597} {arXiv:1909.09597 [hep-th]} \BibitemShut
  {NoStop}%
\bibitem [{\citenamefont {Amariti}\ \emph
  {et~al.}(2021{\natexlab{a}})\citenamefont {Amariti}, \citenamefont
  {Garozzo},\ and\ \citenamefont {Lo~Monaco}}]{Amariti:2019mgp}%
  \BibitemOpen
  \bibfield  {author} {\bibinfo {author} {\bibfnamefont {A.}~\bibnamefont
  {Amariti}}, \bibinfo {author} {\bibfnamefont {I.}~\bibnamefont {Garozzo}}, \
  and\ \bibinfo {author} {\bibfnamefont {G.}~\bibnamefont {Lo~Monaco}},\ }\href
  {\doibase 10.1016/j.nuclphysb.2021.115571} {\bibfield  {journal} {\bibinfo
  {journal} {Nucl. Phys. B}\ }\textbf {\bibinfo {volume} {973}},\ \bibinfo
  {pages} {115571} (\bibinfo {year} {2021}{\natexlab{a}})},\ \Eprint
  {http://arxiv.org/abs/1904.10009} {arXiv:1904.10009 [hep-th]} \BibitemShut
  {NoStop}%
\bibitem [{\citenamefont {Gonz\'alez~Lezcano}\ and\ \citenamefont
  {Pando~Zayas}(2020)}]{GonzalezLezcano:2019nca}%
  \BibitemOpen
  \bibfield  {author} {\bibinfo {author} {\bibfnamefont {A.}~\bibnamefont
  {Gonz\'alez~Lezcano}}\ and\ \bibinfo {author} {\bibfnamefont {L.~A.}\
  \bibnamefont {Pando~Zayas}},\ }\href {\doibase 10.1007/JHEP03(2020)088}
  {\bibfield  {journal} {\bibinfo  {journal} {JHEP}\ }\textbf {\bibinfo
  {volume} {03}},\ \bibinfo {pages} {088} (\bibinfo {year} {2020})},\ \Eprint
  {http://arxiv.org/abs/1907.12841} {arXiv:1907.12841 [hep-th]} \BibitemShut
  {NoStop}%
\bibitem [{\citenamefont {Lanir}\ \emph {et~al.}(2020)\citenamefont {Lanir},
  \citenamefont {Nedelin},\ and\ \citenamefont {Sela}}]{Lanir:2019abx}%
  \BibitemOpen
  \bibfield  {author} {\bibinfo {author} {\bibfnamefont {A.}~\bibnamefont
  {Lanir}}, \bibinfo {author} {\bibfnamefont {A.}~\bibnamefont {Nedelin}}, \
  and\ \bibinfo {author} {\bibfnamefont {O.}~\bibnamefont {Sela}},\ }\href
  {\doibase 10.1007/JHEP04(2020)091} {\bibfield  {journal} {\bibinfo  {journal}
  {JHEP}\ }\textbf {\bibinfo {volume} {04}},\ \bibinfo {pages} {091} (\bibinfo
  {year} {2020})},\ \Eprint {http://arxiv.org/abs/1908.01737} {arXiv:1908.01737
  [hep-th]} \BibitemShut {NoStop}%
\bibitem [{\citenamefont {Goldstein}\ \emph {et~al.}(2020)\citenamefont
  {Goldstein}, \citenamefont {Jejjala}, \citenamefont {Lei}, \citenamefont {van
  Leuven},\ and\ \citenamefont {Li}}]{Goldstein:2019gpz}%
  \BibitemOpen
  \bibfield  {author} {\bibinfo {author} {\bibfnamefont {K.}~\bibnamefont
  {Goldstein}}, \bibinfo {author} {\bibfnamefont {V.}~\bibnamefont {Jejjala}},
  \bibinfo {author} {\bibfnamefont {Y.}~\bibnamefont {Lei}}, \bibinfo {author}
  {\bibfnamefont {S.}~\bibnamefont {van Leuven}}, \ and\ \bibinfo {author}
  {\bibfnamefont {W.}~\bibnamefont {Li}},\ }\href {\doibase
  10.1007/JHEP02(2020)154} {\bibfield  {journal} {\bibinfo  {journal} {JHEP}\
  }\textbf {\bibinfo {volume} {02}},\ \bibinfo {pages} {154} (\bibinfo {year}
  {2020})},\ \Eprint {http://arxiv.org/abs/1910.14293} {arXiv:1910.14293
  [hep-th]} \BibitemShut {NoStop}%
\bibitem [{\citenamefont {Arabi~Ardehali}\ \emph {et~al.}(2020)\citenamefont
  {Arabi~Ardehali}, \citenamefont {Hong},\ and\ \citenamefont
  {Liu}}]{ArabiArdehali:2019orz}%
  \BibitemOpen
  \bibfield  {author} {\bibinfo {author} {\bibfnamefont {A.}~\bibnamefont
  {Arabi~Ardehali}}, \bibinfo {author} {\bibfnamefont {J.}~\bibnamefont
  {Hong}}, \ and\ \bibinfo {author} {\bibfnamefont {J.~T.}\ \bibnamefont
  {Liu}},\ }\href {\doibase 10.1007/JHEP07(2020)073} {\bibfield  {journal}
  {\bibinfo  {journal} {JHEP}\ }\textbf {\bibinfo {volume} {07}},\ \bibinfo
  {pages} {073} (\bibinfo {year} {2020})},\ \Eprint
  {http://arxiv.org/abs/1912.04169} {arXiv:1912.04169 [hep-th]} \BibitemShut
  {NoStop}%
\bibitem [{\citenamefont {Murthy}(2022)}]{Murthy:2020scj}%
  \BibitemOpen
  \bibfield  {author} {\bibinfo {author} {\bibfnamefont {S.}~\bibnamefont
  {Murthy}},\ }\href {\doibase 10.1103/PhysRevD.105.L021903} {\bibfield
  {journal} {\bibinfo  {journal} {Phys. Rev. D}\ }\textbf {\bibinfo {volume}
  {105}},\ \bibinfo {pages} {L021903} (\bibinfo {year} {2022})},\ \Eprint
  {http://arxiv.org/abs/2005.10843} {arXiv:2005.10843 [hep-th]} \BibitemShut
  {NoStop}%
\bibitem [{\citenamefont {Agarwal}\ \emph {et~al.}(2021)\citenamefont
  {Agarwal}, \citenamefont {Choi}, \citenamefont {Kim}, \citenamefont {Kim},\
  and\ \citenamefont {Nahmgoong}}]{Agarwal:2020zwm}%
  \BibitemOpen
  \bibfield  {author} {\bibinfo {author} {\bibfnamefont {P.}~\bibnamefont
  {Agarwal}}, \bibinfo {author} {\bibfnamefont {S.}~\bibnamefont {Choi}},
  \bibinfo {author} {\bibfnamefont {J.}~\bibnamefont {Kim}}, \bibinfo {author}
  {\bibfnamefont {S.}~\bibnamefont {Kim}}, \ and\ \bibinfo {author}
  {\bibfnamefont {J.}~\bibnamefont {Nahmgoong}},\ }\href {\doibase
  10.1103/PhysRevD.103.126006} {\bibfield  {journal} {\bibinfo  {journal}
  {Phys. Rev. D}\ }\textbf {\bibinfo {volume} {103}},\ \bibinfo {pages}
  {126006} (\bibinfo {year} {2021})},\ \Eprint
  {http://arxiv.org/abs/2005.11240} {arXiv:2005.11240 [hep-th]} \BibitemShut
  {NoStop}%
\bibitem [{\citenamefont {Larsen}\ and\ \citenamefont
  {Paranjape}(2021)}]{Larsen:2020lhg}%
  \BibitemOpen
  \bibfield  {author} {\bibinfo {author} {\bibfnamefont {F.}~\bibnamefont
  {Larsen}}\ and\ \bibinfo {author} {\bibfnamefont {S.}~\bibnamefont
  {Paranjape}},\ }\href {\doibase 10.1007/JHEP10(2021)198} {\bibfield
  {journal} {\bibinfo  {journal} {JHEP}\ }\textbf {\bibinfo {volume} {10}},\
  \bibinfo {pages} {198} (\bibinfo {year} {2021})},\ \Eprint
  {http://arxiv.org/abs/2010.04359} {arXiv:2010.04359 [hep-th]} \BibitemShut
  {NoStop}%
\bibitem [{\citenamefont {Benini}\ \emph {et~al.}(2020)\citenamefont {Benini},
  \citenamefont {Colombo}, \citenamefont {Soltani}, \citenamefont {Zaffaroni},\
  and\ \citenamefont {Zhang}}]{Benini:2020gjh}%
  \BibitemOpen
  \bibfield  {author} {\bibinfo {author} {\bibfnamefont {F.}~\bibnamefont
  {Benini}}, \bibinfo {author} {\bibfnamefont {E.}~\bibnamefont {Colombo}},
  \bibinfo {author} {\bibfnamefont {S.}~\bibnamefont {Soltani}}, \bibinfo
  {author} {\bibfnamefont {A.}~\bibnamefont {Zaffaroni}}, \ and\ \bibinfo
  {author} {\bibfnamefont {Z.}~\bibnamefont {Zhang}},\ }\href {\doibase
  10.1088/1361-6382/abb39b} {\bibfield  {journal} {\bibinfo  {journal} {Class.
  Quant. Grav.}\ }\textbf {\bibinfo {volume} {37}},\ \bibinfo {pages} {215021}
  (\bibinfo {year} {2020})},\ \Eprint {http://arxiv.org/abs/2005.12308}
  {arXiv:2005.12308 [hep-th]} \BibitemShut {NoStop}%
\bibitem [{\citenamefont {Cabo-Bizet}\ \emph {et~al.}(2020)\citenamefont
  {Cabo-Bizet}, \citenamefont {Cassani}, \citenamefont {Martelli},\ and\
  \citenamefont {Murthy}}]{Cabo-Bizet:2020nkr}%
  \BibitemOpen
  \bibfield  {author} {\bibinfo {author} {\bibfnamefont {A.}~\bibnamefont
  {Cabo-Bizet}}, \bibinfo {author} {\bibfnamefont {D.}~\bibnamefont {Cassani}},
  \bibinfo {author} {\bibfnamefont {D.}~\bibnamefont {Martelli}}, \ and\
  \bibinfo {author} {\bibfnamefont {S.}~\bibnamefont {Murthy}},\ }\href
  {\doibase 10.1007/JHEP11(2020)150} {\bibfield  {journal} {\bibinfo  {journal}
  {JHEP}\ }\textbf {\bibinfo {volume} {11}},\ \bibinfo {pages} {150} (\bibinfo
  {year} {2020})},\ \Eprint {http://arxiv.org/abs/2005.10654} {arXiv:2005.10654
  [hep-th]} \BibitemShut {NoStop}%
\bibitem [{\citenamefont {Cabo-Bizet}(2023)}]{Cabo-Bizet:2021plf}%
  \BibitemOpen
  \bibfield  {author} {\bibinfo {author} {\bibfnamefont {A.}~\bibnamefont
  {Cabo-Bizet}},\ }\href {\doibase 10.1007/JHEP02(2023)134} {\bibfield
  {journal} {\bibinfo  {journal} {JHEP}\ }\textbf {\bibinfo {volume} {02}},\
  \bibinfo {pages} {134} (\bibinfo {year} {2023})},\ \Eprint
  {http://arxiv.org/abs/2111.14941} {arXiv:2111.14941 [hep-th]} \BibitemShut
  {NoStop}%
\bibitem [{\citenamefont {Cassani}\ and\ \citenamefont
  {Komargodski}(2021)}]{Cassani:2021fyv}%
  \BibitemOpen
  \bibfield  {author} {\bibinfo {author} {\bibfnamefont {D.}~\bibnamefont
  {Cassani}}\ and\ \bibinfo {author} {\bibfnamefont {Z.}~\bibnamefont
  {Komargodski}},\ }\href {\doibase 10.21468/SciPostPhys.11.1.004} {\bibfield
  {journal} {\bibinfo  {journal} {SciPost Phys.}\ }\textbf {\bibinfo {volume}
  {11}},\ \bibinfo {pages} {004} (\bibinfo {year} {2021})},\ \Eprint
  {http://arxiv.org/abs/2104.01464} {arXiv:2104.01464 [hep-th]} \BibitemShut
  {NoStop}%
\bibitem [{\citenamefont {Jejjala}\ \emph {et~al.}(2021)\citenamefont
  {Jejjala}, \citenamefont {Lei}, \citenamefont {van Leuven},\ and\
  \citenamefont {Li}}]{Jejjala:2021hlt}%
  \BibitemOpen
  \bibfield  {author} {\bibinfo {author} {\bibfnamefont {V.}~\bibnamefont
  {Jejjala}}, \bibinfo {author} {\bibfnamefont {Y.}~\bibnamefont {Lei}},
  \bibinfo {author} {\bibfnamefont {S.}~\bibnamefont {van Leuven}}, \ and\
  \bibinfo {author} {\bibfnamefont {W.}~\bibnamefont {Li}},\ }\href {\doibase
  10.1007/JHEP11(2021)047} {\bibfield  {journal} {\bibinfo  {journal} {JHEP}\
  }\textbf {\bibinfo {volume} {11}},\ \bibinfo {pages} {047} (\bibinfo {year}
  {2021})},\ \Eprint {http://arxiv.org/abs/2104.07030} {arXiv:2104.07030
  [hep-th]} \BibitemShut {NoStop}%
\bibitem [{\citenamefont {Jejjala}\ \emph {et~al.}(2023)\citenamefont
  {Jejjala}, \citenamefont {Lei}, \citenamefont {van Leuven},\ and\
  \citenamefont {Li}}]{Jejjala:2022lrm}%
  \BibitemOpen
  \bibfield  {author} {\bibinfo {author} {\bibfnamefont {V.}~\bibnamefont
  {Jejjala}}, \bibinfo {author} {\bibfnamefont {Y.}~\bibnamefont {Lei}},
  \bibinfo {author} {\bibfnamefont {S.}~\bibnamefont {van Leuven}}, \ and\
  \bibinfo {author} {\bibfnamefont {W.}~\bibnamefont {Li}},\ }\href {\doibase
  10.1007/JHEP10(2023)105} {\bibfield  {journal} {\bibinfo  {journal} {JHEP}\
  }\textbf {\bibinfo {volume} {10}},\ \bibinfo {pages} {105} (\bibinfo {year}
  {2023})},\ \Eprint {http://arxiv.org/abs/2210.17551} {arXiv:2210.17551
  [hep-th]} \BibitemShut {NoStop}%
\bibitem [{\citenamefont {Aharony}\ \emph {et~al.}(2021)\citenamefont
  {Aharony}, \citenamefont {Benini}, \citenamefont {Mamroud},\ and\
  \citenamefont {Milan}}]{Aharony:2021zkr}%
  \BibitemOpen
  \bibfield  {author} {\bibinfo {author} {\bibfnamefont {O.}~\bibnamefont
  {Aharony}}, \bibinfo {author} {\bibfnamefont {F.}~\bibnamefont {Benini}},
  \bibinfo {author} {\bibfnamefont {O.}~\bibnamefont {Mamroud}}, \ and\
  \bibinfo {author} {\bibfnamefont {E.}~\bibnamefont {Milan}},\ }\href
  {\doibase 10.1103/PhysRevD.104.086026} {\bibfield  {journal} {\bibinfo
  {journal} {Phys. Rev. D}\ }\textbf {\bibinfo {volume} {104}},\ \bibinfo
  {pages} {086026} (\bibinfo {year} {2021})},\ \Eprint
  {http://arxiv.org/abs/2104.13932} {arXiv:2104.13932 [hep-th]} \BibitemShut
  {NoStop}%
\bibitem [{\citenamefont {Cabo-Bizet}(2022)}]{Cabo-Bizet:2021jar}%
  \BibitemOpen
  \bibfield  {author} {\bibinfo {author} {\bibfnamefont {A.}~\bibnamefont
  {Cabo-Bizet}},\ }\href {\doibase 10.1007/JHEP10(2022)052} {\bibfield
  {journal} {\bibinfo  {journal} {JHEP}\ }\textbf {\bibinfo {volume} {10}},\
  \bibinfo {pages} {052} (\bibinfo {year} {2022})},\ \Eprint
  {http://arxiv.org/abs/2111.14942} {arXiv:2111.14942 [hep-th]} \BibitemShut
  {NoStop}%
\bibitem [{\citenamefont {Goldstein}\ \emph {et~al.}(2021)\citenamefont
  {Goldstein}, \citenamefont {Jejjala}, \citenamefont {Lei}, \citenamefont {van
  Leuven},\ and\ \citenamefont {Li}}]{Goldstein:2020yvj}%
  \BibitemOpen
  \bibfield  {author} {\bibinfo {author} {\bibfnamefont {K.}~\bibnamefont
  {Goldstein}}, \bibinfo {author} {\bibfnamefont {V.}~\bibnamefont {Jejjala}},
  \bibinfo {author} {\bibfnamefont {Y.}~\bibnamefont {Lei}}, \bibinfo {author}
  {\bibfnamefont {S.}~\bibnamefont {van Leuven}}, \ and\ \bibinfo {author}
  {\bibfnamefont {W.}~\bibnamefont {Li}},\ }\href {\doibase
  10.1007/JHEP04(2021)216} {\bibfield  {journal} {\bibinfo  {journal} {JHEP}\
  }\textbf {\bibinfo {volume} {04}},\ \bibinfo {pages} {216} (\bibinfo {year}
  {2021})},\ \Eprint {http://arxiv.org/abs/2011.06605} {arXiv:2011.06605
  [hep-th]} \BibitemShut {NoStop}%
\bibitem [{\citenamefont {Choi}\ \emph {et~al.}(2023)\citenamefont {Choi},
  \citenamefont {Jeong}, \citenamefont {Kim},\ and\ \citenamefont
  {Lee}}]{Choi:2021rxi}%
  \BibitemOpen
  \bibfield  {author} {\bibinfo {author} {\bibfnamefont {S.}~\bibnamefont
  {Choi}}, \bibinfo {author} {\bibfnamefont {S.}~\bibnamefont {Jeong}},
  \bibinfo {author} {\bibfnamefont {S.}~\bibnamefont {Kim}}, \ and\ \bibinfo
  {author} {\bibfnamefont {E.}~\bibnamefont {Lee}},\ }\href {\doibase
  10.1007/JHEP09(2023)138} {\bibfield  {journal} {\bibinfo  {journal} {JHEP}\
  }\textbf {\bibinfo {volume} {09}},\ \bibinfo {pages} {138} (\bibinfo {year}
  {2023})},\ \Eprint {http://arxiv.org/abs/2111.10720} {arXiv:2111.10720
  [hep-th]} \BibitemShut {NoStop}%
\bibitem [{\citenamefont {Choi}\ \emph {et~al.}(2024)\citenamefont {Choi},
  \citenamefont {Kim},\ and\ \citenamefont {Song}}]{Choi:2023tiq}%
  \BibitemOpen
  \bibfield  {author} {\bibinfo {author} {\bibfnamefont {S.}~\bibnamefont
  {Choi}}, \bibinfo {author} {\bibfnamefont {S.}~\bibnamefont {Kim}}, \ and\
  \bibinfo {author} {\bibfnamefont {J.}~\bibnamefont {Song}},\ }\href {\doibase
  10.1007/JHEP08(2024)105} {\bibfield  {journal} {\bibinfo  {journal} {JHEP}\
  }\textbf {\bibinfo {volume} {08}},\ \bibinfo {pages} {105} (\bibinfo {year}
  {2024})},\ \Eprint {http://arxiv.org/abs/2309.07614} {arXiv:2309.07614
  [hep-th]} \BibitemShut {NoStop}%
\bibitem [{\citenamefont {Cassani}\ \emph {et~al.}(2024)\citenamefont
  {Cassani}, \citenamefont {Ruip\'erez},\ and\ \citenamefont
  {Turetta}}]{Cassani:2024tvk}%
  \BibitemOpen
  \bibfield  {author} {\bibinfo {author} {\bibfnamefont {D.}~\bibnamefont
  {Cassani}}, \bibinfo {author} {\bibfnamefont {A.}~\bibnamefont {Ruip\'erez}},
  \ and\ \bibinfo {author} {\bibfnamefont {E.}~\bibnamefont {Turetta}},\ }\href
  {\doibase 10.1007/JHEP05(2024)276} {\bibfield  {journal} {\bibinfo  {journal}
  {JHEP}\ }\textbf {\bibinfo {volume} {05}},\ \bibinfo {pages} {276} (\bibinfo
  {year} {2024})},\ \Eprint {http://arxiv.org/abs/2403.02410} {arXiv:2403.02410
  [hep-th]} \BibitemShut {NoStop}%
\bibitem [{\citenamefont {Ezroura}\ \emph {et~al.}(2022)\citenamefont
  {Ezroura}, \citenamefont {Larsen}, \citenamefont {Liu},\ and\ \citenamefont
  {Zeng}}]{Ezroura:2021vrt}%
  \BibitemOpen
  \bibfield  {author} {\bibinfo {author} {\bibfnamefont {N.}~\bibnamefont
  {Ezroura}}, \bibinfo {author} {\bibfnamefont {F.}~\bibnamefont {Larsen}},
  \bibinfo {author} {\bibfnamefont {Z.}~\bibnamefont {Liu}}, \ and\ \bibinfo
  {author} {\bibfnamefont {Y.}~\bibnamefont {Zeng}},\ }\href {\doibase
  10.1007/JHEP09(2022)033} {\bibfield  {journal} {\bibinfo  {journal} {JHEP}\
  }\textbf {\bibinfo {volume} {09}},\ \bibinfo {pages} {033} (\bibinfo {year}
  {2022})},\ \Eprint {http://arxiv.org/abs/2108.11542} {arXiv:2108.11542
  [hep-th]} \BibitemShut {NoStop}%
\bibitem [{\citenamefont {Almheiri}\ and\ \citenamefont
  {Polchinski}(2015)}]{Almheiri:2014cka}%
  \BibitemOpen
  \bibfield  {author} {\bibinfo {author} {\bibfnamefont {A.}~\bibnamefont
  {Almheiri}}\ and\ \bibinfo {author} {\bibfnamefont {J.}~\bibnamefont
  {Polchinski}},\ }\href {\doibase 10.1007/JHEP11(2015)014} {\bibfield
  {journal} {\bibinfo  {journal} {JHEP}\ }\textbf {\bibinfo {volume} {11}},\
  \bibinfo {pages} {014} (\bibinfo {year} {2015})},\ \Eprint
  {http://arxiv.org/abs/1402.6334} {arXiv:1402.6334 [hep-th]} \BibitemShut
  {NoStop}%
\bibitem [{\citenamefont {Kitaev}(2015)}]{Kitaev}%
  \BibitemOpen
  \bibfield  {author} {\bibinfo {author} {\bibfnamefont {A.}~\bibnamefont
  {Kitaev}},\ }\href {http://online.kitp.ucsb.edu/online/entangled15/}
  {\enquote {\bibinfo {title} {A simple model of quantum holography},}\
  }\bibinfo {howpublished} {KITP Program: Entanglement in Strongly-Correlated
  Quantum Matter} (\bibinfo {year} {2015})\BibitemShut {NoStop}%
\bibitem [{\citenamefont {Maldacena}\ \emph {et~al.}(2016)\citenamefont
  {Maldacena}, \citenamefont {Stanford},\ and\ \citenamefont
  {Yang}}]{Maldacena:2016upp}%
  \BibitemOpen
  \bibfield  {author} {\bibinfo {author} {\bibfnamefont {J.}~\bibnamefont
  {Maldacena}}, \bibinfo {author} {\bibfnamefont {D.}~\bibnamefont {Stanford}},
  \ and\ \bibinfo {author} {\bibfnamefont {Z.}~\bibnamefont {Yang}},\ }\href
  {\doibase 10.1093/ptep/ptw124} {\bibfield  {journal} {\bibinfo  {journal}
  {PTEP}\ }\textbf {\bibinfo {volume} {2016}},\ \bibinfo {pages} {12C104}
  (\bibinfo {year} {2016})},\ \Eprint {http://arxiv.org/abs/1606.01857}
  {arXiv:1606.01857 [hep-th]} \BibitemShut {NoStop}%
\bibitem [{\citenamefont {Jensen}(2016)}]{Jensen:2016pah}%
  \BibitemOpen
  \bibfield  {author} {\bibinfo {author} {\bibfnamefont {K.}~\bibnamefont
  {Jensen}},\ }\href {\doibase 10.1103/PhysRevLett.117.111601} {\bibfield
  {journal} {\bibinfo  {journal} {Phys. Rev. Lett.}\ }\textbf {\bibinfo
  {volume} {117}},\ \bibinfo {pages} {111601} (\bibinfo {year} {2016})},\
  \Eprint {http://arxiv.org/abs/1605.06098} {arXiv:1605.06098 [hep-th]}
  \BibitemShut {NoStop}%
\bibitem [{\citenamefont {Engels\"oy}\ \emph {et~al.}(2016)\citenamefont
  {Engels\"oy}, \citenamefont {Mertens},\ and\ \citenamefont
  {Verlinde}}]{Engelsoy:2016xyb}%
  \BibitemOpen
  \bibfield  {author} {\bibinfo {author} {\bibfnamefont {J.}~\bibnamefont
  {Engels\"oy}}, \bibinfo {author} {\bibfnamefont {T.~G.}\ \bibnamefont
  {Mertens}}, \ and\ \bibinfo {author} {\bibfnamefont {H.}~\bibnamefont
  {Verlinde}},\ }\href {\doibase 10.1007/JHEP07(2016)139} {\bibfield  {journal}
  {\bibinfo  {journal} {JHEP}\ }\textbf {\bibinfo {volume} {07}},\ \bibinfo
  {pages} {139} (\bibinfo {year} {2016})},\ \Eprint
  {http://arxiv.org/abs/1606.03438} {arXiv:1606.03438 [hep-th]} \BibitemShut
  {NoStop}%
\bibitem [{\citenamefont {Larsen}(2019)}]{Larsen:2018iou}%
  \BibitemOpen
  \bibfield  {author} {\bibinfo {author} {\bibfnamefont {F.}~\bibnamefont
  {Larsen}},\ }\href {\doibase 10.1007/JHEP04(2019)055} {\bibfield  {journal}
  {\bibinfo  {journal} {JHEP}\ }\textbf {\bibinfo {volume} {04}},\ \bibinfo
  {pages} {055} (\bibinfo {year} {2019})},\ \Eprint
  {http://arxiv.org/abs/1806.06330} {arXiv:1806.06330 [hep-th]} \BibitemShut
  {NoStop}%
\bibitem [{\citenamefont {Nayak}\ \emph {et~al.}(2018)\citenamefont {Nayak},
  \citenamefont {Shukla}, \citenamefont {Soni}, \citenamefont {Trivedi},\ and\
  \citenamefont {Vishal}}]{Nayak:2018qej}%
  \BibitemOpen
  \bibfield  {author} {\bibinfo {author} {\bibfnamefont {P.}~\bibnamefont
  {Nayak}}, \bibinfo {author} {\bibfnamefont {A.}~\bibnamefont {Shukla}},
  \bibinfo {author} {\bibfnamefont {R.~M.}\ \bibnamefont {Soni}}, \bibinfo
  {author} {\bibfnamefont {S.~P.}\ \bibnamefont {Trivedi}}, \ and\ \bibinfo
  {author} {\bibfnamefont {V.}~\bibnamefont {Vishal}},\ }\href {\doibase
  10.1007/JHEP09(2018)048} {\bibfield  {journal} {\bibinfo  {journal} {JHEP}\
  }\textbf {\bibinfo {volume} {09}},\ \bibinfo {pages} {048} (\bibinfo {year}
  {2018})},\ \Eprint {http://arxiv.org/abs/1802.09547} {arXiv:1802.09547
  [hep-th]} \BibitemShut {NoStop}%
\bibitem [{\citenamefont {Kolekar}\ and\ \citenamefont
  {Narayan}(2018)}]{Kolekar:2018sba}%
  \BibitemOpen
  \bibfield  {author} {\bibinfo {author} {\bibfnamefont {K.~S.}\ \bibnamefont
  {Kolekar}}\ and\ \bibinfo {author} {\bibfnamefont {K.}~\bibnamefont
  {Narayan}},\ }\href {\doibase 10.1103/PhysRevD.98.046012} {\bibfield
  {journal} {\bibinfo  {journal} {Phys. Rev. D}\ }\textbf {\bibinfo {volume}
  {98}},\ \bibinfo {pages} {046012} (\bibinfo {year} {2018})},\ \Eprint
  {http://arxiv.org/abs/1803.06827} {arXiv:1803.06827 [hep-th]} \BibitemShut
  {NoStop}%
\bibitem [{\citenamefont {Moitra}\ \emph {et~al.}(2019)\citenamefont {Moitra},
  \citenamefont {Trivedi},\ and\ \citenamefont {Vishal}}]{Moitra:2018jqs}%
  \BibitemOpen
  \bibfield  {author} {\bibinfo {author} {\bibfnamefont {U.}~\bibnamefont
  {Moitra}}, \bibinfo {author} {\bibfnamefont {S.~P.}\ \bibnamefont {Trivedi}},
  \ and\ \bibinfo {author} {\bibfnamefont {V.}~\bibnamefont {Vishal}},\ }\href
  {\doibase 10.1007/JHEP07(2019)055} {\bibfield  {journal} {\bibinfo  {journal}
  {JHEP}\ }\textbf {\bibinfo {volume} {07}},\ \bibinfo {pages} {055} (\bibinfo
  {year} {2019})},\ \Eprint {http://arxiv.org/abs/1808.08239} {arXiv:1808.08239
  [hep-th]} \BibitemShut {NoStop}%
\bibitem [{\citenamefont {Castro}\ \emph {et~al.}(2021)\citenamefont {Castro},
  \citenamefont {Pedraza}, \citenamefont {Toldo},\ and\ \citenamefont
  {Verheijden}}]{Castro:2021fhc}%
  \BibitemOpen
  \bibfield  {author} {\bibinfo {author} {\bibfnamefont {A.}~\bibnamefont
  {Castro}}, \bibinfo {author} {\bibfnamefont {J.~F.}\ \bibnamefont {Pedraza}},
  \bibinfo {author} {\bibfnamefont {C.}~\bibnamefont {Toldo}}, \ and\ \bibinfo
  {author} {\bibfnamefont {E.}~\bibnamefont {Verheijden}},\ }\href {\doibase
  10.21468/SciPostPhys.11.6.102} {\bibfield  {journal} {\bibinfo  {journal}
  {SciPost Phys.}\ }\textbf {\bibinfo {volume} {11}},\ \bibinfo {pages} {102}
  (\bibinfo {year} {2021})},\ \Eprint {http://arxiv.org/abs/2106.00649}
  {arXiv:2106.00649 [hep-th]} \BibitemShut {NoStop}%
\bibitem [{\citenamefont {Castro}\ and\ \citenamefont
  {Verheijden}(2021)}]{Castro:2021wzn}%
  \BibitemOpen
  \bibfield  {author} {\bibinfo {author} {\bibfnamefont {A.}~\bibnamefont
  {Castro}}\ and\ \bibinfo {author} {\bibfnamefont {E.}~\bibnamefont
  {Verheijden}},\ }\href {\doibase 10.3390/universe7120475} {\bibfield
  {journal} {\bibinfo  {journal} {Universe}\ }\textbf {\bibinfo {volume} {7}},\
  \bibinfo {pages} {475} (\bibinfo {year} {2021})},\ \Eprint
  {http://arxiv.org/abs/2110.04208} {arXiv:2110.04208 [hep-th]} \BibitemShut
  {NoStop}%
\bibitem [{\citenamefont {Larsen}\ and\ \citenamefont
  {Lee}(2021)}]{Larsen:2021wnu}%
  \BibitemOpen
  \bibfield  {author} {\bibinfo {author} {\bibfnamefont {F.}~\bibnamefont
  {Larsen}}\ and\ \bibinfo {author} {\bibfnamefont {S.}~\bibnamefont {Lee}},\
  }\href {\doibase 10.1007/JHEP07(2021)038} {\bibfield  {journal} {\bibinfo
  {journal} {JHEP}\ }\textbf {\bibinfo {volume} {07}},\ \bibinfo {pages} {038}
  (\bibinfo {year} {2021})},\ \Eprint {http://arxiv.org/abs/2101.08497}
  {arXiv:2101.08497 [hep-th]} \BibitemShut {NoStop}%
\bibitem [{\citenamefont {Kubiznak}\ \emph {et~al.}(2017)\citenamefont
  {Kubiznak}, \citenamefont {Mann},\ and\ \citenamefont
  {Teo}}]{Kubiznak:2016qmn}%
  \BibitemOpen
  \bibfield  {author} {\bibinfo {author} {\bibfnamefont {D.}~\bibnamefont
  {Kubiznak}}, \bibinfo {author} {\bibfnamefont {R.~B.}\ \bibnamefont {Mann}},
  \ and\ \bibinfo {author} {\bibfnamefont {M.}~\bibnamefont {Teo}},\ }\href
  {\doibase 10.1088/1361-6382/aa5c69} {\bibfield  {journal} {\bibinfo
  {journal} {Class. Quant. Grav.}\ }\textbf {\bibinfo {volume} {34}},\ \bibinfo
  {pages} {063001} (\bibinfo {year} {2017})},\ \Eprint
  {http://arxiv.org/abs/1608.06147} {arXiv:1608.06147 [hep-th]} \BibitemShut
  {NoStop}%
\bibitem [{\citenamefont {Witten}(1998{\natexlab{a}})}]{Witten:1998zw}%
  \BibitemOpen
  \bibfield  {author} {\bibinfo {author} {\bibfnamefont {E.}~\bibnamefont
  {Witten}},\ }\href {\doibase 10.4310/ATMP.1998.v2.n3.a3} {\bibfield
  {journal} {\bibinfo  {journal} {Adv. Theor. Math. Phys.}\ }\textbf {\bibinfo
  {volume} {2}},\ \bibinfo {pages} {505} (\bibinfo {year}
  {1998}{\natexlab{a}})},\ \Eprint {http://arxiv.org/abs/hep-th/9803131}
  {arXiv:hep-th/9803131} \BibitemShut {NoStop}%
\bibitem [{\citenamefont {Sundborg}(2000)}]{Sundborg:1999ue}%
  \BibitemOpen
  \bibfield  {author} {\bibinfo {author} {\bibfnamefont {B.}~\bibnamefont
  {Sundborg}},\ }\href {\doibase 10.1016/S0550-3213(00)00044-4} {\bibfield
  {journal} {\bibinfo  {journal} {Nucl. Phys. B}\ }\textbf {\bibinfo {volume}
  {573}},\ \bibinfo {pages} {349} (\bibinfo {year} {2000})},\ \Eprint
  {http://arxiv.org/abs/hep-th/9908001} {arXiv:hep-th/9908001} \BibitemShut
  {NoStop}%
\bibitem [{\citenamefont {Aharony}\ \emph {et~al.}(2004)\citenamefont
  {Aharony}, \citenamefont {Marsano}, \citenamefont {Minwalla}, \citenamefont
  {Papadodimas},\ and\ \citenamefont {Van~Raamsdonk}}]{Aharony:2003sx}%
  \BibitemOpen
  \bibfield  {author} {\bibinfo {author} {\bibfnamefont {O.}~\bibnamefont
  {Aharony}}, \bibinfo {author} {\bibfnamefont {J.}~\bibnamefont {Marsano}},
  \bibinfo {author} {\bibfnamefont {S.}~\bibnamefont {Minwalla}}, \bibinfo
  {author} {\bibfnamefont {K.}~\bibnamefont {Papadodimas}}, \ and\ \bibinfo
  {author} {\bibfnamefont {M.}~\bibnamefont {Van~Raamsdonk}},\ }\href {\doibase
  10.4310/ATMP.2004.v8.n4.a1} {\bibfield  {journal} {\bibinfo  {journal} {Adv.
  Theor. Math. Phys.}\ }\textbf {\bibinfo {volume} {8}},\ \bibinfo {pages}
  {603} (\bibinfo {year} {2004})},\ \Eprint
  {http://arxiv.org/abs/hep-th/0310285} {arXiv:hep-th/0310285} \BibitemShut
  {NoStop}%
\bibitem [{\citenamefont {Kinney}\ \emph {et~al.}(2007)\citenamefont {Kinney},
  \citenamefont {Maldacena}, \citenamefont {Minwalla},\ and\ \citenamefont
  {Raju}}]{Kinney:2005ej}%
  \BibitemOpen
  \bibfield  {author} {\bibinfo {author} {\bibfnamefont {J.}~\bibnamefont
  {Kinney}}, \bibinfo {author} {\bibfnamefont {J.~M.}\ \bibnamefont
  {Maldacena}}, \bibinfo {author} {\bibfnamefont {S.}~\bibnamefont {Minwalla}},
  \ and\ \bibinfo {author} {\bibfnamefont {S.}~\bibnamefont {Raju}},\ }\href
  {\doibase 10.1007/s00220-007-0258-7} {\bibfield  {journal} {\bibinfo
  {journal} {Commun. Math. Phys.}\ }\textbf {\bibinfo {volume} {275}},\
  \bibinfo {pages} {209} (\bibinfo {year} {2007})},\ \Eprint
  {http://arxiv.org/abs/hep-th/0510251} {arXiv:hep-th/0510251} \BibitemShut
  {NoStop}%
\bibitem [{\citenamefont {Romelsberger}(2006)}]{Romelsberger:2005eg}%
  \BibitemOpen
  \bibfield  {author} {\bibinfo {author} {\bibfnamefont {C.}~\bibnamefont
  {Romelsberger}},\ }\href {\doibase 10.1016/j.nuclphysb.2006.03.037}
  {\bibfield  {journal} {\bibinfo  {journal} {Nucl. Phys. B}\ }\textbf
  {\bibinfo {volume} {747}},\ \bibinfo {pages} {329} (\bibinfo {year}
  {2006})},\ \Eprint {http://arxiv.org/abs/hep-th/0510060}
  {arXiv:hep-th/0510060} \BibitemShut {NoStop}%
\bibitem [{\citenamefont {Closset}\ \emph {et~al.}(2014)\citenamefont
  {Closset}, \citenamefont {Dumitrescu}, \citenamefont {Festuccia},\ and\
  \citenamefont {Komargodski}}]{Closset:2013vra}%
  \BibitemOpen
  \bibfield  {author} {\bibinfo {author} {\bibfnamefont {C.}~\bibnamefont
  {Closset}}, \bibinfo {author} {\bibfnamefont {T.~T.}\ \bibnamefont
  {Dumitrescu}}, \bibinfo {author} {\bibfnamefont {G.}~\bibnamefont
  {Festuccia}}, \ and\ \bibinfo {author} {\bibfnamefont {Z.}~\bibnamefont
  {Komargodski}},\ }\href {\doibase 10.1007/JHEP01(2014)124} {\bibfield
  {journal} {\bibinfo  {journal} {JHEP}\ }\textbf {\bibinfo {volume} {01}},\
  \bibinfo {pages} {124} (\bibinfo {year} {2014})},\ \Eprint
  {http://arxiv.org/abs/1309.5876} {arXiv:1309.5876 [hep-th]} \BibitemShut
  {NoStop}%
\bibitem [{\citenamefont {Assel}\ \emph {et~al.}(2014)\citenamefont {Assel},
  \citenamefont {Cassani},\ and\ \citenamefont {Martelli}}]{Assel:2014paa}%
  \BibitemOpen
  \bibfield  {author} {\bibinfo {author} {\bibfnamefont {B.}~\bibnamefont
  {Assel}}, \bibinfo {author} {\bibfnamefont {D.}~\bibnamefont {Cassani}}, \
  and\ \bibinfo {author} {\bibfnamefont {D.}~\bibnamefont {Martelli}},\ }\href
  {\doibase 10.1007/JHEP08(2014)123} {\bibfield  {journal} {\bibinfo  {journal}
  {JHEP}\ }\textbf {\bibinfo {volume} {08}},\ \bibinfo {pages} {123} (\bibinfo
  {year} {2014})},\ \Eprint {http://arxiv.org/abs/1405.5144} {arXiv:1405.5144
  [hep-th]} \BibitemShut {NoStop}%
\bibitem [{\citenamefont {Berkooz}\ \emph {et~al.}(2007)\citenamefont
  {Berkooz}, \citenamefont {Reichmann},\ and\ \citenamefont
  {Simon}}]{Berkooz:2006wc}%
  \BibitemOpen
  \bibfield  {author} {\bibinfo {author} {\bibfnamefont {M.}~\bibnamefont
  {Berkooz}}, \bibinfo {author} {\bibfnamefont {D.}~\bibnamefont {Reichmann}},
  \ and\ \bibinfo {author} {\bibfnamefont {J.}~\bibnamefont {Simon}},\ }\href
  {\doibase 10.1088/1126-6708/2007/01/048} {\bibfield  {journal} {\bibinfo
  {journal} {JHEP}\ }\textbf {\bibinfo {volume} {01}},\ \bibinfo {pages} {048}
  (\bibinfo {year} {2007})},\ \Eprint {http://arxiv.org/abs/hep-th/0604023}
  {arXiv:hep-th/0604023} \BibitemShut {NoStop}%
\bibitem [{\citenamefont {Grant}\ \emph {et~al.}(2008)\citenamefont {Grant},
  \citenamefont {Grassi}, \citenamefont {Kim},\ and\ \citenamefont
  {Minwalla}}]{Grant:2008sk}%
  \BibitemOpen
  \bibfield  {author} {\bibinfo {author} {\bibfnamefont {L.}~\bibnamefont
  {Grant}}, \bibinfo {author} {\bibfnamefont {P.~A.}\ \bibnamefont {Grassi}},
  \bibinfo {author} {\bibfnamefont {S.}~\bibnamefont {Kim}}, \ and\ \bibinfo
  {author} {\bibfnamefont {S.}~\bibnamefont {Minwalla}},\ }\href {\doibase
  10.1088/1126-6708/2008/05/049} {\bibfield  {journal} {\bibinfo  {journal}
  {JHEP}\ }\textbf {\bibinfo {volume} {05}},\ \bibinfo {pages} {049} (\bibinfo
  {year} {2008})},\ \Eprint {http://arxiv.org/abs/0803.4183} {arXiv:0803.4183
  [hep-th]} \BibitemShut {NoStop}%
\bibitem [{\citenamefont {Chang}\ and\ \citenamefont
  {Yin}(2013)}]{Chang:2013fba}%
  \BibitemOpen
  \bibfield  {author} {\bibinfo {author} {\bibfnamefont {C.-M.}\ \bibnamefont
  {Chang}}\ and\ \bibinfo {author} {\bibfnamefont {X.}~\bibnamefont {Yin}},\
  }\href {\doibase 10.1103/PhysRevD.88.106005} {\bibfield  {journal} {\bibinfo
  {journal} {Phys. Rev. D}\ }\textbf {\bibinfo {volume} {88}},\ \bibinfo
  {pages} {106005} (\bibinfo {year} {2013})},\ \Eprint
  {http://arxiv.org/abs/1305.6314} {arXiv:1305.6314 [hep-th]} \BibitemShut
  {NoStop}%
\bibitem [{\citenamefont {Liu}\ \emph {et~al.}(2018)\citenamefont {Liu},
  \citenamefont {Pando~Zayas}, \citenamefont {Rathee},\ and\ \citenamefont
  {Zhao}}]{Liu:2017vbl}%
  \BibitemOpen
  \bibfield  {author} {\bibinfo {author} {\bibfnamefont {J.~T.}\ \bibnamefont
  {Liu}}, \bibinfo {author} {\bibfnamefont {L.~A.}\ \bibnamefont
  {Pando~Zayas}}, \bibinfo {author} {\bibfnamefont {V.}~\bibnamefont {Rathee}},
  \ and\ \bibinfo {author} {\bibfnamefont {W.}~\bibnamefont {Zhao}},\ }\href
  {\doibase 10.1103/PhysRevLett.120.221602} {\bibfield  {journal} {\bibinfo
  {journal} {Phys. Rev. Lett.}\ }\textbf {\bibinfo {volume} {120}},\ \bibinfo
  {pages} {221602} (\bibinfo {year} {2018})},\ \Eprint
  {http://arxiv.org/abs/1711.01076} {arXiv:1711.01076 [hep-th]} \BibitemShut
  {NoStop}%
\bibitem [{\citenamefont {Drukker}\ \emph {et~al.}(2011)\citenamefont
  {Drukker}, \citenamefont {Marino},\ and\ \citenamefont
  {Putrov}}]{Drukker:2010nc}%
  \BibitemOpen
  \bibfield  {author} {\bibinfo {author} {\bibfnamefont {N.}~\bibnamefont
  {Drukker}}, \bibinfo {author} {\bibfnamefont {M.}~\bibnamefont {Marino}}, \
  and\ \bibinfo {author} {\bibfnamefont {P.}~\bibnamefont {Putrov}},\ }\href
  {\doibase 10.1007/s00220-011-1253-6} {\bibfield  {journal} {\bibinfo
  {journal} {Commun. Math. Phys.}\ }\textbf {\bibinfo {volume} {306}},\
  \bibinfo {pages} {511} (\bibinfo {year} {2011})},\ \Eprint
  {http://arxiv.org/abs/1007.3837} {arXiv:1007.3837 [hep-th]} \BibitemShut
  {NoStop}%
\bibitem [{\citenamefont {Susskind}(1993)}]{Susskind:1993ws}%
  \BibitemOpen
  \bibfield  {author} {\bibinfo {author} {\bibfnamefont {L.}~\bibnamefont
  {Susskind}},\ }\href@noop {} {\ ,\ \bibinfo {pages} {118} (\bibinfo {year}
  {1993})},\ \Eprint {http://arxiv.org/abs/hep-th/9309145}
  {arXiv:hep-th/9309145} \BibitemShut {NoStop}%
\bibitem [{\citenamefont {Banados}\ \emph {et~al.}(1994)\citenamefont
  {Banados}, \citenamefont {Teitelboim},\ and\ \citenamefont
  {Zanelli}}]{Banados:1993qp}%
  \BibitemOpen
  \bibfield  {author} {\bibinfo {author} {\bibfnamefont {M.}~\bibnamefont
  {Banados}}, \bibinfo {author} {\bibfnamefont {C.}~\bibnamefont {Teitelboim}},
  \ and\ \bibinfo {author} {\bibfnamefont {J.}~\bibnamefont {Zanelli}},\ }\href
  {\doibase 10.1103/PhysRevLett.72.957} {\bibfield  {journal} {\bibinfo
  {journal} {Phys. Rev. Lett.}\ }\textbf {\bibinfo {volume} {72}},\ \bibinfo
  {pages} {957} (\bibinfo {year} {1994})},\ \Eprint
  {http://arxiv.org/abs/gr-qc/9309026} {arXiv:gr-qc/9309026} \BibitemShut
  {NoStop}%
\bibitem [{\citenamefont {Carlip}\ and\ \citenamefont
  {Teitelboim}(1995)}]{Carlip:1993sa}%
  \BibitemOpen
  \bibfield  {author} {\bibinfo {author} {\bibfnamefont {S.}~\bibnamefont
  {Carlip}}\ and\ \bibinfo {author} {\bibfnamefont {C.}~\bibnamefont
  {Teitelboim}},\ }\href {\doibase 10.1088/0264-9381/12/7/011} {\bibfield
  {journal} {\bibinfo  {journal} {Class. Quant. Grav.}\ }\textbf {\bibinfo
  {volume} {12}},\ \bibinfo {pages} {1699} (\bibinfo {year} {1995})},\ \Eprint
  {http://arxiv.org/abs/gr-qc/9312002} {arXiv:gr-qc/9312002} \BibitemShut
  {NoStop}%
\bibitem [{\citenamefont {Teitelboim}(1994)}]{Teitelboim:1994is}%
  \BibitemOpen
  \bibfield  {author} {\bibinfo {author} {\bibfnamefont {C.}~\bibnamefont
  {Teitelboim}},\ }in\ \href@noop {} {\emph {\bibinfo {booktitle} {{Cornelius
  Lanczos International Centenary Conference (NCSU 93)}}}}\ (\bibinfo {year}
  {1994})\ \Eprint {http://arxiv.org/abs/hep-th/9405199} {arXiv:hep-th/9405199}
  \BibitemShut {NoStop}%
\bibitem [{\citenamefont {Fursaev}\ and\ \citenamefont
  {Solodukhin}(1995)}]{Fursaev:1995ef}%
  \BibitemOpen
  \bibfield  {author} {\bibinfo {author} {\bibfnamefont {D.~V.}\ \bibnamefont
  {Fursaev}}\ and\ \bibinfo {author} {\bibfnamefont {S.~N.}\ \bibnamefont
  {Solodukhin}},\ }\href {\doibase 10.1103/PhysRevD.52.2133} {\bibfield
  {journal} {\bibinfo  {journal} {Phys. Rev. D}\ }\textbf {\bibinfo {volume}
  {52}},\ \bibinfo {pages} {2133} (\bibinfo {year} {1995})},\ \Eprint
  {http://arxiv.org/abs/hep-th/9501127} {arXiv:hep-th/9501127} \BibitemShut
  {NoStop}%
\bibitem [{\citenamefont {Mann}\ and\ \citenamefont
  {Solodukhin}(1996)}]{Mann:1996bi}%
  \BibitemOpen
  \bibfield  {author} {\bibinfo {author} {\bibfnamefont {R.~B.}\ \bibnamefont
  {Mann}}\ and\ \bibinfo {author} {\bibfnamefont {S.~N.}\ \bibnamefont
  {Solodukhin}},\ }\href {\doibase 10.1103/PhysRevD.54.3932} {\bibfield
  {journal} {\bibinfo  {journal} {Phys. Rev. D}\ }\textbf {\bibinfo {volume}
  {54}},\ \bibinfo {pages} {3932} (\bibinfo {year} {1996})},\ \Eprint
  {http://arxiv.org/abs/hep-th/9604118} {arXiv:hep-th/9604118} \BibitemShut
  {NoStop}%
\bibitem [{\citenamefont {Solodukhin}(1995{\natexlab{a}})}]{Solodukhin:1994yz}%
  \BibitemOpen
  \bibfield  {author} {\bibinfo {author} {\bibfnamefont {S.~N.}\ \bibnamefont
  {Solodukhin}},\ }\href {\doibase 10.1103/PhysRevD.51.609} {\bibfield
  {journal} {\bibinfo  {journal} {Phys. Rev. D}\ }\textbf {\bibinfo {volume}
  {51}},\ \bibinfo {pages} {609} (\bibinfo {year} {1995}{\natexlab{a}})},\
  \Eprint {http://arxiv.org/abs/hep-th/9407001} {arXiv:hep-th/9407001}
  \BibitemShut {NoStop}%
\bibitem [{\citenamefont {Fursaev}(1995)}]{Fursaev:1994te}%
  \BibitemOpen
  \bibfield  {author} {\bibinfo {author} {\bibfnamefont {D.~V.}\ \bibnamefont
  {Fursaev}},\ }\href {\doibase 10.1103/PhysRevD.51.R5352} {\bibfield
  {journal} {\bibinfo  {journal} {Phys. Rev. D}\ }\textbf {\bibinfo {volume}
  {51}},\ \bibinfo {pages} {5352} (\bibinfo {year} {1995})},\ \Eprint
  {http://arxiv.org/abs/hep-th/9412161} {arXiv:hep-th/9412161} \BibitemShut
  {NoStop}%
\bibitem [{\citenamefont {Solodukhin}(1995{\natexlab{b}})}]{Solodukhin:1995ak}%
  \BibitemOpen
  \bibfield  {author} {\bibinfo {author} {\bibfnamefont {S.~N.}\ \bibnamefont
  {Solodukhin}},\ }\href {\doibase 10.1103/PhysRevD.52.7046} {\bibfield
  {journal} {\bibinfo  {journal} {Phys. Rev. D}\ }\textbf {\bibinfo {volume}
  {52}},\ \bibinfo {pages} {7046} (\bibinfo {year} {1995}{\natexlab{b}})},\
  \Eprint {http://arxiv.org/abs/hep-th/9504022} {arXiv:hep-th/9504022}
  \BibitemShut {NoStop}%
\bibitem [{\citenamefont {Frolov}\ \emph
  {et~al.}(1996{\natexlab{a}})\citenamefont {Frolov}, \citenamefont {Fursaev},\
  and\ \citenamefont {Zelnikov}}]{Frolov:1995xe}%
  \BibitemOpen
  \bibfield  {author} {\bibinfo {author} {\bibfnamefont {V.~P.}\ \bibnamefont
  {Frolov}}, \bibinfo {author} {\bibfnamefont {D.~V.}\ \bibnamefont {Fursaev}},
  \ and\ \bibinfo {author} {\bibfnamefont {A.~I.}\ \bibnamefont {Zelnikov}},\
  }\href {\doibase 10.1103/PhysRevD.54.2711} {\bibfield  {journal} {\bibinfo
  {journal} {Phys. Rev. D}\ }\textbf {\bibinfo {volume} {54}},\ \bibinfo
  {pages} {2711} (\bibinfo {year} {1996}{\natexlab{a}})},\ \Eprint
  {http://arxiv.org/abs/hep-th/9512184} {arXiv:hep-th/9512184} \BibitemShut
  {NoStop}%
\bibitem [{\citenamefont {Frolov}\ \emph
  {et~al.}(1996{\natexlab{b}})\citenamefont {Frolov}, \citenamefont {Israel},\
  and\ \citenamefont {Solodukhin}}]{Frolov:1996hd}%
  \BibitemOpen
  \bibfield  {author} {\bibinfo {author} {\bibfnamefont {V.~P.}\ \bibnamefont
  {Frolov}}, \bibinfo {author} {\bibfnamefont {W.}~\bibnamefont {Israel}}, \
  and\ \bibinfo {author} {\bibfnamefont {S.~N.}\ \bibnamefont {Solodukhin}},\
  }\href {\doibase 10.1103/PhysRevD.54.2732} {\bibfield  {journal} {\bibinfo
  {journal} {Phys. Rev. D}\ }\textbf {\bibinfo {volume} {54}},\ \bibinfo
  {pages} {2732} (\bibinfo {year} {1996}{\natexlab{b}})},\ \Eprint
  {http://arxiv.org/abs/hep-th/9602105} {arXiv:hep-th/9602105} \BibitemShut
  {NoStop}%
\bibitem [{\citenamefont {Bytsenko}\ \emph {et~al.}(1998)\citenamefont
  {Bytsenko}, \citenamefont {Vanzo},\ and\ \citenamefont
  {Zerbini}}]{Bytsenko:1997ru}%
  \BibitemOpen
  \bibfield  {author} {\bibinfo {author} {\bibfnamefont {A.~A.}\ \bibnamefont
  {Bytsenko}}, \bibinfo {author} {\bibfnamefont {L.}~\bibnamefont {Vanzo}}, \
  and\ \bibinfo {author} {\bibfnamefont {S.}~\bibnamefont {Zerbini}},\ }\href
  {\doibase 10.1103/PhysRevD.57.4917} {\bibfield  {journal} {\bibinfo
  {journal} {Phys. Rev. D}\ }\textbf {\bibinfo {volume} {57}},\ \bibinfo
  {pages} {4917} (\bibinfo {year} {1998})},\ \Eprint
  {http://arxiv.org/abs/gr-qc/9710106} {arXiv:gr-qc/9710106} \BibitemShut
  {NoStop}%
\bibitem [{\citenamefont {Solodukhin}(2011)}]{Solodukhin:2011gn}%
  \BibitemOpen
  \bibfield  {author} {\bibinfo {author} {\bibfnamefont {S.~N.}\ \bibnamefont
  {Solodukhin}},\ }\href {\doibase 10.12942/lrr-2011-8} {\bibfield  {journal}
  {\bibinfo  {journal} {Living Rev. Rel.}\ }\textbf {\bibinfo {volume} {14}},\
  \bibinfo {pages} {8} (\bibinfo {year} {2011})},\ \Eprint
  {http://arxiv.org/abs/1104.3712} {arXiv:1104.3712 [hep-th]} \BibitemShut
  {NoStop}%
\bibitem [{\citenamefont {Chaikin}\ and\ \citenamefont
  {Lubensky}(1995)}]{cha95}%
  \BibitemOpen
  \bibfield  {author} {\bibinfo {author} {\bibfnamefont {P.~M.}\ \bibnamefont
  {Chaikin}}\ and\ \bibinfo {author} {\bibfnamefont {T.~C.}\ \bibnamefont
  {Lubensky}},\ }\href {\doibase https://doi.org/10.1017/CBO9780511813467}
  {\emph {\bibinfo {title} {Principles of Condensed Matter Physics}}}\
  (\bibinfo  {publisher} {Cambridge University Press},\ \bibinfo {year}
  {1995})\BibitemShut {NoStop}%
\bibitem [{\citenamefont {Kubo}(1965)}]{kubo}%
  \BibitemOpen
  \bibfield  {author} {\bibinfo {author} {\bibfnamefont {R.}~\bibnamefont
  {Kubo}},\ }\href {https://books.google.co.in/books?id=IphRAAAAMAAJ} {\emph
  {\bibinfo {title} {Statistical Mechanics: An Advanced Course with Problems
  and Solutions}}},\ Ry{\=o}go Kubo\ (\bibinfo  {publisher} {North-Holland
  Publishing Company},\ \bibinfo {year} {1965})\BibitemShut {NoStop}%
\bibitem [{\citenamefont {Bragg}\ and\ \citenamefont {Williams}(1934)}]{bw1}%
  \BibitemOpen
  \bibfield  {author} {\bibinfo {author} {\bibfnamefont {W.~L.}\ \bibnamefont
  {Bragg}}\ and\ \bibinfo {author} {\bibfnamefont {E.~J.}\ \bibnamefont
  {Williams}},\ }\href {\doibase 10.1098/rspa.1934.0132} {\bibfield  {journal}
  {\bibinfo  {journal} {Proceedings of the Royal Society of London. Series A,
  Containing Papers of a Mathematical and Physical Character}\ }\textbf
  {\bibinfo {volume} {145}},\ \bibinfo {pages} {699} (\bibinfo {year}
  {1934})}\BibitemShut {NoStop}%
\bibitem [{\citenamefont {Bragg}\ and\ \citenamefont {Williams}(1935)}]{bw2}%
  \BibitemOpen
  \bibfield  {author} {\bibinfo {author} {\bibfnamefont {W.~L.}\ \bibnamefont
  {Bragg}}\ and\ \bibinfo {author} {\bibfnamefont {E.~J.}\ \bibnamefont
  {Williams}},\ }\href {\doibase 10.1098/rspa.1935.0165} {\bibfield  {journal}
  {\bibinfo  {journal} {Proceedings of the Royal Society of London. Series A -
  Mathematical and Physical Sciences}\ }\textbf {\bibinfo {volume} {151}},\
  \bibinfo {pages} {540} (\bibinfo {year} {1935})}\BibitemShut {NoStop}%
\bibitem [{\citenamefont {Dey}\ \emph {et~al.}(2007{\natexlab{a}})\citenamefont
  {Dey}, \citenamefont {Mukherji}, \citenamefont {Mukhopadhyay},\ and\
  \citenamefont {Sarkar}}]{Dey:2007vt}%
  \BibitemOpen
  \bibfield  {author} {\bibinfo {author} {\bibfnamefont {T.~K.}\ \bibnamefont
  {Dey}}, \bibinfo {author} {\bibfnamefont {S.}~\bibnamefont {Mukherji}},
  \bibinfo {author} {\bibfnamefont {S.}~\bibnamefont {Mukhopadhyay}}, \ and\
  \bibinfo {author} {\bibfnamefont {S.}~\bibnamefont {Sarkar}},\ }\href
  {\doibase 10.1088/1126-6708/2007/09/026} {\bibfield  {journal} {\bibinfo
  {journal} {JHEP}\ }\textbf {\bibinfo {volume} {09}},\ \bibinfo {pages} {026}
  (\bibinfo {year} {2007}{\natexlab{a}})},\ \Eprint
  {http://arxiv.org/abs/0706.3996} {arXiv:0706.3996 [hep-th]} \BibitemShut
  {NoStop}%
\bibitem [{\citenamefont {Dey}\ \emph {et~al.}(2007{\natexlab{b}})\citenamefont
  {Dey}, \citenamefont {Mukherji}, \citenamefont {Mukhopadhyay},\ and\
  \citenamefont {Sarkar}}]{Dey:2006ds}%
  \BibitemOpen
  \bibfield  {author} {\bibinfo {author} {\bibfnamefont {T.~K.}\ \bibnamefont
  {Dey}}, \bibinfo {author} {\bibfnamefont {S.}~\bibnamefont {Mukherji}},
  \bibinfo {author} {\bibfnamefont {S.}~\bibnamefont {Mukhopadhyay}}, \ and\
  \bibinfo {author} {\bibfnamefont {S.}~\bibnamefont {Sarkar}},\ }\href
  {\doibase 10.1088/1126-6708/2007/04/014} {\bibfield  {journal} {\bibinfo
  {journal} {JHEP}\ }\textbf {\bibinfo {volume} {04}},\ \bibinfo {pages} {014}
  (\bibinfo {year} {2007}{\natexlab{b}})},\ \Eprint
  {http://arxiv.org/abs/hep-th/0609038} {arXiv:hep-th/0609038} \BibitemShut
  {NoStop}%
\bibitem [{\citenamefont {Nayak}(2008)}]{nayak2008bragg}%
  \BibitemOpen
  \bibfield  {author} {\bibinfo {author} {\bibfnamefont {B.~P.}\ \bibnamefont
  {Nayak}},\ }\href {https://api.semanticscholar.org/CorpusID:201758770}
  {\bibfield  {journal} {\bibinfo  {journal} {Prayas Students’ Journal of
  Physics}\ } (\bibinfo {year} {2008})}\BibitemShut {NoStop}%
\bibitem [{\citenamefont {Banerjee}\ \emph {et~al.}(2011)\citenamefont
  {Banerjee}, \citenamefont {Chakrabarti}, \citenamefont {Mukherji},\ and\
  \citenamefont {Panda}}]{Banerjee:2010ve}%
  \BibitemOpen
  \bibfield  {author} {\bibinfo {author} {\bibfnamefont {S.}~\bibnamefont
  {Banerjee}}, \bibinfo {author} {\bibfnamefont {S.~K.}\ \bibnamefont
  {Chakrabarti}}, \bibinfo {author} {\bibfnamefont {S.}~\bibnamefont
  {Mukherji}}, \ and\ \bibinfo {author} {\bibfnamefont {B.}~\bibnamefont
  {Panda}},\ }\href {\doibase 10.1142/S0217751X11053845} {\bibfield  {journal}
  {\bibinfo  {journal} {Int. J. Mod. Phys. A}\ }\textbf {\bibinfo {volume}
  {26}},\ \bibinfo {pages} {3469} (\bibinfo {year} {2011})},\ \Eprint
  {http://arxiv.org/abs/1012.3256} {arXiv:1012.3256 [hep-th]} \BibitemShut
  {NoStop}%
\bibitem [{\citenamefont {Banerjee}(2010)}]{Banerjee:2010ng}%
  \BibitemOpen
  \bibfield  {author} {\bibinfo {author} {\bibfnamefont {S.}~\bibnamefont
  {Banerjee}},\ }\href {\doibase 10.1103/PhysRevD.82.106008} {\bibfield
  {journal} {\bibinfo  {journal} {Phys. Rev. D}\ }\textbf {\bibinfo {volume}
  {82}},\ \bibinfo {pages} {106008} (\bibinfo {year} {2010})},\ \Eprint
  {http://arxiv.org/abs/1009.1780} {arXiv:1009.1780 [hep-th]} \BibitemShut
  {NoStop}%
\bibitem [{\citenamefont {Yerra}\ \emph {et~al.}(2022)\citenamefont {Yerra},
  \citenamefont {Bhamidipati},\ and\ \citenamefont {Mukherji}}]{Yerra:2022coh}%
  \BibitemOpen
  \bibfield  {author} {\bibinfo {author} {\bibfnamefont {P.~K.}\ \bibnamefont
  {Yerra}}, \bibinfo {author} {\bibfnamefont {C.}~\bibnamefont {Bhamidipati}},
  \ and\ \bibinfo {author} {\bibfnamefont {S.}~\bibnamefont {Mukherji}},\
  }\href {\doibase 10.1103/PhysRevD.106.064059} {\bibfield  {journal} {\bibinfo
   {journal} {Phys. Rev. D}\ }\textbf {\bibinfo {volume} {106}},\ \bibinfo
  {pages} {064059} (\bibinfo {year} {2022})},\ \Eprint
  {http://arxiv.org/abs/2208.06388} {arXiv:2208.06388 [hep-th]} \BibitemShut
  {NoStop}%
\bibitem [{\citenamefont {Wei}\ \emph {et~al.}(2022)\citenamefont {Wei},
  \citenamefont {Liu},\ and\ \citenamefont {Mann}}]{Wei:2022dzw}%
  \BibitemOpen
  \bibfield  {author} {\bibinfo {author} {\bibfnamefont {S.-W.}\ \bibnamefont
  {Wei}}, \bibinfo {author} {\bibfnamefont {Y.-X.}\ \bibnamefont {Liu}}, \ and\
  \bibinfo {author} {\bibfnamefont {R.~B.}\ \bibnamefont {Mann}},\ }\href
  {\doibase 10.1103/PhysRevLett.129.191101} {\bibfield  {journal} {\bibinfo
  {journal} {Phys. Rev. Lett.}\ }\textbf {\bibinfo {volume} {129}},\ \bibinfo
  {pages} {191101} (\bibinfo {year} {2022})},\ \Eprint
  {http://arxiv.org/abs/2208.01932} {arXiv:2208.01932 [gr-qc]} \BibitemShut
  {NoStop}%
\bibitem [{\citenamefont {Cassani}\ \emph {et~al.}(2022)\citenamefont
  {Cassani}, \citenamefont {Ruip\'erez},\ and\ \citenamefont
  {Turetta}}]{Cassani_2022}%
  \BibitemOpen
  \bibfield  {author} {\bibinfo {author} {\bibfnamefont {D.}~\bibnamefont
  {Cassani}}, \bibinfo {author} {\bibfnamefont {A.}~\bibnamefont {Ruip\'erez}},
  \ and\ \bibinfo {author} {\bibfnamefont {E.}~\bibnamefont {Turetta}},\ }\href
  {\doibase 10.1007/JHEP11(2022)059} {\bibfield  {journal} {\bibinfo  {journal}
  {JHEP}\ }\textbf {\bibinfo {volume} {11}},\ \bibinfo {pages} {059} (\bibinfo
  {year} {2022})},\ \Eprint {http://arxiv.org/abs/2208.01007} {arXiv:2208.01007
  [hep-th]} \BibitemShut {NoStop}%
\bibitem [{\citenamefont {Bobev}\ \emph
  {et~al.}(2022{\natexlab{a}})\citenamefont {Bobev}, \citenamefont {Dimitrov},
  \citenamefont {Reys},\ and\ \citenamefont {Vekemans}}]{Bobev_2022}%
  \BibitemOpen
  \bibfield  {author} {\bibinfo {author} {\bibfnamefont {N.}~\bibnamefont
  {Bobev}}, \bibinfo {author} {\bibfnamefont {V.}~\bibnamefont {Dimitrov}},
  \bibinfo {author} {\bibfnamefont {V.}~\bibnamefont {Reys}}, \ and\ \bibinfo
  {author} {\bibfnamefont {A.}~\bibnamefont {Vekemans}},\ }\href {\doibase
  10.1103/PhysRevD.106.L121903} {\bibfield  {journal} {\bibinfo  {journal}
  {Phys. Rev. D}\ }\textbf {\bibinfo {volume} {106}},\ \bibinfo {pages}
  {L121903} (\bibinfo {year} {2022}{\natexlab{a}})},\ \Eprint
  {http://arxiv.org/abs/2207.10671} {arXiv:2207.10671 [hep-th]} \BibitemShut
  {NoStop}%
\bibitem [{\citenamefont {Adams}\ \emph {et~al.}(2006)\citenamefont {Adams},
  \citenamefont {Arkani-Hamed}, \citenamefont {Dubovsky}, \citenamefont
  {Nicolis},\ and\ \citenamefont {Rattazzi}}]{Adams:2006sv}%
  \BibitemOpen
  \bibfield  {author} {\bibinfo {author} {\bibfnamefont {A.}~\bibnamefont
  {Adams}}, \bibinfo {author} {\bibfnamefont {N.}~\bibnamefont {Arkani-Hamed}},
  \bibinfo {author} {\bibfnamefont {S.}~\bibnamefont {Dubovsky}}, \bibinfo
  {author} {\bibfnamefont {A.}~\bibnamefont {Nicolis}}, \ and\ \bibinfo
  {author} {\bibfnamefont {R.}~\bibnamefont {Rattazzi}},\ }\href {\doibase
  10.1088/1126-6708/2006/10/014} {\bibfield  {journal} {\bibinfo  {journal}
  {JHEP}\ }\textbf {\bibinfo {volume} {10}},\ \bibinfo {pages} {014} (\bibinfo
  {year} {2006})},\ \Eprint {http://arxiv.org/abs/hep-th/0602178}
  {arXiv:hep-th/0602178} \BibitemShut {NoStop}%
\bibitem [{\citenamefont {Arkani-Hamed}\ \emph {et~al.}(2007)\citenamefont
  {Arkani-Hamed}, \citenamefont {Motl}, \citenamefont {Nicolis},\ and\
  \citenamefont {Vafa}}]{Arkani-Hamed:2006emk}%
  \BibitemOpen
  \bibfield  {author} {\bibinfo {author} {\bibfnamefont {N.}~\bibnamefont
  {Arkani-Hamed}}, \bibinfo {author} {\bibfnamefont {L.}~\bibnamefont {Motl}},
  \bibinfo {author} {\bibfnamefont {A.}~\bibnamefont {Nicolis}}, \ and\
  \bibinfo {author} {\bibfnamefont {C.}~\bibnamefont {Vafa}},\ }\href {\doibase
  10.1088/1126-6708/2007/06/060} {\bibfield  {journal} {\bibinfo  {journal}
  {JHEP}\ }\textbf {\bibinfo {volume} {06}},\ \bibinfo {pages} {060} (\bibinfo
  {year} {2007})},\ \Eprint {http://arxiv.org/abs/hep-th/0601001}
  {arXiv:hep-th/0601001} \BibitemShut {NoStop}%
\bibitem [{\citenamefont {Baggio}\ \emph {et~al.}(2014)\citenamefont {Baggio},
  \citenamefont {Halmagyi}, \citenamefont {Mayerson}, \citenamefont {Robbins},\
  and\ \citenamefont {Wecht}}]{Baggio:2014hua}%
  \BibitemOpen
  \bibfield  {author} {\bibinfo {author} {\bibfnamefont {M.}~\bibnamefont
  {Baggio}}, \bibinfo {author} {\bibfnamefont {N.}~\bibnamefont {Halmagyi}},
  \bibinfo {author} {\bibfnamefont {D.~R.}\ \bibnamefont {Mayerson}}, \bibinfo
  {author} {\bibfnamefont {D.}~\bibnamefont {Robbins}}, \ and\ \bibinfo
  {author} {\bibfnamefont {B.}~\bibnamefont {Wecht}},\ }\href {\doibase
  10.1007/JHEP12(2014)042} {\bibfield  {journal} {\bibinfo  {journal} {JHEP}\
  }\textbf {\bibinfo {volume} {12}},\ \bibinfo {pages} {042} (\bibinfo {year}
  {2014})},\ \Eprint {http://arxiv.org/abs/1408.2538} {arXiv:1408.2538
  [hep-th]} \BibitemShut {NoStop}%
\bibitem [{\citenamefont {Bobev}\ \emph
  {et~al.}(2022{\natexlab{b}})\citenamefont {Bobev}, \citenamefont {Hristov},\
  and\ \citenamefont {Reys}}]{Bobev:2021qxx}%
  \BibitemOpen
  \bibfield  {author} {\bibinfo {author} {\bibfnamefont {N.}~\bibnamefont
  {Bobev}}, \bibinfo {author} {\bibfnamefont {K.}~\bibnamefont {Hristov}}, \
  and\ \bibinfo {author} {\bibfnamefont {V.}~\bibnamefont {Reys}},\ }\href
  {\doibase 10.1007/JHEP04(2022)088} {\bibfield  {journal} {\bibinfo  {journal}
  {JHEP}\ }\textbf {\bibinfo {volume} {04}},\ \bibinfo {pages} {088} (\bibinfo
  {year} {2022}{\natexlab{b}})},\ \Eprint {http://arxiv.org/abs/2112.06961}
  {arXiv:2112.06961 [hep-th]} \BibitemShut {NoStop}%
\bibitem [{\citenamefont {Liu}\ and\ \citenamefont
  {Saskowski}(2022)}]{Liu:2022sew}%
  \BibitemOpen
  \bibfield  {author} {\bibinfo {author} {\bibfnamefont {J.~T.}\ \bibnamefont
  {Liu}}\ and\ \bibinfo {author} {\bibfnamefont {R.~J.}\ \bibnamefont
  {Saskowski}},\ }\href {\doibase 10.1007/JHEP05(2022)171} {\bibfield
  {journal} {\bibinfo  {journal} {JHEP}\ }\textbf {\bibinfo {volume} {05}},\
  \bibinfo {pages} {171} (\bibinfo {year} {2022})},\ \Eprint
  {http://arxiv.org/abs/2201.04690} {arXiv:2201.04690 [hep-th]} \BibitemShut
  {NoStop}%
\bibitem [{\citenamefont {Melo}\ and\ \citenamefont
  {Santos}(2021)}]{Melo:2020amq}%
  \BibitemOpen
  \bibfield  {author} {\bibinfo {author} {\bibfnamefont {J.~a.~F.}\
  \bibnamefont {Melo}}\ and\ \bibinfo {author} {\bibfnamefont {J.~E.}\
  \bibnamefont {Santos}},\ }\href {\doibase 10.1103/PhysRevD.103.066008}
  {\bibfield  {journal} {\bibinfo  {journal} {Phys. Rev. D}\ }\textbf {\bibinfo
  {volume} {103}},\ \bibinfo {pages} {066008} (\bibinfo {year} {2021})},\
  \Eprint {http://arxiv.org/abs/2007.06582} {arXiv:2007.06582 [hep-th]}
  \BibitemShut {NoStop}%
\bibitem [{\citenamefont {Bobev}\ \emph {et~al.}(2020)\citenamefont {Bobev},
  \citenamefont {Charles}, \citenamefont {Hristov},\ and\ \citenamefont
  {Reys}}]{Bobev:2020egg}%
  \BibitemOpen
  \bibfield  {author} {\bibinfo {author} {\bibfnamefont {N.}~\bibnamefont
  {Bobev}}, \bibinfo {author} {\bibfnamefont {A.~M.}\ \bibnamefont {Charles}},
  \bibinfo {author} {\bibfnamefont {K.}~\bibnamefont {Hristov}}, \ and\
  \bibinfo {author} {\bibfnamefont {V.}~\bibnamefont {Reys}},\ }\href {\doibase
  10.1103/PhysRevLett.125.131601} {\bibfield  {journal} {\bibinfo  {journal}
  {Phys. Rev. Lett.}\ }\textbf {\bibinfo {volume} {125}},\ \bibinfo {pages}
  {131601} (\bibinfo {year} {2020})},\ \Eprint
  {http://arxiv.org/abs/2006.09390} {arXiv:2006.09390 [hep-th]} \BibitemShut
  {NoStop}%
\bibitem [{\citenamefont {Bobev}\ \emph {et~al.}(2021)\citenamefont {Bobev},
  \citenamefont {Charles}, \citenamefont {Hristov},\ and\ \citenamefont
  {Reys}}]{Bobev:2021oku}%
  \BibitemOpen
  \bibfield  {author} {\bibinfo {author} {\bibfnamefont {N.}~\bibnamefont
  {Bobev}}, \bibinfo {author} {\bibfnamefont {A.~M.}\ \bibnamefont {Charles}},
  \bibinfo {author} {\bibfnamefont {K.}~\bibnamefont {Hristov}}, \ and\
  \bibinfo {author} {\bibfnamefont {V.}~\bibnamefont {Reys}},\ }\href {\doibase
  10.1007/JHEP08(2021)173} {\bibfield  {journal} {\bibinfo  {journal} {JHEP}\
  }\textbf {\bibinfo {volume} {08}},\ \bibinfo {pages} {173} (\bibinfo {year}
  {2021})},\ \Eprint {http://arxiv.org/abs/2106.04581} {arXiv:2106.04581
  [hep-th]} \BibitemShut {NoStop}%
\bibitem [{\citenamefont {Genolini}\ and\ \citenamefont
  {Richmond}(2021)}]{Genolini:2021urf}%
  \BibitemOpen
  \bibfield  {author} {\bibinfo {author} {\bibfnamefont {P.~B.}\ \bibnamefont
  {Genolini}}\ and\ \bibinfo {author} {\bibfnamefont {P.}~\bibnamefont
  {Richmond}},\ }\href {\doibase 10.1103/PhysRevD.104.L061902} {\bibfield
  {journal} {\bibinfo  {journal} {Phys. Rev. D}\ }\textbf {\bibinfo {volume}
  {104}},\ \bibinfo {pages} {L061902} (\bibinfo {year} {2021})},\ \Eprint
  {http://arxiv.org/abs/2107.04590} {arXiv:2107.04590 [hep-th]} \BibitemShut
  {NoStop}%
\bibitem [{\citenamefont {Arabi~Ardehali}\ and\ \citenamefont
  {Murthy}(2021)}]{ArabiArdehali:2021nsx}%
  \BibitemOpen
  \bibfield  {author} {\bibinfo {author} {\bibfnamefont {A.}~\bibnamefont
  {Arabi~Ardehali}}\ and\ \bibinfo {author} {\bibfnamefont {S.}~\bibnamefont
  {Murthy}},\ }\href {\doibase 10.1007/JHEP10(2021)207} {\bibfield  {journal}
  {\bibinfo  {journal} {JHEP}\ }\textbf {\bibinfo {volume} {10}},\ \bibinfo
  {pages} {207} (\bibinfo {year} {2021})},\ \Eprint
  {http://arxiv.org/abs/2104.02051} {arXiv:2104.02051 [hep-th]} \BibitemShut
  {NoStop}%
\bibitem [{\citenamefont {Gonz\'alez~Lezcano}\ \emph
  {et~al.}(2021)\citenamefont {Gonz\'alez~Lezcano}, \citenamefont {Hong},
  \citenamefont {Liu},\ and\ \citenamefont
  {Pando~Zayas}}]{GonzalezLezcano:2020yeb}%
  \BibitemOpen
  \bibfield  {author} {\bibinfo {author} {\bibfnamefont {A.}~\bibnamefont
  {Gonz\'alez~Lezcano}}, \bibinfo {author} {\bibfnamefont {J.}~\bibnamefont
  {Hong}}, \bibinfo {author} {\bibfnamefont {J.~T.}\ \bibnamefont {Liu}}, \
  and\ \bibinfo {author} {\bibfnamefont {L.~A.}\ \bibnamefont {Pando~Zayas}},\
  }\href {\doibase 10.1007/JHEP01(2021)001} {\bibfield  {journal} {\bibinfo
  {journal} {JHEP}\ }\textbf {\bibinfo {volume} {01}},\ \bibinfo {pages} {001}
  (\bibinfo {year} {2021})},\ \Eprint {http://arxiv.org/abs/2007.12604}
  {arXiv:2007.12604 [hep-th]} \BibitemShut {NoStop}%
\bibitem [{\citenamefont {Amariti}\ \emph
  {et~al.}(2021{\natexlab{b}})\citenamefont {Amariti}, \citenamefont {Fazzi},\
  and\ \citenamefont {Segati}}]{Amariti:2021ubd}%
  \BibitemOpen
  \bibfield  {author} {\bibinfo {author} {\bibfnamefont {A.}~\bibnamefont
  {Amariti}}, \bibinfo {author} {\bibfnamefont {M.}~\bibnamefont {Fazzi}}, \
  and\ \bibinfo {author} {\bibfnamefont {A.}~\bibnamefont {Segati}},\ }\href
  {\doibase 10.1007/JHEP07(2021)141} {\bibfield  {journal} {\bibinfo  {journal}
  {JHEP}\ }\textbf {\bibinfo {volume} {07}},\ \bibinfo {pages} {141} (\bibinfo
  {year} {2021}{\natexlab{b}})},\ \Eprint {http://arxiv.org/abs/2103.15853}
  {arXiv:2103.15853 [hep-th]} \BibitemShut {NoStop}%
\bibitem [{\citenamefont {Ohmori}\ and\ \citenamefont
  {Tizzano}(2022)}]{Ohmori_2022}%
  \BibitemOpen
  \bibfield  {author} {\bibinfo {author} {\bibfnamefont {K.}~\bibnamefont
  {Ohmori}}\ and\ \bibinfo {author} {\bibfnamefont {L.}~\bibnamefont
  {Tizzano}},\ }\href {\doibase 10.1007/JHEP12(2022)027} {\bibfield  {journal}
  {\bibinfo  {journal} {JHEP}\ }\textbf {\bibinfo {volume} {12}},\ \bibinfo
  {pages} {027} (\bibinfo {year} {2022})},\ \Eprint
  {http://arxiv.org/abs/2112.13445} {arXiv:2112.13445 [hep-th]} \BibitemShut
  {NoStop}%
\bibitem [{\citenamefont {Chong}\ \emph
  {et~al.}(2005{\natexlab{a}})\citenamefont {Chong}, \citenamefont {Cvetic},
  \citenamefont {Lu},\ and\ \citenamefont {Pope}}]{Chong:2005hr}%
  \BibitemOpen
  \bibfield  {author} {\bibinfo {author} {\bibfnamefont {Z.~W.}\ \bibnamefont
  {Chong}}, \bibinfo {author} {\bibfnamefont {M.}~\bibnamefont {Cvetic}},
  \bibinfo {author} {\bibfnamefont {H.}~\bibnamefont {Lu}}, \ and\ \bibinfo
  {author} {\bibfnamefont {C.~N.}\ \bibnamefont {Pope}},\ }\href {\doibase
  10.1103/PhysRevLett.95.161301} {\bibfield  {journal} {\bibinfo  {journal}
  {Phys. Rev. Lett.}\ }\textbf {\bibinfo {volume} {95}},\ \bibinfo {pages}
  {161301} (\bibinfo {year} {2005}{\natexlab{a}})},\ \Eprint
  {http://arxiv.org/abs/hep-th/0506029} {arXiv:hep-th/0506029} \BibitemShut
  {NoStop}%
\bibitem [{\citenamefont {Reall}\ and\ \citenamefont
  {Santos}(2019)}]{Reall:2019sah}%
  \BibitemOpen
  \bibfield  {author} {\bibinfo {author} {\bibfnamefont {H.~S.}\ \bibnamefont
  {Reall}}\ and\ \bibinfo {author} {\bibfnamefont {J.~E.}\ \bibnamefont
  {Santos}},\ }\href {\doibase 10.1007/JHEP04(2019)021} {\bibfield  {journal}
  {\bibinfo  {journal} {JHEP}\ }\textbf {\bibinfo {volume} {04}},\ \bibinfo
  {pages} {021} (\bibinfo {year} {2019})},\ \Eprint
  {http://arxiv.org/abs/1901.11535} {arXiv:1901.11535 [hep-th]} \BibitemShut
  {NoStop}%
\bibitem [{\citenamefont {Gauntlett}\ and\ \citenamefont
  {Varela}(2007)}]{Gauntlett:2007ma}%
  \BibitemOpen
  \bibfield  {author} {\bibinfo {author} {\bibfnamefont {J.~P.}\ \bibnamefont
  {Gauntlett}}\ and\ \bibinfo {author} {\bibfnamefont {O.}~\bibnamefont
  {Varela}},\ }\href {\doibase 10.1103/PhysRevD.76.126007} {\bibfield
  {journal} {\bibinfo  {journal} {Phys. Rev. D}\ }\textbf {\bibinfo {volume}
  {76}},\ \bibinfo {pages} {126007} (\bibinfo {year} {2007})},\ \Eprint
  {http://arxiv.org/abs/0707.2315} {arXiv:0707.2315 [hep-th]} \BibitemShut
  {NoStop}%
\bibitem [{\citenamefont {Cassani}\ \emph {et~al.}(2019)\citenamefont
  {Cassani}, \citenamefont {Josse}, \citenamefont {Petrini},\ and\
  \citenamefont {Waldram}}]{Cassani:2019vcl}%
  \BibitemOpen
  \bibfield  {author} {\bibinfo {author} {\bibfnamefont {D.}~\bibnamefont
  {Cassani}}, \bibinfo {author} {\bibfnamefont {G.}~\bibnamefont {Josse}},
  \bibinfo {author} {\bibfnamefont {M.}~\bibnamefont {Petrini}}, \ and\
  \bibinfo {author} {\bibfnamefont {D.}~\bibnamefont {Waldram}},\ }\href
  {\doibase 10.1007/JHEP11(2019)017} {\bibfield  {journal} {\bibinfo  {journal}
  {JHEP}\ }\textbf {\bibinfo {volume} {11}},\ \bibinfo {pages} {017} (\bibinfo
  {year} {2019})},\ \Eprint {http://arxiv.org/abs/1907.06730} {arXiv:1907.06730
  [hep-th]} \BibitemShut {NoStop}%
\bibitem [{\citenamefont {Gibbons}\ and\ \citenamefont
  {Hawking}(1977)}]{PhysRevD.15.2752}%
  \BibitemOpen
  \bibfield  {author} {\bibinfo {author} {\bibfnamefont {G.~W.}\ \bibnamefont
  {Gibbons}}\ and\ \bibinfo {author} {\bibfnamefont {S.~W.}\ \bibnamefont
  {Hawking}},\ }\href {\doibase 10.1103/PhysRevD.15.2752} {\bibfield  {journal}
  {\bibinfo  {journal} {Phys. Rev. D}\ }\textbf {\bibinfo {volume} {15}},\
  \bibinfo {pages} {2752} (\bibinfo {year} {1977})}\BibitemShut {NoStop}%
\bibitem [{\citenamefont {Chen}\ \emph {et~al.}(2006)\citenamefont {Chen},
  \citenamefont {Lu},\ and\ \citenamefont {Pope}}]{chen2006mass}%
  \BibitemOpen
  \bibfield  {author} {\bibinfo {author} {\bibfnamefont {W.}~\bibnamefont
  {Chen}}, \bibinfo {author} {\bibfnamefont {H.}~\bibnamefont {Lu}}, \ and\
  \bibinfo {author} {\bibfnamefont {C.~N.}\ \bibnamefont {Pope}},\ }\href
  {\doibase 10.1103/PhysRevD.73.104036} {\bibfield  {journal} {\bibinfo
  {journal} {Phys. Rev. D}\ }\textbf {\bibinfo {volume} {73}},\ \bibinfo
  {pages} {104036} (\bibinfo {year} {2006})},\ \Eprint
  {http://arxiv.org/abs/hep-th/0510081} {arXiv:hep-th/0510081} \BibitemShut
  {NoStop}%
\bibitem [{\citenamefont {de~Haro}\ \emph {et~al.}(2001)\citenamefont
  {de~Haro}, \citenamefont {Solodukhin},\ and\ \citenamefont
  {Skenderis}}]{de_Haro_2001}%
  \BibitemOpen
  \bibfield  {author} {\bibinfo {author} {\bibfnamefont {S.}~\bibnamefont
  {de~Haro}}, \bibinfo {author} {\bibfnamefont {S.~N.}\ \bibnamefont
  {Solodukhin}}, \ and\ \bibinfo {author} {\bibfnamefont {K.}~\bibnamefont
  {Skenderis}},\ }\href {\doibase 10.1007/s002200100381} {\bibfield  {journal}
  {\bibinfo  {journal} {Commun. Math. Phys.}\ }\textbf {\bibinfo {volume}
  {217}},\ \bibinfo {pages} {595} (\bibinfo {year} {2001})},\ \Eprint
  {http://arxiv.org/abs/hep-th/0002230} {arXiv:hep-th/0002230} \BibitemShut
  {NoStop}%
\bibitem [{\citenamefont {Bianchi}\ \emph {et~al.}(2002)\citenamefont
  {Bianchi}, \citenamefont {Freedman},\ and\ \citenamefont
  {Skenderis}}]{Bianchi_2002}%
  \BibitemOpen
  \bibfield  {author} {\bibinfo {author} {\bibfnamefont {M.}~\bibnamefont
  {Bianchi}}, \bibinfo {author} {\bibfnamefont {D.~Z.}\ \bibnamefont
  {Freedman}}, \ and\ \bibinfo {author} {\bibfnamefont {K.}~\bibnamefont
  {Skenderis}},\ }\href {\doibase 10.1016/S0550-3213(02)00179-7} {\bibfield
  {journal} {\bibinfo  {journal} {Nucl. Phys. B}\ }\textbf {\bibinfo {volume}
  {631}},\ \bibinfo {pages} {159} (\bibinfo {year} {2002})},\ \Eprint
  {http://arxiv.org/abs/hep-th/0112119} {arXiv:hep-th/0112119} \BibitemShut
  {NoStop}%
\bibitem [{\citenamefont {Cassani}\ and\ \citenamefont
  {Papini}(2019)}]{Cassani_2019}%
  \BibitemOpen
  \bibfield  {author} {\bibinfo {author} {\bibfnamefont {D.}~\bibnamefont
  {Cassani}}\ and\ \bibinfo {author} {\bibfnamefont {L.}~\bibnamefont
  {Papini}},\ }\href {\doibase 10.1007/JHEP09(2019)079} {\bibfield  {journal}
  {\bibinfo  {journal} {JHEP}\ }\textbf {\bibinfo {volume} {09}},\ \bibinfo
  {pages} {079} (\bibinfo {year} {2019})},\ \Eprint
  {http://arxiv.org/abs/1906.10148} {arXiv:1906.10148 [hep-th]} \BibitemShut
  {NoStop}%
\bibitem [{\citenamefont {Yerra}\ \emph {et~al.}(2024)\citenamefont {Yerra},
  \citenamefont {Bhamidipati},\ and\ \citenamefont {Mukherji}}]{Yerra_2024}%
  \BibitemOpen
  \bibfield  {author} {\bibinfo {author} {\bibfnamefont {P.~K.}\ \bibnamefont
  {Yerra}}, \bibinfo {author} {\bibfnamefont {C.}~\bibnamefont {Bhamidipati}},
  \ and\ \bibinfo {author} {\bibfnamefont {S.}~\bibnamefont {Mukherji}},\
  }\href {\doibase 10.1007/JHEP03(2024)138} {\bibfield  {journal} {\bibinfo
  {journal} {JHEP}\ }\textbf {\bibinfo {volume} {03}},\ \bibinfo {pages} {138}
  (\bibinfo {year} {2024})},\ \Eprint {http://arxiv.org/abs/2304.14988}
  {arXiv:2304.14988 [hep-th]} \BibitemShut {NoStop}%
\bibitem [{\citenamefont {Dijkgraaf}\ \emph {et~al.}(2000)\citenamefont
  {Dijkgraaf}, \citenamefont {Maldacena}, \citenamefont {Moore},\ and\
  \citenamefont {Verlinde}}]{Dijkgraaf:2000fq}%
  \BibitemOpen
  \bibfield  {author} {\bibinfo {author} {\bibfnamefont {R.}~\bibnamefont
  {Dijkgraaf}}, \bibinfo {author} {\bibfnamefont {J.~M.}\ \bibnamefont
  {Maldacena}}, \bibinfo {author} {\bibfnamefont {G.~W.}\ \bibnamefont
  {Moore}}, \ and\ \bibinfo {author} {\bibfnamefont {E.~P.}\ \bibnamefont
  {Verlinde}},\ }\href@noop {} {\  (\bibinfo {year} {2000})},\ \Eprint
  {http://arxiv.org/abs/hep-th/0005003} {arXiv:hep-th/0005003} \BibitemShut
  {NoStop}%
\bibitem [{\citenamefont {Witten}(1998{\natexlab{b}})}]{Witten:1998qj}%
  \BibitemOpen
  \bibfield  {author} {\bibinfo {author} {\bibfnamefont {E.}~\bibnamefont
  {Witten}},\ }\href {\doibase 10.4310/ATMP.1998.v2.n2.a2} {\bibfield
  {journal} {\bibinfo  {journal} {Adv. Theor. Math. Phys.}\ }\textbf {\bibinfo
  {volume} {2}},\ \bibinfo {pages} {253} (\bibinfo {year}
  {1998}{\natexlab{b}})},\ \Eprint {http://arxiv.org/abs/hep-th/9802150}
  {arXiv:hep-th/9802150} \BibitemShut {NoStop}%
\bibitem [{\citenamefont {Natsuume}(2015)}]{Natsuume:2014sfa}%
  \BibitemOpen
  \bibfield  {author} {\bibinfo {author} {\bibfnamefont {M.}~\bibnamefont
  {Natsuume}},\ }\href {\doibase 10.1007/978-4-431-55441-7} {\emph {\bibinfo
  {title} {{AdS/CFT Duality User Guide}}}},\ Vol.\ \bibinfo {volume} {903}\
  (\bibinfo {year} {2015})\ \Eprint {http://arxiv.org/abs/1409.3575}
  {arXiv:1409.3575 [hep-th]} \BibitemShut {NoStop}%
\bibitem [{\citenamefont {Landau}\ and\ \citenamefont
  {Lifshitz}(1980)}]{Landau:1980mil}%
  \BibitemOpen
  \bibfield  {author} {\bibinfo {author} {\bibfnamefont {L.~D.}\ \bibnamefont
  {Landau}}\ and\ \bibinfo {author} {\bibfnamefont {E.~M.}\ \bibnamefont
  {Lifshitz}},\ }\href@noop {} {\emph {\bibinfo {title} {{Statistical Physics,
  Part 1}}}},\ \bibinfo {series} {Course of Theoretical Physics}, Vol.~\bibinfo
  {volume} {5}\ (\bibinfo  {publisher} {Butterworth-Heinemann},\ \bibinfo
  {address} {Oxford},\ \bibinfo {year} {1980})\BibitemShut {NoStop}%
\bibitem [{\citenamefont {Cappiello}\ and\ \citenamefont
  {Mueck}(2001)}]{Cappiello:2001tf}%
  \BibitemOpen
  \bibfield  {author} {\bibinfo {author} {\bibfnamefont {L.}~\bibnamefont
  {Cappiello}}\ and\ \bibinfo {author} {\bibfnamefont {W.}~\bibnamefont
  {Mueck}},\ }\href {\doibase 10.1016/S0370-2693(01)01283-7} {\bibfield
  {journal} {\bibinfo  {journal} {Phys. Lett. B}\ }\textbf {\bibinfo {volume}
  {522}},\ \bibinfo {pages} {139} (\bibinfo {year} {2001})},\ \Eprint
  {http://arxiv.org/abs/hep-th/0107238} {arXiv:hep-th/0107238} \BibitemShut
  {NoStop}%
\bibitem [{\citenamefont {Basu}\ \emph {et~al.}(2010)\citenamefont {Basu},
  \citenamefont {Bhattacharya}, \citenamefont {Bhattacharyya}, \citenamefont
  {Loganayagam}, \citenamefont {Minwalla},\ and\ \citenamefont
  {Umesh}}]{Basu:2010uz}%
  \BibitemOpen
  \bibfield  {author} {\bibinfo {author} {\bibfnamefont {P.}~\bibnamefont
  {Basu}}, \bibinfo {author} {\bibfnamefont {J.}~\bibnamefont {Bhattacharya}},
  \bibinfo {author} {\bibfnamefont {S.}~\bibnamefont {Bhattacharyya}}, \bibinfo
  {author} {\bibfnamefont {R.}~\bibnamefont {Loganayagam}}, \bibinfo {author}
  {\bibfnamefont {S.}~\bibnamefont {Minwalla}}, \ and\ \bibinfo {author}
  {\bibfnamefont {V.}~\bibnamefont {Umesh}},\ }\href {\doibase
  10.1007/JHEP10(2010)045} {\bibfield  {journal} {\bibinfo  {journal} {JHEP}\
  }\textbf {\bibinfo {volume} {10}},\ \bibinfo {pages} {045} (\bibinfo {year}
  {2010})},\ \Eprint {http://arxiv.org/abs/1003.3232} {arXiv:1003.3232
  [hep-th]} \BibitemShut {NoStop}%
\bibitem [{\citenamefont {Bhattacharyya}\ \emph {et~al.}(2011)\citenamefont
  {Bhattacharyya}, \citenamefont {Minwalla},\ and\ \citenamefont
  {Papadodimas}}]{Bhattacharyya:2010yg}%
  \BibitemOpen
  \bibfield  {author} {\bibinfo {author} {\bibfnamefont {S.}~\bibnamefont
  {Bhattacharyya}}, \bibinfo {author} {\bibfnamefont {S.}~\bibnamefont
  {Minwalla}}, \ and\ \bibinfo {author} {\bibfnamefont {K.}~\bibnamefont
  {Papadodimas}},\ }\href {\doibase 10.1007/JHEP11(2011)035} {\bibfield
  {journal} {\bibinfo  {journal} {JHEP}\ }\textbf {\bibinfo {volume} {11}},\
  \bibinfo {pages} {035} (\bibinfo {year} {2011})},\ \Eprint
  {http://arxiv.org/abs/1005.1287} {arXiv:1005.1287 [hep-th]} \BibitemShut
  {NoStop}%
\bibitem [{\citenamefont {Markeviciute}\ and\ \citenamefont
  {Santos}(2016)}]{Markeviciute:2016ivy}%
  \BibitemOpen
  \bibfield  {author} {\bibinfo {author} {\bibfnamefont {J.}~\bibnamefont
  {Markeviciute}}\ and\ \bibinfo {author} {\bibfnamefont {J.~E.}\ \bibnamefont
  {Santos}},\ }\href {\doibase 10.1007/JHEP06(2016)096} {\bibfield  {journal}
  {\bibinfo  {journal} {JHEP}\ }\textbf {\bibinfo {volume} {06}},\ \bibinfo
  {pages} {096} (\bibinfo {year} {2016})},\ \Eprint
  {http://arxiv.org/abs/1602.03893} {arXiv:1602.03893 [hep-th]} \BibitemShut
  {NoStop}%
\bibitem [{\citenamefont {Markeviciute}\ and\ \citenamefont
  {Santos}(2019)}]{Markeviciute:2018yal}%
  \BibitemOpen
  \bibfield  {author} {\bibinfo {author} {\bibfnamefont {J.}~\bibnamefont
  {Markeviciute}}\ and\ \bibinfo {author} {\bibfnamefont {J.~E.}\ \bibnamefont
  {Santos}},\ }\href {\doibase 10.1088/1361-6382/aaf680} {\bibfield  {journal}
  {\bibinfo  {journal} {Class. Quant. Grav.}\ }\textbf {\bibinfo {volume}
  {36}},\ \bibinfo {pages} {02LT01} (\bibinfo {year} {2019})},\ \Eprint
  {http://arxiv.org/abs/1806.01849} {arXiv:1806.01849 [hep-th]} \BibitemShut
  {NoStop}%
\bibitem [{\citenamefont {Markeviciute}(2019)}]{Markeviciute:2018cqs}%
  \BibitemOpen
  \bibfield  {author} {\bibinfo {author} {\bibfnamefont {J.}~\bibnamefont
  {Markeviciute}},\ }\href {\doibase 10.1007/JHEP03(2019)110} {\bibfield
  {journal} {\bibinfo  {journal} {JHEP}\ }\textbf {\bibinfo {volume} {03}},\
  \bibinfo {pages} {110} (\bibinfo {year} {2019})},\ \Eprint
  {http://arxiv.org/abs/1809.04084} {arXiv:1809.04084 [hep-th]} \BibitemShut
  {NoStop}%
\bibitem [{\citenamefont {Dias}\ \emph {et~al.}(2025)\citenamefont {Dias},
  \citenamefont {Mitra},\ and\ \citenamefont {Santos}}]{Dias:2024edd}%
  \BibitemOpen
  \bibfield  {author} {\bibinfo {author} {\bibfnamefont {O.~J.~C.}\
  \bibnamefont {Dias}}, \bibinfo {author} {\bibfnamefont {P.}~\bibnamefont
  {Mitra}}, \ and\ \bibinfo {author} {\bibfnamefont {J.~E.}\ \bibnamefont
  {Santos}},\ }\href {\doibase 10.1007/JHEP06(2025)051} {\bibfield  {journal}
  {\bibinfo  {journal} {JHEP}\ }\textbf {\bibinfo {volume} {06}},\ \bibinfo
  {pages} {051} (\bibinfo {year} {2025})},\ \Eprint
  {http://arxiv.org/abs/2411.18712} {arXiv:2411.18712 [hep-th]} \BibitemShut
  {NoStop}%
\bibitem [{\citenamefont {Sen}(1995)}]{Sen:1995in}%
  \BibitemOpen
  \bibfield  {author} {\bibinfo {author} {\bibfnamefont {A.}~\bibnamefont
  {Sen}},\ }\href {\doibase 10.1142/S0217732395002234} {\bibfield  {journal}
  {\bibinfo  {journal} {Mod. Phys. Lett. A}\ }\textbf {\bibinfo {volume}
  {10}},\ \bibinfo {pages} {2081} (\bibinfo {year} {1995})},\ \Eprint
  {http://arxiv.org/abs/hep-th/9504147} {arXiv:hep-th/9504147} \BibitemShut
  {NoStop}%
\bibitem [{\citenamefont {Pestun}(2012)}]{Pestun:2007rz}%
  \BibitemOpen
  \bibfield  {author} {\bibinfo {author} {\bibfnamefont {V.}~\bibnamefont
  {Pestun}},\ }\href {\doibase 10.1007/s00220-012-1485-0} {\bibfield  {journal}
  {\bibinfo  {journal} {Commun. Math. Phys.}\ }\textbf {\bibinfo {volume}
  {313}},\ \bibinfo {pages} {71} (\bibinfo {year} {2012})},\ \Eprint
  {http://arxiv.org/abs/0712.2824} {arXiv:0712.2824 [hep-th]} \BibitemShut
  {NoStop}%
\bibitem [{\citenamefont {Nian}\ and\ \citenamefont
  {Pando~Zayas}(2020)}]{Nian:2019pxj}%
  \BibitemOpen
  \bibfield  {author} {\bibinfo {author} {\bibfnamefont {J.}~\bibnamefont
  {Nian}}\ and\ \bibinfo {author} {\bibfnamefont {L.~A.}\ \bibnamefont
  {Pando~Zayas}},\ }\href {\doibase 10.1007/JHEP03(2020)081} {\bibfield
  {journal} {\bibinfo  {journal} {JHEP}\ }\textbf {\bibinfo {volume} {03}},\
  \bibinfo {pages} {081} (\bibinfo {year} {2020})},\ \Eprint
  {http://arxiv.org/abs/1909.07943} {arXiv:1909.07943 [hep-th]} \BibitemShut
  {NoStop}%
\bibitem [{\citenamefont {Chong}\ \emph
  {et~al.}(2005{\natexlab{b}})\citenamefont {Chong}, \citenamefont {Cvetic},
  \citenamefont {Lu},\ and\ \citenamefont {Pope}}]{Chong:2005da}%
  \BibitemOpen
  \bibfield  {author} {\bibinfo {author} {\bibfnamefont {Z.~W.}\ \bibnamefont
  {Chong}}, \bibinfo {author} {\bibfnamefont {M.}~\bibnamefont {Cvetic}},
  \bibinfo {author} {\bibfnamefont {H.}~\bibnamefont {Lu}}, \ and\ \bibinfo
  {author} {\bibfnamefont {C.~N.}\ \bibnamefont {Pope}},\ }\href {\doibase
  10.1103/PhysRevD.72.041901} {\bibfield  {journal} {\bibinfo  {journal} {Phys.
  Rev. D}\ }\textbf {\bibinfo {volume} {72}},\ \bibinfo {pages} {041901}
  (\bibinfo {year} {2005}{\natexlab{b}})},\ \Eprint
  {http://arxiv.org/abs/hep-th/0505112} {arXiv:hep-th/0505112} \BibitemShut
  {NoStop}%
\bibitem [{\citenamefont {Kunduri}\ \emph {et~al.}(2006)\citenamefont
  {Kunduri}, \citenamefont {Lucietti},\ and\ \citenamefont
  {Reall}}]{Kunduri:2006ek}%
  \BibitemOpen
  \bibfield  {author} {\bibinfo {author} {\bibfnamefont {H.~K.}\ \bibnamefont
  {Kunduri}}, \bibinfo {author} {\bibfnamefont {J.}~\bibnamefont {Lucietti}}, \
  and\ \bibinfo {author} {\bibfnamefont {H.~S.}\ \bibnamefont {Reall}},\ }\href
  {\doibase 10.1088/1126-6708/2006/04/036} {\bibfield  {journal} {\bibinfo
  {journal} {JHEP}\ }\textbf {\bibinfo {volume} {04}},\ \bibinfo {pages} {036}
  (\bibinfo {year} {2006})},\ \Eprint {http://arxiv.org/abs/hep-th/0601156}
  {arXiv:hep-th/0601156} \BibitemShut {NoStop}%
\bibitem [{\citenamefont {'t~Hooft}(1978)}]{tHooft:1977nqb}%
  \BibitemOpen
  \bibfield  {author} {\bibinfo {author} {\bibfnamefont {G.}~\bibnamefont
  {'t~Hooft}},\ }\href {\doibase 10.1016/0550-3213(78)90153-0} {\bibfield
  {journal} {\bibinfo  {journal} {Nucl. Phys. B}\ }\textbf {\bibinfo {volume}
  {138}},\ \bibinfo {pages} {1} (\bibinfo {year} {1978})}\BibitemShut {NoStop}%
\bibitem [{\citenamefont {Polyakov}(1978)}]{Polyakov:1978vu}%
  \BibitemOpen
  \bibfield  {author} {\bibinfo {author} {\bibfnamefont {A.~M.}\ \bibnamefont
  {Polyakov}},\ }\href {\doibase 10.1016/0370-2693(78)90737-2} {\bibfield
  {journal} {\bibinfo  {journal} {Phys. Lett. B}\ }\textbf {\bibinfo {volume}
  {72}},\ \bibinfo {pages} {477} (\bibinfo {year} {1978})}\BibitemShut
  {NoStop}%
\bibitem [{\citenamefont {Susskind}(1979)}]{Susskind:1979up}%
  \BibitemOpen
  \bibfield  {author} {\bibinfo {author} {\bibfnamefont {L.}~\bibnamefont
  {Susskind}},\ }\href {\doibase 10.1103/PhysRevD.20.2610} {\bibfield
  {journal} {\bibinfo  {journal} {Phys. Rev. D}\ }\textbf {\bibinfo {volume}
  {20}},\ \bibinfo {pages} {2610} (\bibinfo {year} {1979})}\BibitemShut
  {NoStop}%
\bibitem [{\citenamefont {Maldacena}(1998)}]{Maldacena:1997re}%
  \BibitemOpen
  \bibfield  {author} {\bibinfo {author} {\bibfnamefont {J.~M.}\ \bibnamefont
  {Maldacena}},\ }\href {\doibase 10.4310/ATMP.1998.v2.n2.a1} {\bibfield
  {journal} {\bibinfo  {journal} {Adv. Theor. Math. Phys.}\ }\textbf {\bibinfo
  {volume} {2}},\ \bibinfo {pages} {231} (\bibinfo {year} {1998})},\ \Eprint
  {http://arxiv.org/abs/hep-th/9711200} {arXiv:hep-th/9711200} \BibitemShut
  {NoStop}%
\bibitem [{\citenamefont {Gubser}\ \emph {et~al.}(1998)\citenamefont {Gubser},
  \citenamefont {Klebanov},\ and\ \citenamefont {Polyakov}}]{Gubser:1998bc}%
  \BibitemOpen
  \bibfield  {author} {\bibinfo {author} {\bibfnamefont {S.~S.}\ \bibnamefont
  {Gubser}}, \bibinfo {author} {\bibfnamefont {I.~R.}\ \bibnamefont
  {Klebanov}}, \ and\ \bibinfo {author} {\bibfnamefont {A.~M.}\ \bibnamefont
  {Polyakov}},\ }\href {\doibase 10.1016/S0370-2693(98)00377-3} {\bibfield
  {journal} {\bibinfo  {journal} {Phys. Lett. B}\ }\textbf {\bibinfo {volume}
  {428}},\ \bibinfo {pages} {105} (\bibinfo {year} {1998})},\ \Eprint
  {http://arxiv.org/abs/hep-th/9802109} {arXiv:hep-th/9802109} \BibitemShut
  {NoStop}%
\bibitem [{\citenamefont {Alvarez-Gaume}\ \emph {et~al.}(2005)\citenamefont
  {Alvarez-Gaume}, \citenamefont {Gomez}, \citenamefont {Liu},\ and\
  \citenamefont {Wadia}}]{Alvarez-Gaume:2005dvb}%
  \BibitemOpen
  \bibfield  {author} {\bibinfo {author} {\bibfnamefont {L.}~\bibnamefont
  {Alvarez-Gaume}}, \bibinfo {author} {\bibfnamefont {C.}~\bibnamefont
  {Gomez}}, \bibinfo {author} {\bibfnamefont {H.}~\bibnamefont {Liu}}, \ and\
  \bibinfo {author} {\bibfnamefont {S.}~\bibnamefont {Wadia}},\ }\href
  {\doibase 10.1103/PhysRevD.71.124023} {\bibfield  {journal} {\bibinfo
  {journal} {Phys. Rev. D}\ }\textbf {\bibinfo {volume} {71}},\ \bibinfo
  {pages} {124023} (\bibinfo {year} {2005})},\ \Eprint
  {http://arxiv.org/abs/hep-th/0502227} {arXiv:hep-th/0502227} \BibitemShut
  {NoStop}%
\bibitem [{\citenamefont {Gross}\ and\ \citenamefont
  {Witten}(1980)}]{Gross:1980he}%
  \BibitemOpen
  \bibfield  {author} {\bibinfo {author} {\bibfnamefont {D.~J.}\ \bibnamefont
  {Gross}}\ and\ \bibinfo {author} {\bibfnamefont {E.}~\bibnamefont {Witten}},\
  }\href {\doibase 10.1103/PhysRevD.21.446} {\bibfield  {journal} {\bibinfo
  {journal} {Phys. Rev. D}\ }\textbf {\bibinfo {volume} {21}},\ \bibinfo
  {pages} {446} (\bibinfo {year} {1980})}\BibitemShut {NoStop}%
\bibitem [{\citenamefont {Basu}\ and\ \citenamefont
  {Wadia}(2006)}]{Basu:2005pj}%
  \BibitemOpen
  \bibfield  {author} {\bibinfo {author} {\bibfnamefont {P.}~\bibnamefont
  {Basu}}\ and\ \bibinfo {author} {\bibfnamefont {S.~R.}\ \bibnamefont
  {Wadia}},\ }\href {\doibase 10.1103/PhysRevD.73.045022} {\bibfield  {journal}
  {\bibinfo  {journal} {Phys. Rev. D}\ }\textbf {\bibinfo {volume} {73}},\
  \bibinfo {pages} {045022} (\bibinfo {year} {2006})},\ \Eprint
  {http://arxiv.org/abs/hep-th/0506203} {arXiv:hep-th/0506203} \BibitemShut
  {NoStop}%
\bibitem [{\citenamefont {Dey}\ \emph {et~al.}(2009)\citenamefont {Dey},
  \citenamefont {Mukherji}, \citenamefont {Mukhopadhyay},\ and\ \citenamefont
  {Sarkar}}]{Dey:2008bw}%
  \BibitemOpen
  \bibfield  {author} {\bibinfo {author} {\bibfnamefont {T.~K.}\ \bibnamefont
  {Dey}}, \bibinfo {author} {\bibfnamefont {S.}~\bibnamefont {Mukherji}},
  \bibinfo {author} {\bibfnamefont {S.}~\bibnamefont {Mukhopadhyay}}, \ and\
  \bibinfo {author} {\bibfnamefont {S.}~\bibnamefont {Sarkar}},\ }\href
  {\doibase 10.1142/S0217751X09046266} {\bibfield  {journal} {\bibinfo
  {journal} {Int. J. Mod. Phys. A}\ }\textbf {\bibinfo {volume} {24}},\
  \bibinfo {pages} {5235} (\bibinfo {year} {2009})},\ \Eprint
  {http://arxiv.org/abs/0806.4562} {arXiv:0806.4562 [hep-th]} \BibitemShut
  {NoStop}%
\bibitem [{\citenamefont {Chandrasekhar}\ \emph {et~al.}(2012)\citenamefont
  {Chandrasekhar}, \citenamefont {Mukherji}, \citenamefont {Sahay},\ and\
  \citenamefont {Sarkar}}]{Chandrasekhar:2012vh}%
  \BibitemOpen
  \bibfield  {author} {\bibinfo {author} {\bibfnamefont {B.}~\bibnamefont
  {Chandrasekhar}}, \bibinfo {author} {\bibfnamefont {S.}~\bibnamefont
  {Mukherji}}, \bibinfo {author} {\bibfnamefont {A.}~\bibnamefont {Sahay}}, \
  and\ \bibinfo {author} {\bibfnamefont {S.}~\bibnamefont {Sarkar}},\ }\href
  {\doibase 10.1007/JHEP05(2012)004} {\bibfield  {journal} {\bibinfo  {journal}
  {JHEP}\ }\textbf {\bibinfo {volume} {05}},\ \bibinfo {pages} {004} (\bibinfo
  {year} {2012})},\ \Eprint {http://arxiv.org/abs/1202.4059} {arXiv:1202.4059
  [hep-th]} \BibitemShut {NoStop}%
\bibitem [{\citenamefont {Yerra}\ \emph {et~al.}(2025)\citenamefont {Yerra},
  \citenamefont {Bhamidipati},\ and\ \citenamefont {Mukherji}}]{Yerra:2023wjo}%
  \BibitemOpen
  \bibfield  {author} {\bibinfo {author} {\bibfnamefont {P.~K.}\ \bibnamefont
  {Yerra}}, \bibinfo {author} {\bibfnamefont {C.}~\bibnamefont {Bhamidipati}},
  \ and\ \bibinfo {author} {\bibfnamefont {S.}~\bibnamefont {Mukherji}},\
  }\href {\doibase 10.1016/j.nuclphysb.2024.116783} {\bibfield  {journal}
  {\bibinfo  {journal} {Nucl. Phys. B}\ }\textbf {\bibinfo {volume} {1010}},\
  \bibinfo {pages} {116783} (\bibinfo {year} {2025})},\ \Eprint
  {http://arxiv.org/abs/2312.10783} {arXiv:2312.10783 [hep-th]} \BibitemShut
  {NoStop}%
\end{thebibliography}%
	
\end{document}